\documentclass[sigconf,nonacm]{acmart}

\usepackage{subfigure}
\usepackage{siunitx}
\usepackage{enumitem}

\newtheorem{theorem}{Conclusion}[section]
\newtheorem{lemma}[theorem]{Observation}

\usepackage{algorithm}
\usepackage[noend]{algpseudocode}
\let\ReturnInline\Return
\renewcommand{\Return}{\State\ReturnInline}
\algrenewcommand\algorithmicrequire{$\rhd$}
\algrenewcommand\algorithmicensure{$\square$}

\AtBeginDocument{%
  \providecommand\BibTeX{{%
    \normalfont B\kern-0.5em{\scshape i\kern-0.25em b}\kern-0.8em\TeX}}}

\setcopyright{acmcopyright}
\copyrightyear{2018}
\acmYear{2018}
\acmDOI{XXXXXXX.XXXXXXX}

\acmConference[Conference acronym 'XX]{Make sure to enter the correct
  conference title from your rights confirmation emai}{June 03--05,
  2018}{Woodstock, NY}

\newcommand{\ignore}[1]{}

\newcommand{\FroLou}{$\text{P-DF}_\text{L}$}

\begin{document}

\title{DF Louvain: Fast Incrementally Expanding Approach for Community Detection on Dynamic Graphs}


\author{Subhajit Sahu}
\email{subhajit.sahu@research.iiit.ac.in}
\affiliation{%
  \institution{IIIT Hyderabad}
  \streetaddress{Professor CR Rao Rd, Gachibowli}
  \city{Hyderabad}
  \state{Telangana}
  \country{India}
  \postcode{500032}
}


\settopmatter{printfolios=true}

\begin{abstract}
Community detection is the problem of recognizing natural divisions in networks. A relevant challenge in this problem is to find communities on rapidly evolving graphs. In this report we present our Parallel Dynamic Frontier (DF) Louvain algorithm, which given a batch update of edge deletions and insertions, incrementally identifies and processes an approximate set of affected vertices in the graph with minimal overhead, while using a novel approach of incrementally updating weighted-degrees of vertices and total edge weights of communities. We also present our parallel implementations of Naive-dynamic (ND) and Delta-screening (DS) Louvain. On a server with a 64-core AMD EPYC-7742 processor, our experiments show that DF Louvain obtains speedups of $179\times$, $7.2\times$, and $5.3\times$ on real-world dynamic graphs, compared to Static, ND, and DS Louvain, respectively, and is $183\times$, $13.8\times$, and $8.7\times$ faster, respectively, on large graphs with random batch updates. Moreover, DF Louvain improves its performance by $1.6\times$ for every doubling of threads.
\end{abstract}



\maketitle

\section{Introduction}
\label{sec:introduction}
Identifying hidden communities within networks is a crucial graph analytics problem that arises in various domains such as drug discovery, disease prediction, protein annotation, topic discovery, inferring land use, and criminal identification. Here, we want to identify groups of vertices that exhibit dense internal connections but sparse connections with the rest of the graph.\ignore{One of the difficulties in the community detection problem is the lack of apriori knowledge on the number and size distribution of communities.} These communities are intrinsic when identified based on network topology alone, without external attributes, and they are disjoint when each vertex belongs to only one community \cite{com-gregory10}. However, the problem is NP-hard, and there is a lack of apriori knowledge on the number and size distribution of communities \cite{com-blondel08}. To solve this issue, researchers have come up with a number of heuristics for finding communities \cite{com-guimera05, com-derenyi05, com-newman06, com-reichardt06, com-raghavan07, com-blondel08, com-rosvall08, infomap-rosvall09, com-fortunato10, com-gregory10, com-kloster14, com-come15, com-ruan15, com-newman16, com-ghoshal19, com-rita20, com-lu20, com-gupta22}. To measure the quality of communities identified, fitness metrics such as the modularity score proposed by Newman et al. \cite{com-newman06} are used.\ignore{Normalized Mutual Information index (NMI) \cite{com-jain17, com-chopade17}, and Jaccard Index \cite{com-jain17} are also employed.}

The \textit{Louvain method}, proposed by Blondel et al. \cite{com-blondel08}, is one of the most popular community detection algorithms \cite{com-lancichinetti09}. It is a greedy, modularity-based optimization algorithm, that hierarchically agglomerates vertices in a graph to obtain communities \cite{com-blondel08}. It has a time complexity of $O(KM)$\ignore{and an average time complexity of $\Theta (N \log N)$} (where $M$ represents the number of edges in the graph, $K$ the total number of iterations performed across all passes), and it efficiently identifies communities with resulting high modularity. A number of algorithmic improvements to the Louvain algorithm have been proposed \cite{com-rotta11, com-waltman13, com-gach14, com-ryu16, com-ozaki16, com-traag15, com-lu15, com-naim17, com-halappanavar17, com-ghosh18, com-traag19, com-shi21, com-zhang21, com-you22, com-aldabobi22}. To parallelize the algorithm on multicore CPUs \cite{staudt2015engineering, staudt2016networkit, com-fazlali17, com-halappanavar17, qie2022isolate}, GPUs \cite{com-naim17}, CPU-GPU hybrids \cite{com-bhowmik19, com-mohammadi20}, multi-GPUs \cite{com-cheong13, hricik2020using, chou2022batched, com-gawande22}, and multi-node systems \cite{com-ghosh18, ghosh2018scalable, sattar2022scalable, com-bhowmick22}, a number of strategies have been attempted \cite{com-cheong13, com-wickramaarachchi14, com-zeng15, com-que15, com-fazlali17, com-naim17, com-halappanavar17, com-ghosh18, com-bhowmik19, com-mohammadi20, com-shi21, com-bhowmick22}.

However, many real-world graphs rapidly evolve with time, through the insertion/deletion of edges/vertices. A growing number of research efforts have focused on detecting communities in dynamic networks \cite{com-aynaud10, com-chong13, com-shang14, com-aktunc15, com-yin16, com-held16, com-cordeiro16, com-meng16, com-zhuang19, com-zarayeneh21}. A suitable dynamic approach should identify and process only a subset of\ignore{(affected)} vertices that are likely to change their community membership. Note that if the subset of the graph identified as \textit{affected} is too small, we may end up with inaccurate communities, and if it is too large, we incur significant computation time. Hence, one should look to identify an approximate affected set of vertices that covers most of the true affected set of vertices as possible. In addition, determining the vertices to be processed should have low overhead \cite{incr-ramalingam96}.

\ignore{Further, real-world dynamic graphs are usually immense in scale, stemming from applications such as machine learning and social networks, and are gradually becoming ubiquitous.\ignore{With this data deluge, newer challenges are emerging. For efficiency reasons, one needs algorithms that update the results without re-computing from scratch. Such algorithms are known as \textit{dynamic algorithms}.} Parallel algorithms for graph analytics on dynamic graphs have thus become a subject of considerable research interest.\ignore{Examples of parallel dynamic algorithms include those for dynamic graph coloring \cite{color-yuan17, color-bhattacharya18}, maintaining shortest routes \cite{path-zhang17, path-khanda21}, and updating centrality scores \cite{cent-shao20, cent-regunta21}.}}

However, a critical examination of the extant literature on dynamic community detection algorithms indicates a few shortcomings. Some of these algorithms \cite{com-cordeiro16, com-meng16} do not outperform static algorithms even for modest-sized batch updates. Aynaud et al. \cite{com-aynaud10} and Chong et al. \cite{com-chong13} adapt the existing community labels and run an algorithm, such as Louvain, on the entire graph. Often, this is unwarranted since not every vertex would need to change its community on the insertion/deletion of a few edges. Cordeiro et al. \cite{com-cordeiro16} do not consider the cascading impact of changes in community labels, where the community label of a vertex changes because of a change in the community label of its neighbor. Zarayeneh et al. \cite{com-zarayeneh21} identify a subset of vertices whose community labels are likely to change on the insertion/deletion of a few edges. However, as this set of vertices identified is large, the algorithm of Zaranayeh et al. incurs a significant computation time. Moreover, most of the reported algorithms \cite{com-aynaud10, com-chong13, com-meng16, com-cordeiro16, com-zhuang19, com-zarayeneh21} are sequential. There is thus a pressing need for efficient parallel algorithms for community detection on large dynamic graphs. Further, none of the works recommend reusing the previous \textit{total edge weight} of each vertex/community (required for local-moving phase of Louvain algorithm) as auxiliary information to the dynamic algorithm. Recomputing it from scratch is expensive and becomes a bottleneck for dynamic Louvain algorithm. Table \ref{tab:compare-properties} summarizes the above discussion.

\ignore{The simplest approach for dynamic community detection is to use the community membership of vertices from the previous snapshot of the graph \cite{com-aynaud10, com-chong13, com-shang14, com-zhuang19} (which we call \textit{Naive-dynamic}). Alternatively, more advanced techniques have been employed to minimize computation by identifying a smaller subset of the graph that is affected by changes, such as moving only changed vertices \cite{com-aktunc15, com-yin16}, recomputing vertices close to an updated edge (below a given threshold distance) \cite{com-held16}, disbanding affected communities to lower-level network \cite{com-cordeiro16}, or using a dynamic modularity metric to compute community membership of vertices from scratch \cite{com-meng16}. \textit{Delta-Screening} (or \textit{$\Delta$-screening}) is a recently proposed technique that finds a subset of vertices impacted by changes in a graph using delta-modularity \cite{com-zarayeneh21}.}

\begin{table*}[hbtp]
  \centering
  \caption{A comparison of the properties of dynamic community detection approaches.}
  \label{tab:compare-properties}
  \begin{tabular}{|c|c||c|c|c|c|c|}
    \toprule
    \textbf{Approach} &
    \textbf{Published} &
    \textbf{Fully dynamic} &
    \textbf{Batch update} &
    \textbf{Process subset} &
    \textbf{Use auxiliary info.} &
    \textbf{Parallel algorithm} \\
    \midrule
    Aynaud et al. \cite{com-aynaud10} & 2010 & $\checkmark$ & $\checkmark$ & $\times$ & $\times$ & $\times$ \\ \hline
    Chong et al. \cite{com-chong13} & 2013 & $\checkmark$ & $\checkmark$ & $\times$ & $\times$ & $\times$ \\ \hline
    Meng et al. \cite{com-meng16} & 2016 & $\checkmark$ & $\checkmark$ & $\times$ & $\times$ & $\times$ \\ \hline
    Cordeiro et al. \cite{com-cordeiro16} & 2016 & $\checkmark$ & $\checkmark$ & $\times$ & $\times$ & $\times$ \\ \hline
    Zarayeneh et al. \cite{com-zarayeneh21} & 2021 & $\checkmark$ & $\checkmark$ & $\checkmark$ & $\times$ & $\times$ \\ \hline
    Ours & 2024 & $\checkmark$ & $\checkmark$ & $\checkmark$ & $\checkmark$ & $\checkmark$ \\ \hline
  \bottomrule
  \end{tabular}
\end{table*}

\ignore{\begin{table}[hbtp]
  \centering
  \caption{Speedup of our multicore implementation of Leiden algorithm compared to other state-of-the-art implementations. Direct comparisons entail running the given implementation on our server, while indirect comparisons (marked with a $*$, explained in Section \ref{sec:comparison-indirect}) involve comparing results relative to a common reference\ignore{(original Leiden)}.\ignore{Notably, the Leiden implementations vary in their classification, with some being multi-core and others multi-node.}}
  \label{tab:compare}
  \begin{tabular}{|c|c||c|}
    \toprule
    \textbf{Leiden implementation} &
    \textbf{Published} &
    \textbf{Our Speedup} \\
    \midrule
    Static \cite{sahu2023gvelouvain} & 2023 & $22\times$ \\ \hline
    Naive-dynamic \cite{csardi2006igraph} & 2006 & $\gg 50\times$ \\ \hline
    Delta-screening \cite{com-zarayeneh21} & 2021 & $\gg 20\times$ \\ \hline
    DynaMo \cite{com-zhuang19} & 2019 & $\gg 166\times^*$ \\ \hline
    Batch \cite{com-chong13} & 2013 & $\gg 22\times^*$ \\ \hline
  \bottomrule
  \end{tabular}
\end{table}
}

\ignore{\subsection{Our Contributions}}

This technical report introduces our Parallel Dynamic Frontier (DF) Louvain,\footnote{\url{https://github.com/puzzlef/louvain-communities-openmp-dynamic}} which given a batch update, efficiently identifies and processes affected vertices with minimal overhead, while incrementally updating weighted-degrees of vertices and total edge weights of communities, which significantly improves its performance.

\section{Related work}
\label{sec:related}
A core idea for dynamic community detection, among most approaches, is to use the community membership of each vertex from the previous snapshot of the graph, instead of initializing each vertex into singleton communities \cite{com-aynaud10, com-chong13, com-cordeiro16, com-zarayeneh21}. Aynaud et al. \cite{com-aynaud10} simply run the Louvain algorithm after assigning the community membership of each vertex as its previous community membership. Chong et al. \cite{com-chong13} reset the community membership of vertices linked to an updated edge, in addition to the steps performed by Aynaud et al., and process all vertices with the Louvain algorithm.

Meng et al. \cite{com-meng16} present a dynamic Louvain algorithm with an objective of obtaining temporally smoothed community structures as the graphs evolve over time. To avoid dramatic changes in community structure, they use an approximate version of delta-modularity optimization. This approximate formulation relies on both graphs in the previous and the current snapshot with a user-defined ratio. Their algorithm demonstrates improvement over Dynamic Spectral Clustering (DSC) \cite{chi2009evolutionary} and Multi-Step Community Detection (MSCD) \cite{aynaud2011multi} in terms of runtime. In terms of quality of communities obtained (using modularity score), their algorithm outperforms DSC, and is on par with MSCD.

Cordeiro et al. \cite{com-cordeiro16} propose a dynamic algorithm with a similar objective, i.e., tracking communities over time. Their algorithm performs a local modularity optimization that maximizes the modularity gain function only for those communities where the editing of nodes and edges was performed by disbanding such communities to a lower-level network, keeping the rest of the network unchanged. They confirm supremacy of their algorithm over LabelRank \cite{xie2013labelrank}, LabelRankT \cite{com-xie13}, Speaker–Listener Label Propagation (SLPA) \cite{com-xie11}, and Adaptive Finding Overlapping Community Structure (AFOCS) \cite{nguyen2011overlapping} in terms of runtime --- and over LabelRank, LabelRankT and AFOCS in terms of modularity score.

\ignore{Yin et al. \cite{com-yin16} propose an incremental community detection method based on modularity optimization for node-grained streaming networks. This method takes one vertex and its connecting edges as a processing unit, and equally treats edges involved by same node.}

\ignore{Zhuang et al. \cite{com-zhuang19} propose DynaMo, a modularity-based dynamic community detection algorithm, aimed to detect communities of dynamic networks as effective as repeatedly applying static algorithms but in a more efficient way. DynaMo is an adaptive and incremental algorithm, which is designed for incrementally maximizing the modularity gain while updating the community structure of dynamic networks.}

Zarayeneh et al. \cite{com-zarayeneh21} propose the Delta screening approach for updating communities in a dynamic graph. This technique examines edge deletions and insertions to the original graph, and identifies a subset of vertices that are likely to be impacted by the change using the modularity objective. Subsequently, only the identified subsets are processed for community state updates --- using two modularity optimization based community detection algorithms, Louvain and Smart Local-Moving (SLM). Their results demonstrate performance improvement over the Static Louvain algorithm, and a dynamic baseline version of Louvain algorithm \cite{com-aynaud10}. They also compare the performance of their algorithm against two other state-of-the-art methods DynaMo \cite{com-zhuang19} and Batch \cite{com-chong13}, and observe improvement over the methods both in terms of runtime and modularity.

\section{Preliminaries}
\label{sec:preliminaries}
Let $G(V, E, w)$ be an undirected graph, with $V$ as the set of vertices, $E$ as the set of edges, and $w_{ij} = w_{ji}$ a positive weight associated with each edge in the graph. If the graph is unweighted, we assume each edge to be associated with unit weight ($w_{ij} = 1$). Further, we denote the neighbors of each vertex $i$ as $J_i = \{j\ |\ (i, j) \in E\}$, the weighted degree of each vertex $i$ as $K_i = \sum_{j \in J_i} w_{ij}$, the total number of vertices in the graph as $N = |V|$, the total number of edges in the graph as $M = |E|$, and the sum of edge weights in the undirected graph as $m = \sum_{i, j \in V} w_{ij}/2$.

\subsection{Community detection}

Disjoint community detection is the process of arriving at a community membership mapping, $C: V \rightarrow \Gamma$, which maps each vertex $i \in V$ to a community-id $c \in \Gamma$, where $\Gamma$ is the set of community-ids. We denote the vertices of a community $c \in \Gamma$ as $V_c$. We denote the community that a vertex $i$ belongs to as $C_i$. Further, we denote the neighbors of vertex $i$ belonging to a community $c$ as $J_{i \rightarrow c} = \{j\ |\ j \in J_i\ and\ C_j = c\}$, the sum of those edge weights as $K_{i \rightarrow c} = \{w_{ij}\ |\ j \in J_{i \rightarrow c}\}$, the sum of weights of edges within a community $c$ as $\sigma_c = \sum_{(i, j) \in E\ and\ C_i = C_j = c} w_{ij}$, and the total edge weight of a community $c$ as $\Sigma_c = \sum_{(i, j) \in E\ \mbox{and}\ C_i = c} w_{ij}$ \cite{com-zarayeneh21, com-leskovec21}.

\subsection{Modularity}

Modularity is a fitness metric that is used to evaluate the quality of communities obtained by community detection algorithms (as they are heuristic based). It is calculated as the difference between the fraction of edges within communities and the expected fraction of edges if the edges were distributed randomly. It lies in the range $[-0.5, 1]$ (higher is better) \cite{com-brandes07}. Optimizing this function theoretically leads to the best possible grouping \cite{com-newman04, com-traag11}.

We can calculate the modularity $Q$ of obtained communities using Equation \ref{eq:modularity}, where $\delta$ is the Kronecker delta function ($\delta (x,y)=1$ if $x=y$, $0$ otherwise). The \textit{delta modularity} of moving a vertex $i$ from community $d$ to community $c$, denoted as $\Delta Q_{i: d \rightarrow c}$, can be calculated using Equation \ref{eq:delta-modularity}.

\begin{equation}
\label{eq:modularity}
  Q
  = \frac{1}{2m} \sum_{(i, j) \in E} \left[w_{ij} - \frac{K_i K_j}{2m}\right] \delta(C_i, C_j)
  = \sum_{c \in \Gamma} \left[\frac{\sigma_c}{2m} - \left(\frac{\Sigma_c}{2m}\right)^2\right]
\end{equation}

\begin{equation}
\label{eq:delta-modularity}
  \Delta Q_{i: d \rightarrow c}
  = \frac{1}{m} (K_{i \rightarrow c} - K_{i \rightarrow d}) - \frac{K_i}{2m^2} (K_i + \Sigma_c - \Sigma_d)
\end{equation}

\subsection{Louvain algorithm}
\label{sec:about-louvain}

The Louvain method is a greedy, modularity optimization based agglomerative algorithm that finds high quality communities within a graph, with a time complexity of $O (KM)$ (where $K$ is the number of iterations performed across all passes), and a space complexity of $O(N + M)$ \cite{com-lancichinetti09}. It consists of two phases: the \textit{local-moving phase}, where each vertex $i$ greedily decides to move to the community of one of its neighbors $j \in J_i$ that gives the greatest increase in modularity $\Delta Q_{i:C_i \rightarrow C_j}$ (using Equation \ref{eq:delta-modularity}), and the \textit{aggregation phase}, where all the vertices in a community are collapsed into a single super-vertex. The two phases make up one pass, which repeats until there is no further increase in modularity. As a result, we have a hierarchy of community memberships for each vertex as a dendrogram. The top-level hierarchy is the final result \cite{com-leskovec21}.

\subsection{Dynamic approaches}
\label{sec:dynamic-graphs}

A dynamic graph can be denoted as a sequence of graphs, where $G^t(V^t, E^t, w^t)$ denotes the graph at time step $t$. The changes between graphs $G^{t-1}(V^{t-1}, E^{t-1}, w^{t-1})$ and $G^t(V^t, E^t, w^t)$ at consecutive time steps $t-1$ and $t$ can be denoted as a batch update $\Delta^t$ at time step $t$ which consists of a set of edge deletions $\Delta^{t-} = \{(i, j)\ |\ i, j \in V\} = E^{t-1} \setminus E^t$ and a set of edge insertions $\Delta^{t+} = \{(i, j, w_{ij})\ |\ i, j \in V; w_{ij} > 0\} = E^t \setminus E^{t-1}$ \cite{com-zarayeneh21}. We refer to the setting where $\Delta^t$ consists of multiple edges being deleted and inserted as a \textit{batch update}.

\subsubsection{Naive-dynamic (ND) approach}
\label{sec:naive-dynamic}

The \textit{Naive-dynamic} approach, originally presented by Aynaud et al. \cite{com-aynaud10}, is a simple approach for identifying communities in dynamic networks. Here, one assigns vertices to communities from the previous snapshot of the graph and processes all the vertices, regardless of the edge deletions and insertions in the batch update (hence the prefix \textit{naive}). This is demonstrated in Figure \ref{fig:about-cases--naive}, where all vertices are marked as affected, highlighted in yellow. Since all communities are also marked as affected, they are all shown as hatched. Note that within the figure, edge deletions are shown in the top row (denoted by dotted lines), edge insertions are shown in the middle row (also denoted by dotted lines), and the migration of a vertex during the community detection algorithm is shown in the bottom row. The community membership obtained through this approach is guaranteed to be at least as accurate as the static algorithm. The psuedocode for our parallel version of this approach is given in Algorithm \ref{alg:naive}, with its explanation given in Section \ref{sec:our-naive}. It uses weighted-degrees of vertices and total edge weights of communities as auxiliary information.

\subsubsection{Delta-screening (DS) approach}
\label{sec:delta-screening}

\textit{Delta-screening}, proposed by Zarayeneh et al. \cite{com-zarayeneh21}, is a dynamic community detection approach that uses modularity-based scoring to determine an approximate region of the graph in which vertices are likely to change their community membership. Figure \ref{fig:about-cases--delta} presents a high-level overview of the vertices (and communities), linked to a single source vertex $i$, that are identified as affected using the DS approach in response to a batch update involving both edge deletions and insertions.\ignore{As mentioned above, in this figure, edge deletions are shown in the top row (denoted by dotted lines), edge insertions are shown in the middle row (also denoted by dotted lines), and the migration of a vertex during the community detection algorithm is shown in the bottom row. Further, vertices marked as affected are highlighted in yellow, while entire communities marked as affected are hatched (in addition to its vertices being highlighted in yellow).} In the DS approach, Zarayeneh et al. first separately sort the batch update consisting of edge deletions $(i, j) \in \Delta^{t-}$ and insertions $(i, j, w) \in \Delta^{t+}$ by their source vertex-id. For edge deletions within the same community, they mark $i$'s neighbors and $j$'s community as affected. For edge insertions across communities, they pick the highest modularity changing vertex $j*$ among all the insertions linked to vertex $i$ and mark $i$'s neighbors and $j*$'s community as affected. Edge deletions between different communities and edge insertions between the same community are unlikely to affect the community membership of either of the vertices or any other vertices and hence ignored. The affected vertices identified by the DS approach apply to the first pass of Louvain algorithm, and the community membership of each vertex is initialized at the start of the community detection algorithm to that obtained in the previous snapshot of the graph. We note that the DS approach, is \textit{not} guaranteed to explore all vertices that have the potential to change their membership \cite{com-zarayeneh21}.

The DS approach, proposed by Zarayeneh et al. \cite{com-zarayeneh21} is not parallel. In this paper, we translate their approach into a\ignore{multicore} parallel algorithm. To this end, we scan sorted edge deletions and insertions in parallel, apply the DS approach, as mentioned above, and mark vertices, neighbors of a vertex, and the community of a vertex using three separate flag vectors. Finally, we use the neighbors and community flag vectors to mark appropriate vertices. We also use per-thread collision-free hash tables. The psuedocode for our parallel version of the DS approach is given in Algorithm \ref{alg:delta}, with its explanation given in Section \ref{sec:our-delta}. Similar to our parallel implementation of ND Louvain, it utilizes the weighted-degrees of vertices and the total edge weights of communities as auxiliary information.

\section{Approach}
\label{sec:approach}
Given a batch update on the original graph, it is likely that only a small subset of vertices in the graph would change their community membership. Selection of the appropriate set of affected vertices to be processed (that are likely to change their community), in addition to the overhead of finding them, plays a significant role in the overall accuracy and efficiency of a dynamic batch parallel algorithm. Too small a subset may result in poor-quality communities, while a too-large subset will increase computation time. However, the Naive-dynamic (ND) approach processes all the vertices, while the Delta-screening (DS) approach generally overestimates the set of affected vertices and has a high overhead. Our proposed Dynamic Frontier (DF) approach addresses these issues.

\subsection{Our Dynamic Frontier (DF) approach}
\label{sec:frontier}

\begin{figure*}[hbtp]
  \centering
  \subfigure[Naive-dynamic (ND) approach]{
    \label{fig:about-cases--naive}
    \includegraphics[width=0.3\linewidth]{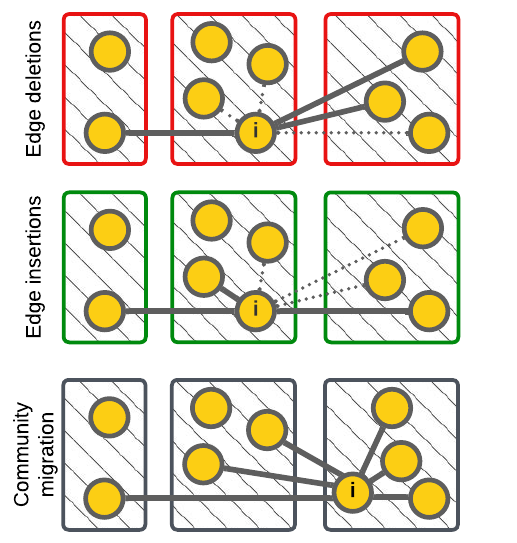}
  }
  \subfigure[Delta-screening (DS) approach]{
    \label{fig:about-cases--delta}
    \includegraphics[width=0.3\linewidth]{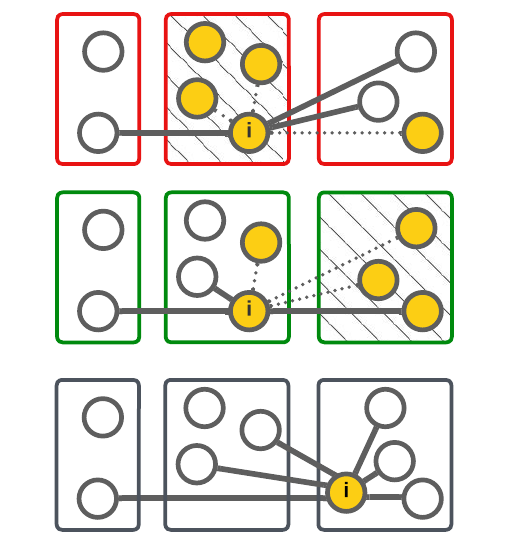}
  }
  \subfigure[Dynamic Frontier (DF) approach]{
    \label{fig:about-cases--frontier}
    \includegraphics[width=0.3\linewidth]{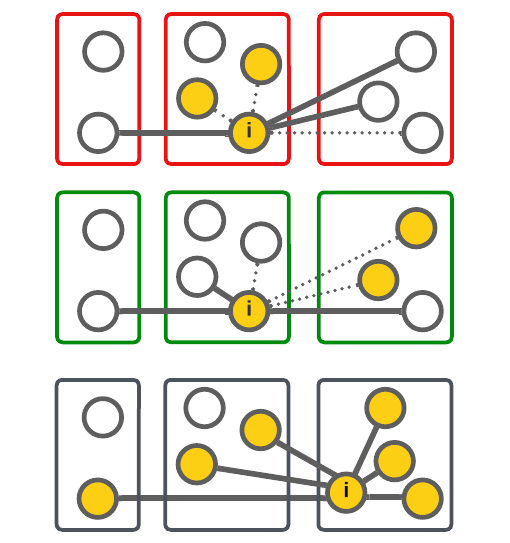}
  } \\[-2ex]
  \caption{Comparison of dynamic community detection approaches: \textit{Naive-dynamic (ND)}, \textit{Delta-screening (DS)}, and our \textit{Dynamic Frontier (DF)} approach. Edge deletions/insertions are indicated with dotted lines. Vertices marked as affected (initially) with each approach are highlighted in yellow, and when entire communities are marked as affected, they are hatched.}
  \label{fig:about-cases}
\end{figure*}

We now explain the \textit{Dynamic Frontier} approach. Consider a batch update consisting of edge deletions $(i, j, w) \in \Delta^{t-}$ and insertions $(i, j, w) \in \Delta^{t+}$, both shown with dotted lines, with respect to a single source vertex $i$, in Figure \ref{fig:about-cases--frontier}. At the start of the community detection algorithm, we initialize the community membership of each vertex to that obtained in the previous snapshot of the graph.

\begin{figure*}[!hbtp]
  \centering
  \subfigure[Initial communities]{
    \label{fig:frontier-example-01}
    \includegraphics[width=0.18\linewidth]{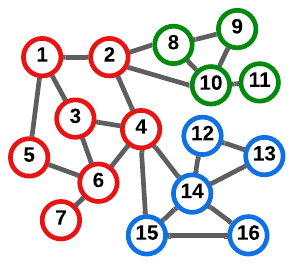}
  }
  \subfigure[Marking affected (initial)]{
    \label{fig:frontier-example-03}
    \includegraphics[width=0.18\linewidth]{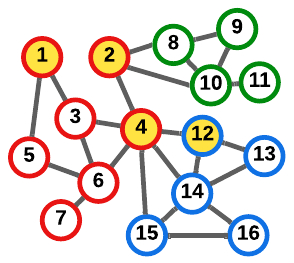}
  }
  \subfigure[After $1^{st}$ iteration]{
    \label{fig:frontier-example-04}
    \includegraphics[width=0.18\linewidth]{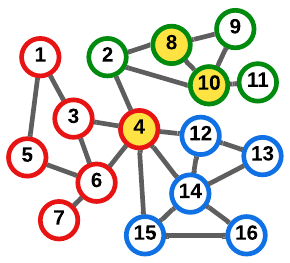}
  }
  \subfigure[After $2^{nd}$ iteration]{
    \label{fig:frontier-example-05}
    \includegraphics[width=0.18\linewidth]{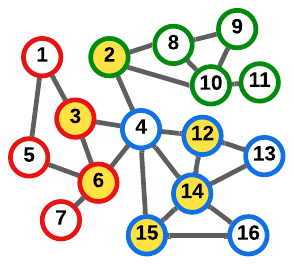}
  }
  \subfigure[Communities converged]{
    \label{fig:frontier-example-06}
    \includegraphics[width=0.18\linewidth]{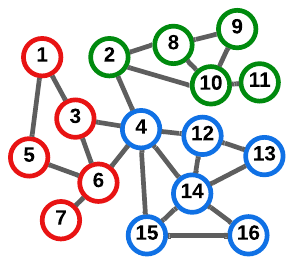}
  } \\[-2ex]
  \caption{An example explaining our \textit{Dynamic Frontier (DF)} approach. The community membership of each vertex is shown with border color (red, green, or blue), and the algorithm proceeds from left to right. A batch update arrives, affecting vertices $1$, $2$, $4$, and $12$. In the first iteration, vertex $2$ switches from red to green, impacting neighbors $8$ and $10$. In the second iteration, vertex $4$ changes from red to blue, affecting neighbors $3$, $6$, $12$, $14$, and $15$. Afterward, there are no more community changes.}
  \label{fig:frontier-example}
\end{figure*}

\paragraph{Initial marking of affected vertices upon edge deletion/insertion}

For edge deletions between vertices belonging to the same community and edge insertions between vertices belonging to different communities, we mark the source vertex $i$ as affected, as shown with vertices highlighted in yellow, in Figure \ref{fig:about-cases--frontier}. Note that batch updates are undirected, so we effectively mark both the endpoints $i$ and $j$. Edge deletions between vertices lying across communities and edge insertions for vertices lying within the same community are ignored (for reasons stated before, in Section \ref{sec:delta-screening}).

\paragraph{Incremental marking of affected vertices on vertex migration to another community}

When a vertex $i$ changes its community during the community detection algorithm (shown by moving $i$ from its original community in the center to its new community on the right), we mark all its neighbor vertices $j \in J_i$ as affected, as shown in Figure \ref{fig:about-cases--frontier} (highlighted in yellow), and mark $i$ as not affected. If $i$ does not change its community, it is also marked as unaffected --- this is known as vertex pruning optimization \cite{com-ozaki16, com-ryu16, com-shi21, com-zhang21}. The process is akin to a graph traversal and continues until the community assignments of the vertices have converged.

\paragraph{Application to the first pass of Louvain algorithm}

We apply the DF approach to the first pass of Louvain algorithm\ignore{(see line \ref{alg:louvain--remark-pass} in Algorithm \ref{alg:louvain})}, as with the DS approach. In subsequent passes\ignore{, if the aggregation tolerance condition is not met (line \ref{alg:louvain--aggregation-tolerance} in Algorithm \ref{alg:louvain}),} all super-vertices are\ignore{marked as affected and} processed, according to the Louvain algorithm. This takes less than $15\%$ of total time, on average, so we don't use the DF approach to find affected super-vertices.\ignore{The tolerance condition only fails in the case of large batch updates.} The psuedocode for our parallel DF Louvain is given in Algorithm \ref{alg:frontier}, with its explanation in Section \ref{sec:our-frontier}. Similar to our parallel ND/DS Louvain, it utilizes\ignore{weighted-degrees of vertices and total edge weights of communities as} auxiliary information.

\subsubsection{An example of DF approach}

See Figure \ref{fig:frontier-example}\ignore{shows an example of our DF approach}.

\paragraph{Initial communities}

The original graph consists of $16$ vertices, which are divided into three communities, distinguished by the border colors of \textit{red}, \textit{green}, and \textit{blue} (see Figure \ref{fig:frontier-example-01}). This community membership\ignore{information} of each vertex could have been obtained by executing either the static or dynamic version of Louvain algorithm.

\paragraph{Batch update and marking affected (initial)}

Subsequently a batch update is applied to the original graph (see Figure \ref{fig:frontier-example-03}), involving and edge deletion between vertices $1$ and $2$, and an insertion vertices $4$ and $12$. Following the batch update, we perform the initial step of the DF approach, marking endpoints $1$, $2$, $4$, and $12$ as affected. Now, we are ready to execute the first iteration of Louvain algorithm.

\paragraph{After first iteration}

During the first iteration (see Figure \ref{fig:frontier-example-04}), the community membership of vertex $2$ changes from \textit{red} to \textit{green} because it exhibits stronger connections with vertices in the \textit{green} community. In response to this change, the DF approach incrementally marks the neighboring vertices of $2$ as affected, specifically vertices $8$ and $10$. Vertex $2$ is no longer\ignore{marked as} affected due to pruning.

\paragraph{After second iteration}

Let us now consider the second iteration (see Figure \ref{fig:frontier-example-05}). Vertex $4$ is now more strongly connected to the \textit{blue} community, resulting in a change of its community membership from \textit{red} to \textit{blue}. As before, we mark the neighbors of vertex $4$ as affected, namely vertices $12$, $14$, and $15$. Vertex $4$, once again, no longer marked as affected due to vertex pruning.

\paragraph{Communities converged}

In the subsequent iteration (see Figure \ref{fig:frontier-example-06}), no other vertices have a strong enough reason to change their community membership. At this point\ignore{, when employing the Louvain algorithm}, the aggregation phase commences, consolidating communities into super-vertices to prepare for the subsequent pass of Louvain algorithm.

\subsection{Utilizing Auxiliary information}
\label{sec:auxiliary}

\begin{figure}[hbtp]
  \centering
  \subfigure{
    \label{fig:about-auxiliary--with}
    \includegraphics[width=0.98\linewidth]{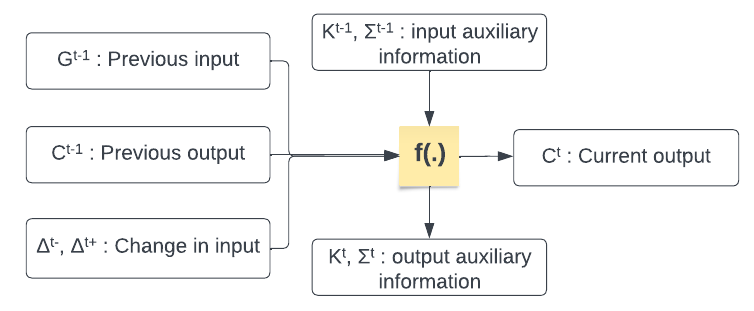}
  } \\[-2ex]
  \caption{A dynamic community detection algorithm $f(.)$ accepts as input the previous graph $G^{t-1}$, community memberships $C^{t-1}$, and the batch update $\Delta^{t-}$, $\Delta^{t+}$, and returns the updated community memberships $C^t$. However\ignore{, as shown in (b)}, it may also accept weighted degree of vertices $K^{t-1}$ and total edge weights of communities $\Sigma^{t-1}$ as auxiliary information, and generate updated auxiliary information $K^t$, $\Sigma^t$.}
  \label{fig:about-auxiliary}
\end{figure}

We note that computing the weighted-degree of each vertex $K^t$ and the total edge weight of each community $\Sigma^t$ incurs considerable runtime, in comparison to the time required for the local-moving and aggregation phases of the Louvain algorithm\ignore{to converge, especially for small batch updates}. It would be more efficient to incrementally update the previous weighted-degrees of vertices $K^{t-1}$ and total edge weights of communities $\Sigma^{t-1}$ by taking into account edge deletions $\Delta^{t-}$ and insertions $\Delta^{t+}$ within the batch update, instead of recomputing from scratch. We refer to $K^t$ and $\Sigma^t$ (and $K^{t-1}$, $\Sigma^{t-1}$) as auxiliary information as they must be maintained by the dynamic algorithm, but do not represent the output of the algorithm. Figure \ref{fig:about-auxiliary} illustrates this concept.

\begin{figure}[hbtp]
  \centering
  \subfigure{
    \label{fig:adjust-auxiliary--8020}
    \includegraphics[width=0.98\linewidth]{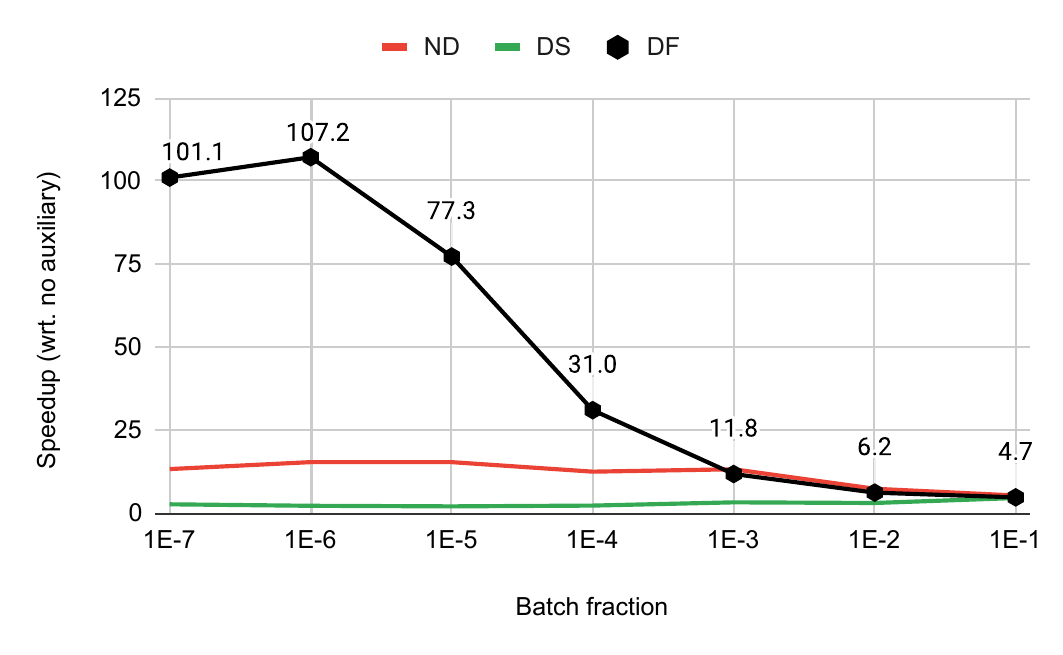}
  } \\[-2ex]
  \caption{Speedup of \textit{Naive-dynamic (ND)}, \textit{Delta-screening (DS)}, and \textit{Dynamic Frontier (DF)} Louvain when reusing the previous \textit{weighted-degrees of vertices} and \textit{total edge weight of communities} as auxiliary information to the dynamic algorithm, compared to the same dynamic algorithm when both are recomputed from scratch. This is done on large graphs with generated random batch updates of size $10^{-7} |E|$ to $0.1 |E|$\ignore{, consisting of $80\%$ edge insertions and $20\%$ deletions, to simulate realistic dynamic graph updates}.}
  \label{fig:adjust-auxiliary}
\end{figure}

In Figure \ref{fig:adjust-auxiliary}, we present the mean speedup observed for ND, DS, and DF Louvain when making use of auxiliary information $K^{t-1}$ and $\Sigma^{t-1}$, in contrast to the same dynamic algorithm when calculating from scratch. This is done on graphs from Table \ref{tab:dataset-large} with random batch updates of size $10^{-7} |E|$ to $0.1 |E|$, consisting of $80\%$ edge insertions and $20\%$ edge deletions, to simulate realistic dynamic graph updates. Results indicate that employing auxiliary information enables ND, DS, and DF Louvain to achieve average speedups of $11.8\times$, $2.9\times$, and $48.5\times$, respectively. Moreover, DF Louvain achieves remarkable speedups, reaching up to $107\times$ for smaller batch sizes. Incrementally updating $K^{t-1}$ and $\Sigma^{t-1}$ to obtain $K^t$ and $\Sigma^t$, thus, significantly speeds up DF Louvain. To the best of our knowledge, none of the existing dynamic algorithms for Louvain algorithm make such use of auxiliary information.

\subsection{Our DF Louvain implementation}
\label{sec:our-frontier}

Algorithm \ref{alg:frontier} shows the psuedocode of our Parallel Dynamic Frontier (DF) Louvain. It takes as input the previous $G^{t-1}$ and current graph snapshot $G^t$, edge deletions $\Delta^{t-}$ and insertions $\Delta^{t+}$ in the batch update, the previous community assignments $C^{t-1}$ for each vertex, the previous weighted-degrees $K^{t-1}$ of vertices, and the previous total edge weights $\Sigma^{t-1}$ of communities. It returns the updated community memberships $C^t$ of vertices, weighted-degrees $K^t$, and total edge weights $\Sigma^t$ of communities.

\begin{algorithm}[hbtp]
\caption{Our Parallel \textit{Dynamic Frontier (DF)} Louvain.}
\label{alg:frontier}
\begin{algorithmic}[1]
\Require{$G^{t-1}, G^t$: Previous, current input graph}
\Require{$\Delta^{t-}, \Delta^{t+}$: Edge deletions and insertions (batch update)}
\Require{$C^{t-1}, C^t$: Previous, current community of each vertex}
\Require{$K^{t-1}, K^t$: Previous, current weighted-degree of vertices}
\Require{$\Sigma^{t-1}, \Sigma^t$: Previous, current total edge weight of communities}
\Ensure{$\delta V$: Flag vector indicating if each vertex is affected}
\Ensure{$isAffected(i)$: Is vertex $i$ is marked as affected?}
\Ensure{$inAffectedRange(i)$: Can $i$ be incrementally marked?}
\Ensure{$onChange(i)$: What happens if $i$ changes its community?}
\Ensure{$F$: Lambda functions passed to parallel Louvain (Alg. \ref{alg:louvain})}

\Statex

\Function{dynamicFrontier}{$G^{t-1}, G^t, \Delta^{t-}, \Delta^{t+}, C^{t-1}, K^{t-1}, \Sigma^{t-1}$}
  \State $\rhd$ Mark initial affected vertices
  \ForAll{$(i, j) \in \Delta^{t-}$ \textbf{in parallel}} \label{alg:frontier--loopdel-begin}
    \If{$C^{t-1}[i] = C^{t-1}[j]$} $\delta V[i] \gets 1$
    \EndIf
  \EndFor \label{alg:frontier--loopdel-end}
  \ForAll{$(i, j, w) \in \Delta^{t+}$ \textbf{in parallel}} \label{alg:frontier--loopins-begin}
    \If{$C^{t-1}[i] \neq C^{t-1}[j]$} $\delta V[i] \gets 1$
    \EndIf
  \EndFor \label{alg:frontier--loopins-end}
  \Function{isAffected}{$i$} \label{alg:frontier--isaff-begin}
    \Return{$\delta V[i]$}
  \EndFunction \label{alg:frontier--isaff-end}
  \Function{inAffectedRange}{$i$} \label{alg:frontier--isaffrng-begin}
    \Return{$1$}
  \EndFunction \label{alg:frontier--isaffrng-end}
  \Function{onChange}{$i$} \label{alg:frontier--onchg-begin}
    \ForAll{$j \in G^t.neighbors(i)$} $\delta V[j] \gets 1$
    \EndFor
  \EndFunction \label{alg:frontier--onchg-end}
  \State $F \gets \{isAffected, inAffectedRange, onChange\}$ \label{alg:frontier--lambdas}
  \State $\rhd$ Use $K^{t-1}$, $\Sigma^{t-1}$ as auxiliary information (Alg. \ref{alg:update})
  \State $\{K^t, \Sigma^t\} \gets updateWeights(G^t, \Delta^{t-}, \Delta^{t+}, C^{t-1}, K^{t-1}, \Sigma^{t-1})$\label{alg:frontier--auxiliary}
  \State $\rhd$ Obtain updated communities (Alg. \ref{alg:louvain})
  \State $C^t \gets louvain(G^t, C^{t-1}, K^t, \Sigma^t, F)$ \label{alg:frontier--louvain}
  \Return{$\{C^t, K^t, \Sigma^t\}$} \label{alg:frontier--return}
\EndFunction
\end{algorithmic}
\end{algorithm}

In the algorithm, we first identify an initial set of affected vertices, whose communities may directly change due to the batch updates, by marking them in the flag vector $\delta V$. We do this by marking the endpoints of edge deletions $\Delta^{t-}$ which lie in the same community (lines \ref{alg:frontier--loopdel-begin}-\ref{alg:frontier--loopdel-end}), and by marking the endpoints of edge insertions $\Delta^{t+}$ which lie in disjoint communities (lines \ref{alg:frontier--loopins-begin}-\ref{alg:frontier--loopins-end}). We then define three lambda functions for the Louvain algorithm, \texttt{isAffected()} (lines \ref{alg:frontier--isaff-begin}-\ref{alg:frontier--isaff-end}), \texttt{inAffectedRange()} (lines \ref{alg:frontier--isaffrng-begin}-\ref{alg:frontier--isaffrng-end}), and \texttt{onChange()} (lines \ref{alg:frontier--onchg-begin}-\ref{alg:frontier--onchg-end}), which indicate that a set of vertices are (initially) marked as affected, that all vertices in the graph can be incrementally marked as affected, and that the neighbors of a vertex are marked as affected if it changes its community membership, respectively. Note that the set of affected vertices will expand automatically due to vertex pruning optimization used in our Parallel Louvain algorithm (Algorithm \ref{alg:louvain}). Thus, \texttt{onChange()} reflects what the DF approach would do in the absence of vertex pruning. Further, unlike existing approaches, we leverage $K^{t-1}$ and $\Sigma^{t-1}$, alongside the batch updates $\Delta^{t-}$ and $\Delta^{t+}$, to efficiently compute $K^t$ and $\Sigma^t$ required for the local-moving phase of the Louvain algorithm (line \ref{alg:frontier--auxiliary}). These lambda functions and the total vertex/edge weights are then utilized to execute the Louvain algorithm and obtain the updated community assignments $C^t$ (line \ref{alg:frontier--louvain}). Finally, we return $C^t$, alongside $K^t$ and $\Sigma^t$\ignore{, serving as the updated auxiliary information} (line \ref{alg:frontier--return}).

\ignore{\subsection{Time and Space complexity}}
\ignore{\label{sec:complexity}}

\ignore{To discuss the time complexity of DF Louvain, we use $N_B$ to denote the number of vertices marked as affected (which is dependent on the size and nature of batch update) by it on a batch $B$ of edge updates, $M_B$ to denote the number of edges with one endpoint in $N_B$, and $K$ to denote the total number of iterations performed. Then, the time complexity of DF Louvain is $O(K M_B)$.\ignore{In the worst case, the time complexity of our algorithms would be the same as that of the respective static algorithms, i.e., $O(KM)$.} The space complexity of DF Louvain is the same as that of Static Louvain, i.e., $O(N + M)$.}

\section{Evaluation}
\label{sec:evaluation}
\subsection{Experimental setup}
\label{sec:setup}

\subsubsection{System}
\label{sec:system}

For our experiments, we use a server that has an x86-based 64 core AMD EPYC-7742 processor. This processor has a clock frequency of $2.25$ GHz and $512$ GB of DDR4 system memory. Each core has an L1 cache of $4$ MB, an L2 cache of $32$ MB, and a shared L3 cache of $256$ MB. The machine uses Ubuntu 20.04.

\subsubsection{Configuration}
\label{sec:configuration}

We use 32-bit unsigned integer for vertex IDs, 32-bit floating point for edge weights, but use 64-bit floating point for hashtable values, total edge weight, modularity calculation, and all places where performing an aggregation/sum of floating point values. Affected vertices are represented with an 8-bit integer vector. Computing the weighted degree of each vertex, the local moving phase, and aggregating the edges for the super-vertex graph, employ OpenMP's \textit{dynamic schedule} with a chunk size of $2048$ for dynamic workload balancing among threads. We set the iteration tolerance $\tau$ to $10^{-2}$, the tolerance drop per pass $TOLERANCE\_DECLINE\_FACTOR$ to $10$ (threshold-scaling optimization), maximum number of iterations per pass $MAX\_ITERATI$ $ONS$ to $20$, and the maximum number of passes $MAX\_PASSES$ to $10$ \cite{sahu2023gvelouvain}. Further, we set the aggregation tolerance $\tau_{agg}$ to $0.8$ for large (static) graphs with generated random batch updates, but keep it disabled, i.e., set $\tau_{agg}$ to $1$, for real-world dynamic graphs. Unless mentioned otherwise, we execute all parallel implementations with a default of $64$ threads (to match the number of cores available on the system). We use GCC 9.4 and OpenMP 5.0 \cite{openmp18} for compilation.

\subsubsection{Dataset}
\label{sec:dataset}

To experiment with real-world dynamic graphs, we employ five temporal networks sourced from the Stanford Large Network Dataset Collection \cite{snapnets}, detailed in Table \ref{tab:dataset}. Here, the number of vertices range from $24.8$ thousand to $2.60$ million, temporal edges from $507$ thousand to $63.4$ million, and static edges from $240$ thousand to $36.2$ million. For experiments involving large static graphs with random batch updates, we utilize $12$ graphs as listed in Table \ref{tab:dataset-large}, obtained from the SuiteSparse Matrix Collection \cite{suite19}. Here, the number of vertices in the graphs varies from $3.07$ to $214$ million, and the number of edges varies from $25.4$ million to $3.80$ billion. With each graph, we ensure that all edges are undirected and weighted with a default weight of $1$.

\begin{table}[hbtp]
  \centering
  \caption{List of $5$ real-world dynamic graphs\ignore{, i.e., temporal networks}, sourced from the Stanford Large Network Dataset Collection \cite{snapnets}. Here, $|V|$ denotes the number of vertices, $|E_T|$ represents the count of temporal edges (inclusive of duplicates), and $|E|$ indicates the number of static edges (without duplicates).}
  \label{tab:dataset}
  \begin{tabular}{|c||c|c|c|c|}
    \toprule
    \textbf{Graph} &
    \textbf{\textbf{$|V|$}} &
    \textbf{\textbf{$|E_T|$}} &
    \textbf{\textbf{$|E|$}} \\
    \midrule
    sx-mathoverflow & 24.8K & 507K & 240K \\ \hline
    sx-askubuntu & 159K & 964K & 597K \\ \hline
    sx-superuser & 194K & 1.44M & 925K \\ \hline
    wiki-talk-temporal & 1.14M & 7.83M & 3.31M \\ \hline
    sx-stackoverflow & 2.60M & 63.4M & 36.2M \\ \hline
  \bottomrule
  \end{tabular}
\end{table}

\begin{table}[hbtp]
  \centering
  \caption{List of $12$ graphs obtained from the SuiteSparse Matrix Collection \cite{suite19} (directed graphs are marked with $*$). Here, $|V|$ is the total number of vertices, $|E|$ is the total number of edges (after making the graph undirected), and $|\Gamma|$ is the number of communities obtained using \textit{Static Louvain} \cite{sahu2023gvelouvain}.}
  \label{tab:dataset-large}
  \begin{tabular}{|c||c|c|c|}
    \toprule
    \textbf{Graph} &
    \textbf{\textbf{$|V|$}} &
    \textbf{\textbf{$|E|$}} &
    \textbf{\textbf{$|\Gamma|$}} \\
    \midrule
    \multicolumn{4}{|c|}{\textbf{Web Graphs (LAW)}} \\ \hline
    indochina-2004$^*$ & 7.41M & 341M & 4.24K \\ \hline
    arabic-2005$^*$ & 22.7M & 1.21B & 3.66K \\ \hline
    uk-2005$^*$ & 39.5M & 1.73B & 20.8K \\ \hline
    webbase-2001$^*$ & 118M & 1.89B & 2.76M \\ \hline
    it-2004$^*$ & 41.3M & 2.19B & 5.28K \\ \hline
    sk-2005$^*$ & 50.6M & 3.80B & 3.47K \\ \hline
    \multicolumn{4}{|c|}{\textbf{Social Networks (SNAP)}} \\ \hline
    com-LiveJournal & 4.00M & 69.4M & 2.54K \\ \hline
    com-Orkut & 3.07M & 234M & 29 \\ \hline
    \multicolumn{4}{|c|}{\textbf{Road Networks (DIMACS10)}} \\ \hline
    asia\_osm & 12.0M & 25.4M & 2.38K \\ \hline
    europe\_osm & 50.9M & 108M & 3.05K \\ \hline
    \multicolumn{4}{|c|}{\textbf{Protein k-mer Graphs (GenBank)}} \\ \hline
    kmer\_A2a & 171M & 361M & 21.2K \\ \hline
    kmer\_V1r & 214M & 465M & 6.17K \\ \hline
  \bottomrule
  \end{tabular}
\end{table}

\subsubsection{Batch generation}
\label{sec:batch-generation}

For experiments involving real-world dynamic graphs, we first load $90\%$ of the temporal edges of each graph from Table \ref{tab:dataset}, and ensure that all edges are weighted with a default weight of $1$, and that they are undirected by adding the reverse edges. Subsequently, we load $B$ edges in $100$ batch updates. Here, $B$ denotes the desired batch size, specified as a fraction of the total number of temporal edges $|E_T|$ in the graph, and ensure that the batch update is undirected. For experiments on large graphs with random batch updates, we take each base graph from Table \ref{tab:dataset-large} and generate random batch updates \cite{com-zarayeneh21} comprising an $80\% : 20\%$ mix of edge insertions and deletions to emulate realistic batch updates, each with an edge weight of $1$. To prepare the set of edges for insertion, we select vertex pairs with equal probability. For edge deletions, we remove the desired number of existing edges, with equal probability of selecting each edge. To simplify, we ensure no new vertices are added or removed from the graph. The batch size is measured as a fraction of edges in the original graph, ranging from $10^{-7}$ to $0.1$ (i.e., $10^{-7}|E|$ to $0.1|E|$), with multiple batches generated for each size for averaging. For a billion-edge graph, this amounts to a batch size of $100$ to $100$ million edges.\ignore{Keep in mind that dynamic graph algorithms are helpful for small batch sizes in interactive applications. For large batches, it is usually more efficient to run the static algorithm.} All batch updates are undirected, i.e., for every edge insertion $(i, j, w)$ in the batch update, the edge $(j, i, w)$ is also a part of the batch update. We employ five distinct random batch updates for each batch size, and report average across these runs in our experiments.

\subsubsection{Measurement}
\label{sec:measurement}

We evaluate the runtime of each approach on the entire updated graph, including the local-moving phase, aggregation phase, the initial and incremental marking of affected vertices, convergence detection, and all the necessary intermediary steps, but excluding only the initial memory allocation/deallocation time. We assume that the total edge weight of the graphs is known, and can be tracked upon each batch update.

\begin{figure*}[!hbt]
  \centering
  \subfigure[Overall Runtime]{
    \label{fig:temporal-summary--runtime-overall}
    \includegraphics[width=0.48\linewidth]{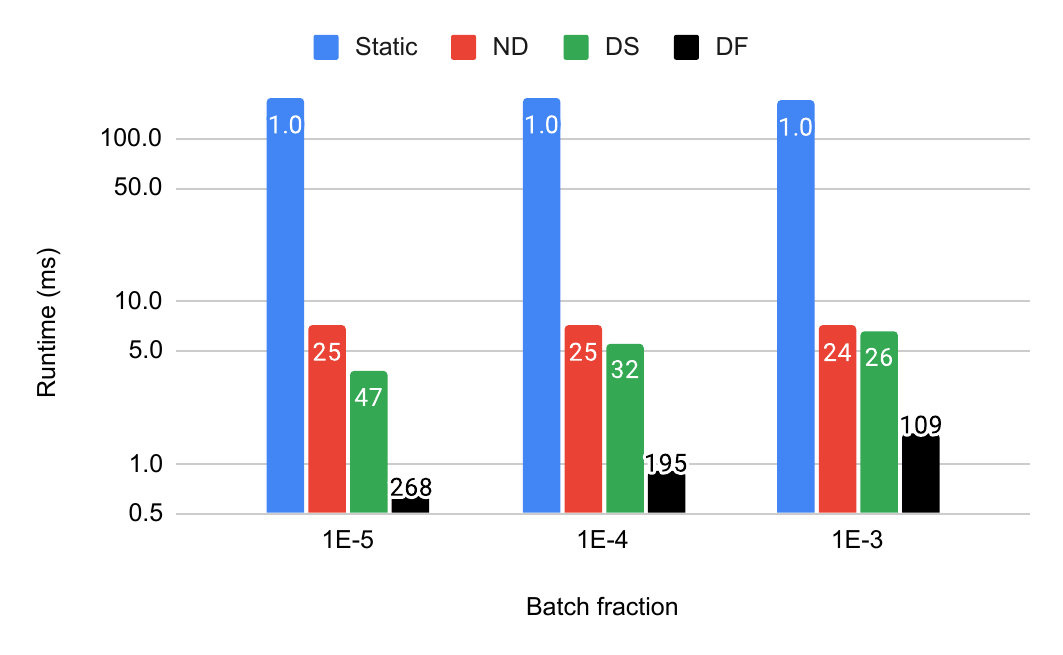}
  }
  \subfigure[Overall Modularity of communities obtained]{
    \label{fig:temporal-summary--modularity-overall}
    \includegraphics[width=0.48\linewidth]{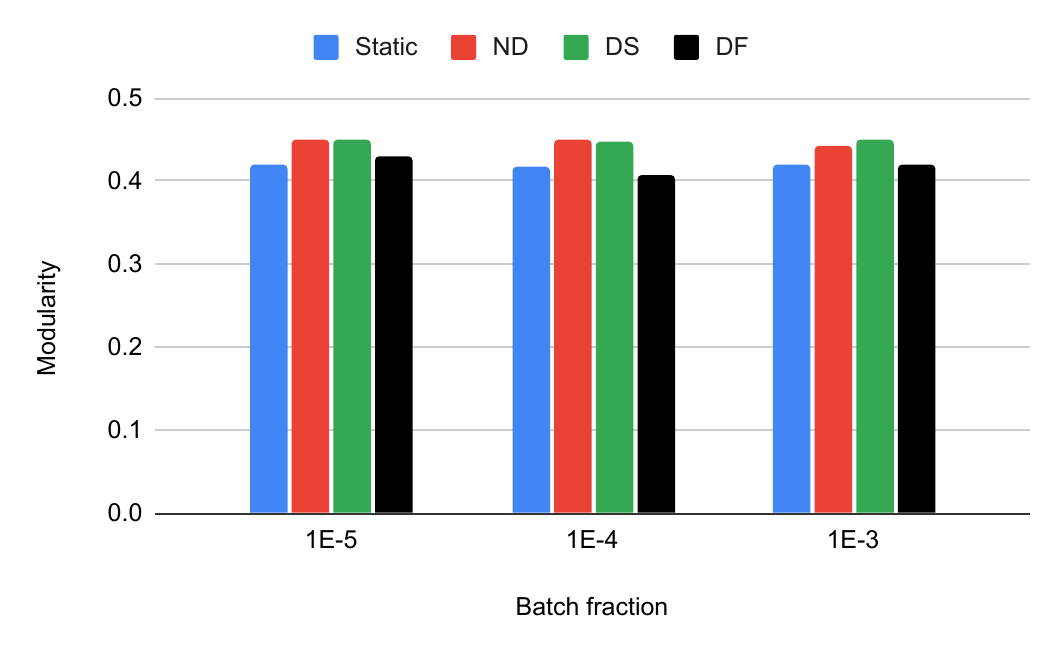}
  } \\[2ex]
  \includegraphics[width=0.48\linewidth]{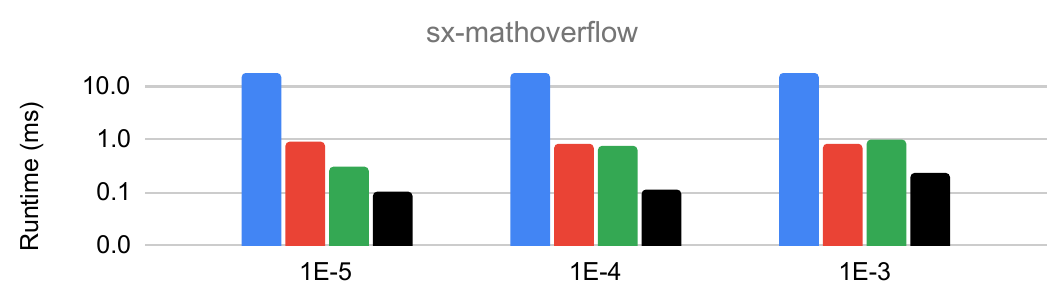}
  \includegraphics[width=0.48\linewidth]{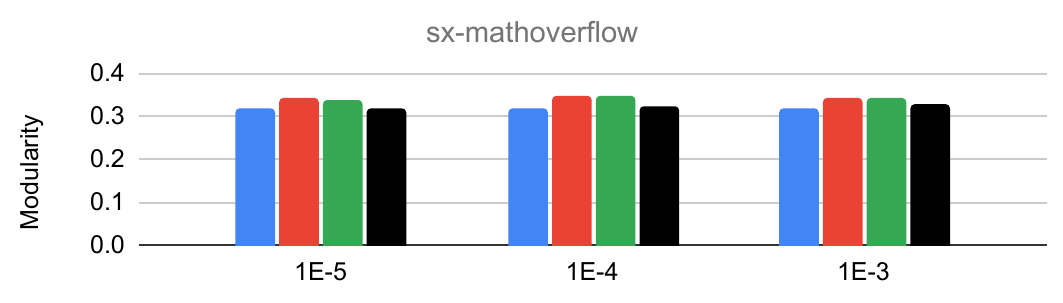}
  \includegraphics[width=0.48\linewidth]{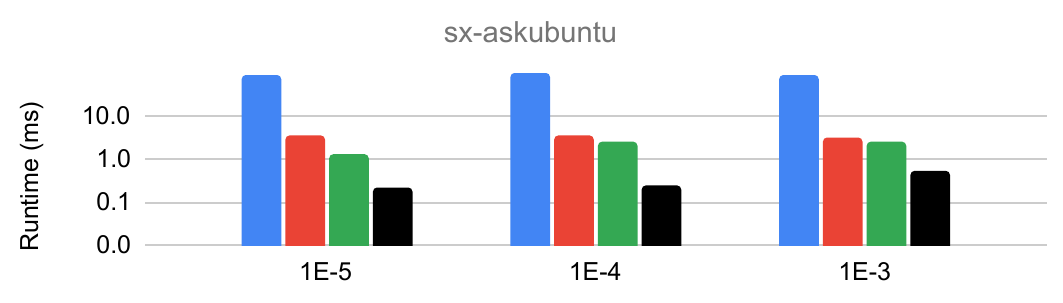}
  \includegraphics[width=0.48\linewidth]{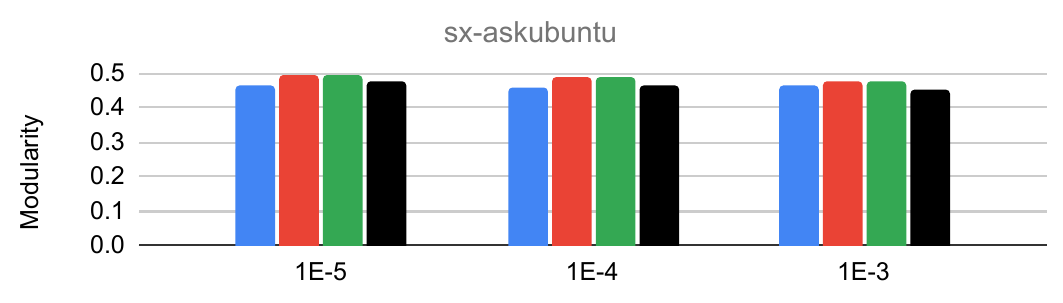}
  \includegraphics[width=0.48\linewidth]{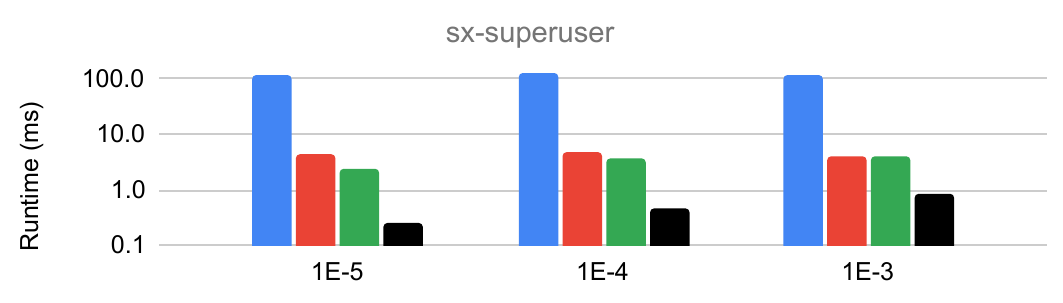}
  \includegraphics[width=0.48\linewidth]{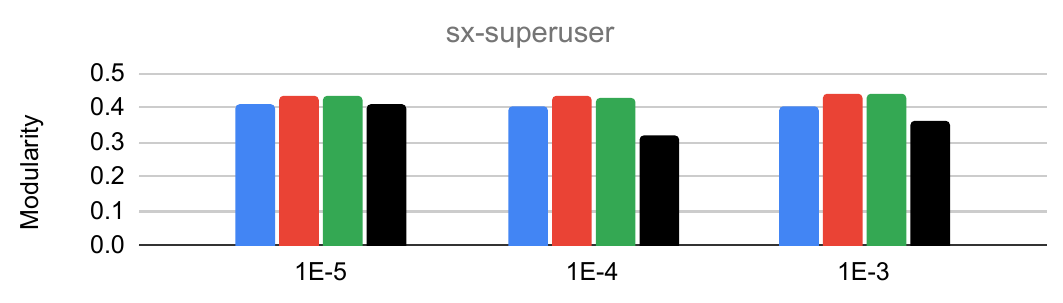}
  \includegraphics[width=0.48\linewidth]{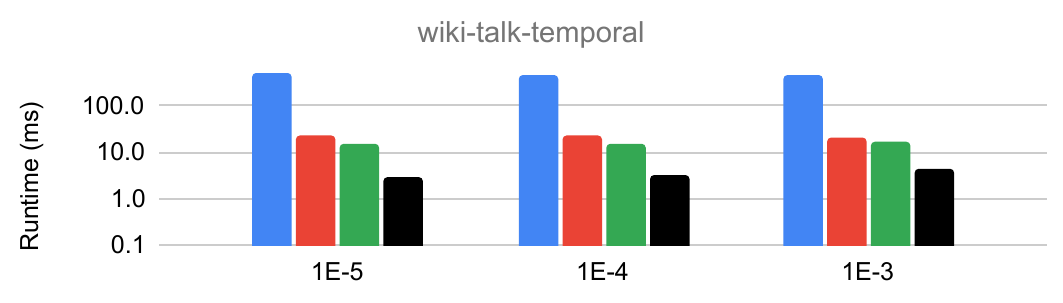}
  \includegraphics[width=0.48\linewidth]{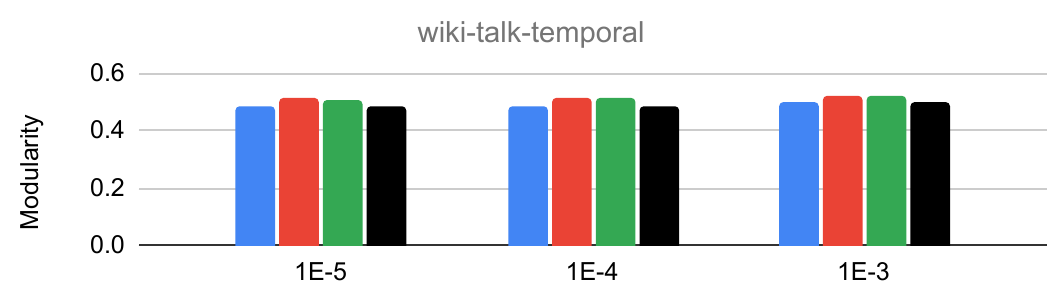}
  \subfigure[Runtime on each dynamic graph]{
    \label{fig:temporal-summary--runtime-graph}
    \includegraphics[width=0.48\linewidth]{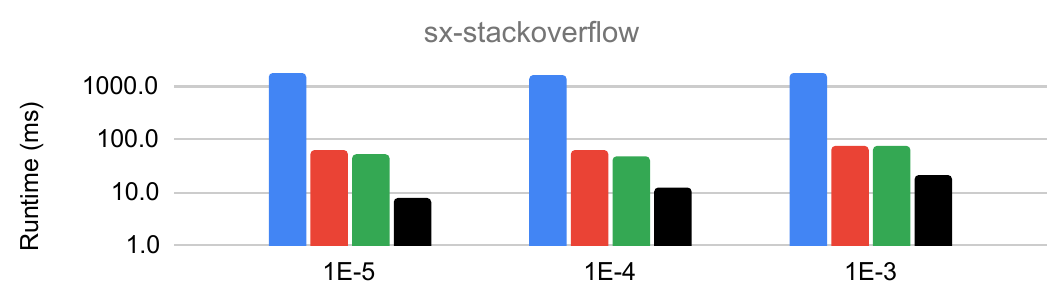}
  }
  \subfigure[Modularity in communities obtained on each dynamic graph]{
    \label{fig:temporal-summary--modularity-graph}
    \includegraphics[width=0.48\linewidth]{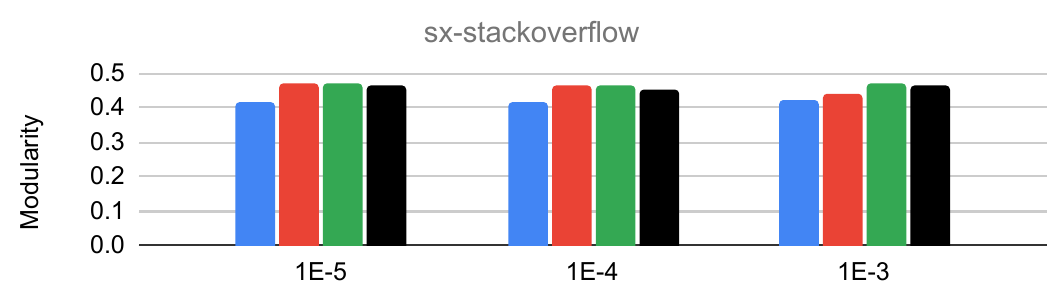}
  } \\[-2ex]
  \caption{Mean Runtime and Modularity of communities obtained with our multicore implementation of \textit{Static}, \textit{Naive-dynamic (ND)}, \textit{Delta-screening (DS)}, and \textit{Dynamic Frontier (DF)} Louvain on real-world dynamic graphs, with batch updates of size $10^{-5}|E_T|$ to $10^{-3}|E_T|$. Here, (a) and (b) show the overall runtime and modularity across all temporal graphs, while (c) and (d) show the runtime and modularity for each graph. In (a), the speedup of each approach with respect to Static Louvain is labeled.}
  \label{fig:temporal-summary}
\end{figure*}

\begin{figure*}[hbtp]
  \centering
  \subfigure[Overall result]{
    \label{fig:8020-runtime--mean}
    \includegraphics[width=0.38\linewidth]{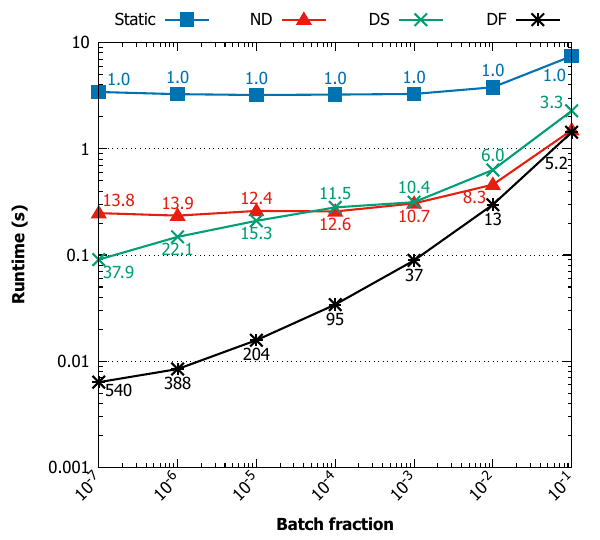}
  }
  \subfigure[Results on each graph]{
    \label{fig:8020-runtime--all}
    \includegraphics[width=0.58\linewidth]{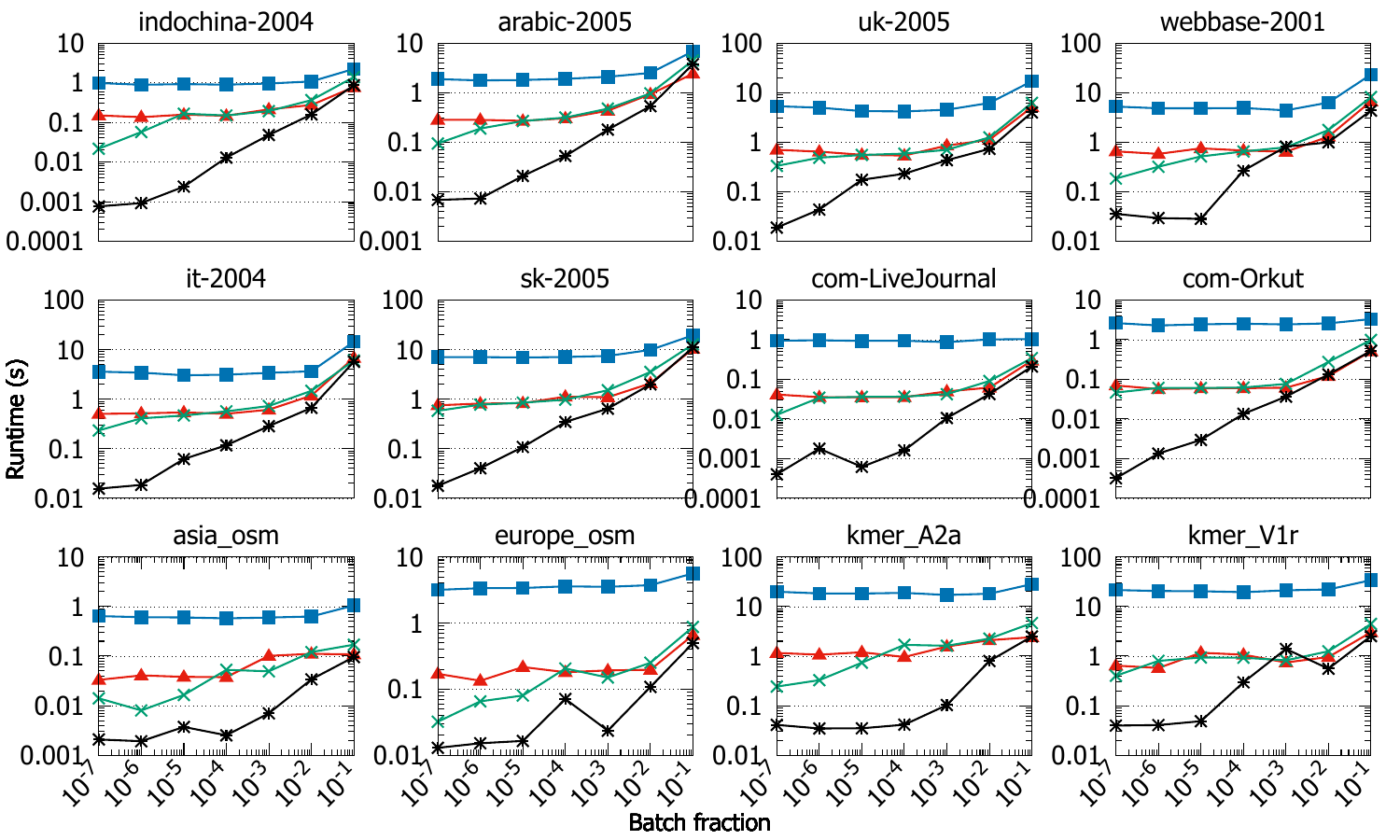}
  } \\[-2ex]
  \caption{Runtime (logarithmic scale) of our multicore implementation of \textit{Static}, \textit{Naive-dynamic (ND)}, \textit{Delta-screening (DS)}, and \textit{Dynamic Frontier (DF)} Louvain on large (static) graphs with generated random batch updates. Batch updates range in size from $10^{-7}|E|$ to $0.1|E|$ in multiples of $10$. These updates consist of $80\%$ edge insertions and $20\%$ edge deletions, mimicking realistic changes in a dynamic graph scenario. The right subfigure illustrates the runtime of each approach for individual graphs in the dataset, while the left subfigure presents overall runtimes (using geometric mean for consistent scaling across graphs). Additionally, the speedup of each approach relative to Static Louvain is labeled on respective lines.}
  \label{fig:8020-runtime}
\end{figure*}

\begin{figure*}[hbtp]
  \centering
  \subfigure[Overall result]{
    \label{fig:8020-modularity--mean}
    \includegraphics[width=0.38\linewidth]{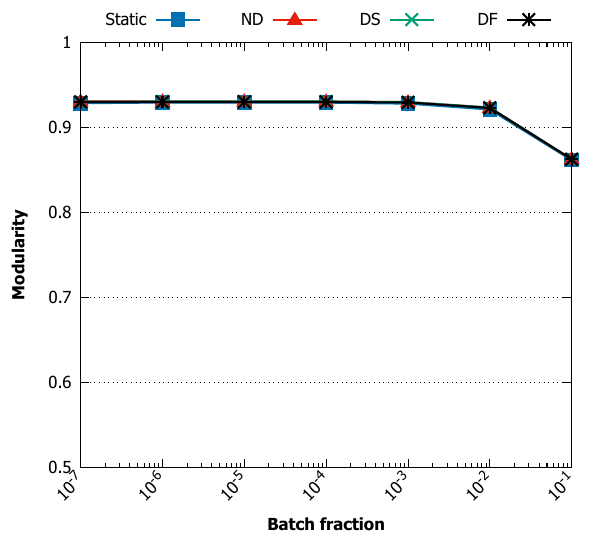}
  }
  \subfigure[Results on each graph]{
    \label{fig:8020-modularity--all}
    \includegraphics[width=0.58\linewidth]{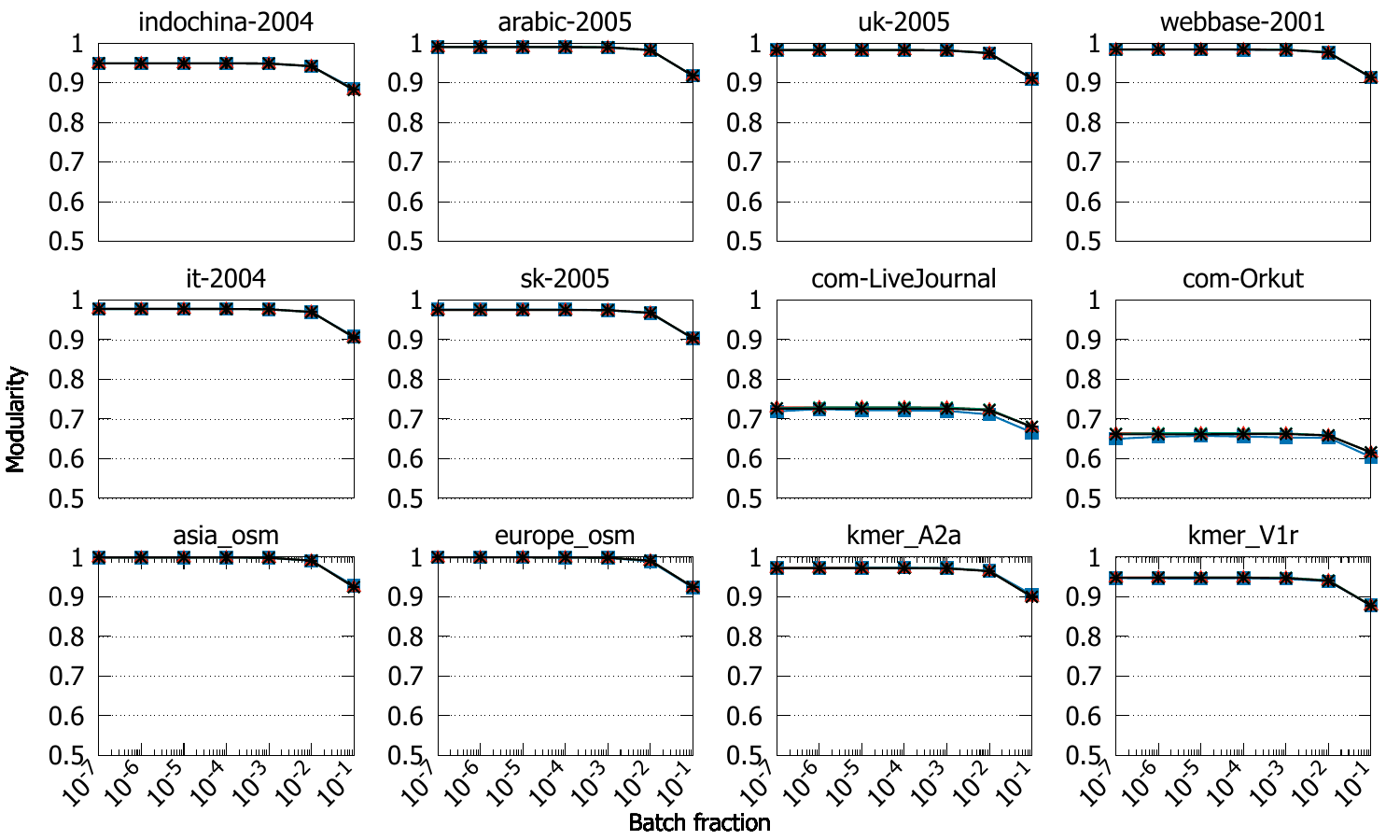}
  } \\[-2ex]
  \caption{Modularity comparison of our multicore implementation of \textit{Static}, \textit{Naive-dynamic (ND)}, \textit{Delta-screening (DS)}, and \textit{Dynamic Frontier (DF)} Louvain on large (static) graphs with generated random batch updates. The size of batch updates range from $10^{-7} |E|$ to $0.1 |E|$ in multiples of $10$ (logarithmic scale), consisting of $80\%$ edge insertions and $20\%$ edge deletions to simulate realistic dynamic graph updates. The right subfigure depicts the modularity for each approach in relation to each graph, while the left subfigure showcases overall modularity using arithmetic mean.}
  \label{fig:8020-modularity}
\end{figure*}

\ignore{\subsubsection{Determining optimality of result}}
\ignore{\label{sec:evaluation--optimality}}

\ignore{Community detection is an NP-hard problem and existing polynomial algorithms are \textit{heuristic}. We study correctness in terms of \textit{modularity score} of communities identified (higher is better), similar to previous works in the area \cite{com-traag19, com-zarayeneh21}. As Figures \ref{fig:louvain}-\ref{fig:hybrid} show, modularity of communities detected by our proposed dynamic algorithms is close to the modularity of communities detected by corresponding static algorithms.}

\begin{figure*}[hbtp]
  \centering
  \subfigure[Mean percentage of affected vertices on real-world dynamic graphs]{
    \label{fig:measure-affected--temporal}
    \includegraphics[width=0.48\linewidth]{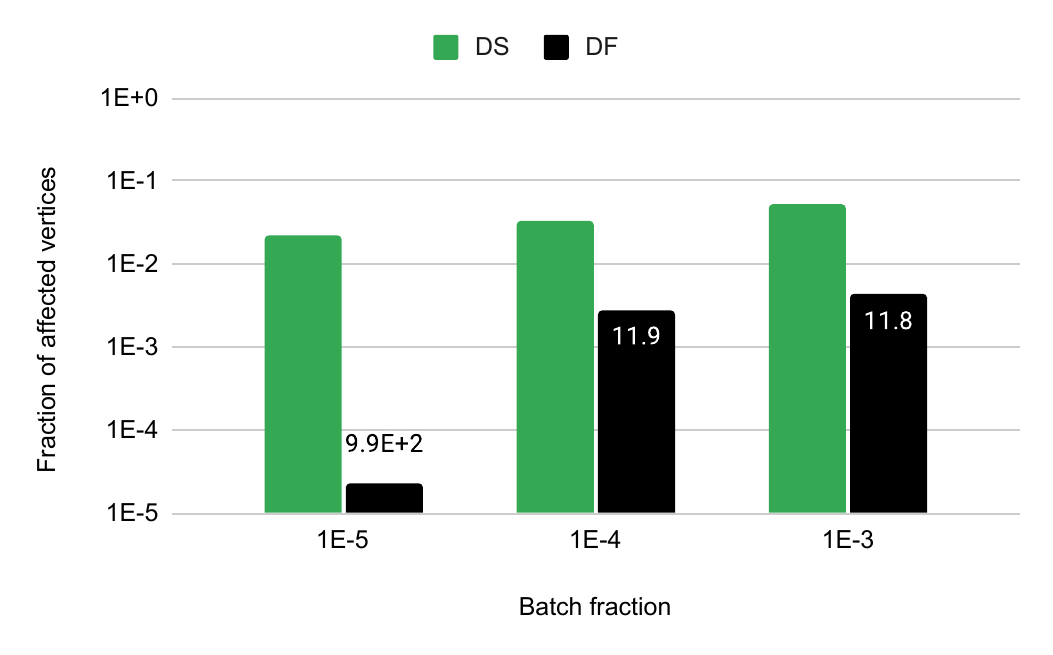}
  }
  \subfigure[Mean percentage of affected vertices on large graphs with random batch updates]{
    \label{fig:measure-affected--8020}
    \includegraphics[width=0.48\linewidth]{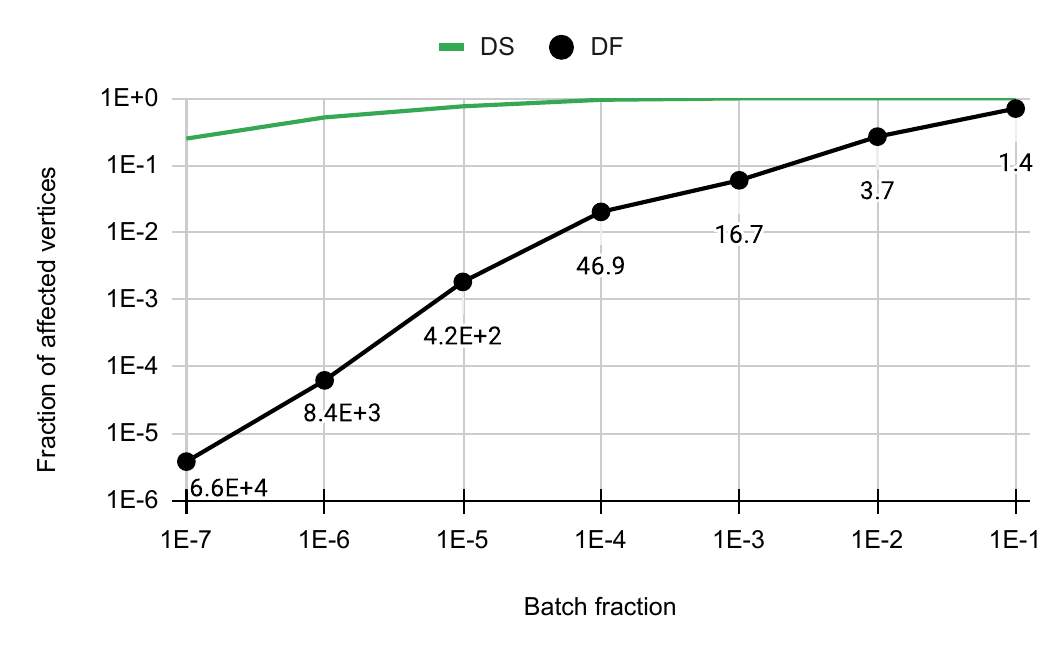}
  } \\[-2ex]
  \caption{Mean percentage of vertices marked as affected by \textit{Delta-screening (DS)} and our \textit{Dynamic Frontier (DF)} Louvain, on real-world dynamic graphs (with batch updates of size $10^{-5}|E|$ to $10^{-3}|E|$), and on large graphs with random batch updates ($80\%$ edge insertions and $20\%$ edge deletions with batch updates of size $10^{-7}|E|$ to $0.1|E|$).}
  \label{fig:measure-affected}
\end{figure*}

\ignore{\begin{figure}[hbtp]
  \centering
  \subfigure{
    \label{fig:measure-stability--5050}
    \includegraphics[width=0.98\linewidth]{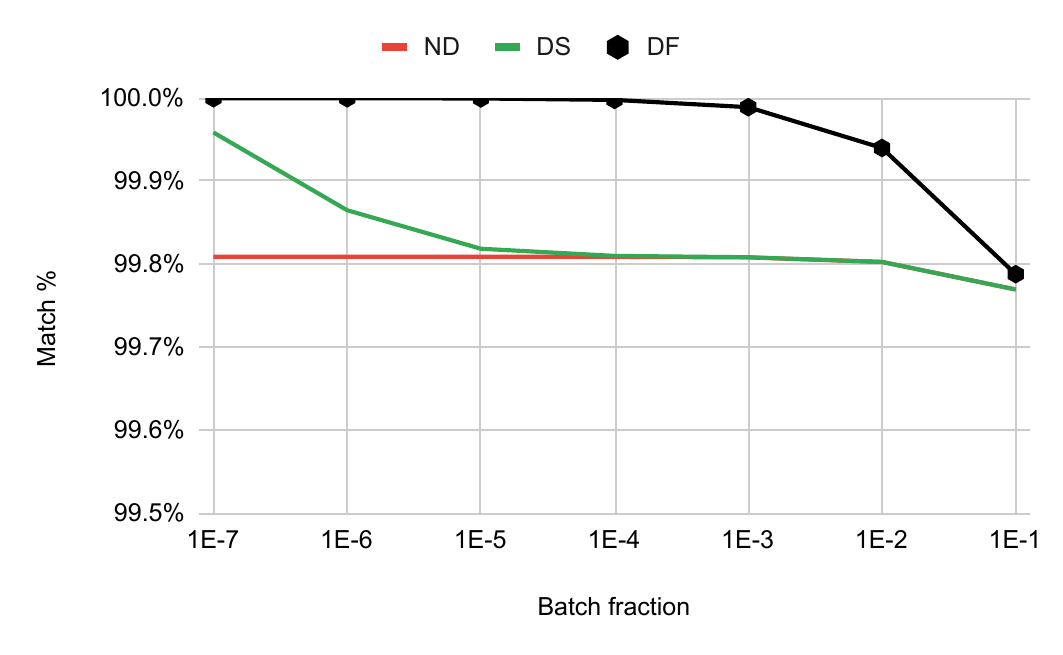}
  } \\[-2ex]
  \caption{Test.}
  \label{fig:measure-stability}
\end{figure}
}

\subsection{Performance Comparison}
\label{sec:performance-comparison}

\subsubsection{Results on real-world dynamic graphs}

We now compare the performance of our Parallel Dynamic Frontier (DF) Louvain with our parallel implementation of Static, Naive-dynamic (ND), and Delta-screening (DS) Louvain on real-world dynamic graphs from Table \ref{tab:dataset}. These evaluations are conducted on batch updates of size $10^{-5}|E_T|$ to $10^{-3}|E_T|$ in multiples of $10$. For each batch size, as mentioned in Section \ref{sec:batch-generation}, we load $90\%$ of the graph, add reverse edges to make all edges in the graph undirected, and then load $B$ edges (where $B$ is the batch size) consecutively in $100$ batch updates. The work of Zarayeneh et al. \cite{com-zarayeneh21} demonstrates improved performance of the DS approach compared to DynaMo \cite{com-zhuang19} and Batch \cite{com-chong13}. Thus, we limit our comparison to DS Louvain. Figure \ref{fig:temporal-summary--runtime-overall} displays the overall runtime of each approach across all graphs for each batch size, while Figure \ref{fig:temporal-summary--modularity-overall} illustrates the overall modularity of obtained communities. Additionally, Figures \ref{fig:temporal-summary--runtime-graph} and \ref{fig:temporal-summary--modularity-graph} present the mean runtime and modularity of communities obtained with the approaches on each dynamic graph in the dataset. Finally, Figures \ref{fig:temporal-sx-mathoverflow}, \ref{fig:temporal-sx-askubuntu}, \ref{fig:temporal-sx-superuser}, \ref{fig:temporal-wiki-talk-temporal}, and \ref{fig:temporal-sx-stackoverflow} show the runtime and modularity of communities obtained with Static, ND, DS, and DF Louvain on each dynamic graph in Table \ref{tab:dataset}, upon each consecutive batch update.

Figure \ref{fig:temporal-summary--runtime-overall} shows that DF Louvain is, on average, $268\times$, $195\times$, and $109\times$ faster than Static Louvain on batch updates of size $10^{-5}|E_T|$, $10^{-4}|E_T|$, and $10^{-3}|E_T|$, respectively. In contrast, DS Louvain demonstrates average speedups of $47\times$, $32\times$, and $26\times$ over Static Louvain for batch update sizes of $10^{-5}|E_T|$ to $10^{-3}|E_T|$, while ND Louvain obtains a mean speedup of $25\times$\ignore{over Static Louvain}.\ignore{DF Louvain thus achieves average speedups of $5.7\times$, $6.1\times$, and $4.2\times$ compared to DS Louvain for the same batch updates.} DF Louvain is thus, overall, $179\times$, $7.2\times$, and $5.3\times$ faster than Static, ND, and DS Louvain. This speedup is particularly pronounced on the \textit{sx-superuser} graph, as indicated by Figure \ref{fig:temporal-summary--runtime-graph}. Regarding modularity, Figures \ref{fig:temporal-summary--modularity-overall} and \ref{fig:temporal-summary--modularity-graph} illustrate that DF Louvain generally exhibits slightly lower modularity on average compared to ND and DS Louvain but is on par with the modularity obtained by Static Louvain, except for the \textit{sx-superuser} graph (because it fails to mark certain vertices as affected, likely because they are outlier vertices and were not directly reached from the expanding frontier). This makes the communities obtained with DF Louvain generally acceptable. However, if lower modularity is observed (during intermediate empirical tests), transitioning to our parallel implementation of DS Louvain is advisable. It is also worth noting that ND Louvain produces higher-quality communities than Static Louvain, as it builds upon the community memberships obtained from Static Louvain and further optimizes them (we limit the number of passes for Static Louvain to achieve a large speed gain, even if it means sacrificing some quality).

\subsubsection{Results on large graphs with random batch updates}

We also assess the performance of our parallel DF Louvain alongside our parallel implementation of Static, ND, and DS Louvain on large (static) graphs listed in Table \ref{tab:dataset-large}, with randomly generated batch updates. As elaborated in Section \ref{sec:batch-generation}, the batch updates vary in size from $10^{-7}|E|$ to $0.1|E|$ (in multiples of $10$), comprising $80\%$ edge insertions and $20\%$ edge deletions to mimic realistic scenarios. Reverse edges are added with each batch update, to ensure that the graph is undirected. As mentioned in Section \ref{sec:batch-generation}, we generate $5$ different random batch updates for each batch size to minimize measurement noise. Figure \ref{fig:8020-runtime} illustrates the runtime of Static, ND, DS, and DF Louvain, while Figure \ref{fig:8020-modularity} displays the modularity of communities obtained with each approach.

Figure \ref{fig:8020-runtime--mean} illustrates that DF Louvain achieves a mean speedup of $183\times$, $13.8\times$, and $8.7\times$ compared to Static, ND, and DS Louvain, respectively. On a batch update of size $10^{-7}|E|$, DF Louvain is significantly faster, attaining speedups of $540\times$, $39\times$, and $14\times$, with respect to Static, ND, and DS Louvain, respectively. The speedup is particularly pronounced on web graphs, social networks, and protein k-mer graphs, characterized by a large number of vertices (as depicted in Figure \ref{fig:8020-runtime--all}). It may be noted that DS Louvain exhibits slower performance than ND Louvain on large batch updates, i.e., on batch updates of size $0.01|E|$ and $0.1|E|$. This is attributed to the cost of initial marking of affected vertices with DS Louvain --- since the updates are scattered randomly across the graph, DS Louvain ends up marking a significant number of vertices as affected, making it almost equivalent to ND Louvain, but with the added cost of the marking of affected vertices (particularly on web graphs and social networks, characterized by a high average degree and a small diameter). Figures \ref{fig:8020-modularity--mean} and \ref{fig:8020-modularity--all} indicate that DF Louvain obtains communities with the roughly same modularity as Static, ND, and DS Louvain. Hence, for large graphs with random batch updates, DF Louvain is the best dynamic community detection method.

In Figure \ref{fig:8020-runtime}, also note that runtime of Static Louvain increases for larger batch updates. This more likely due to the random batch updates arbitrarily disrupting the original community structure --- which results in Static Louvain needing more iterations to converge --- than due to the increased number of edges in the graph.

\subsubsection{Comparison of vertices marked as affected}

Figure \ref{fig:measure-affected--temporal} displays the mean percentage of vertices marked as affected by DS and DF Louvain on real-world dynamic graphs from Table \ref{tab:dataset}, with batch updates of size $10^{-5}|E_T|$ to $10^{-3}|E_T|$ in multiples of $10$ (see Section \ref{sec:batch-generation} for details) --- while Figure \ref{fig:measure-affected--8020} displays the mean percentage of vertices marked as affected by DS and DF Louvain on large (static) graphs with generated random batch updates (consisting of $80\%$ edge insertions and $20\%$ deletions), on batch updates of size $10^{-7}|E|$ to $0.1|E|$. For DS Louvain, the affected vertices are marked at the start of the algorithm, while, for DF Louvain, affected vertices are marked incrementally --- therefore, we count all vertices that were ever flagged as affected with DF Louvain.

Figure \ref{fig:measure-affected--temporal} shows that the proportion of vertices marked as affected by DF Louvain is $990\times$, $11.9\times$, and $11.8\times$ lower than DS Louvain for batch updates of size $10^{-5}|E_T|$, $10^{-4}|E_T|$, and $10^{-3}|E_T|$, respectively. Figure \ref{fig:measure-affected--8020} also paints a similar picture, with significantly fewer vertices being marked as affected by DF Louvain for smaller batch updates (i.e., batch updates of size $10^{-7}|E|$ and $10^{-6}|E|$). Therefore, the performance improvement with DF Louvain can be attributed to both marking fewer vertices as affected, and to the incremental marking of affected vertices. Additionally, it is worth noting that, on real-world dynamic graphs, the fraction of vertices marked as affected is generally low across both approaches. This is likely because updates in such graphs tend to be concentrated in specific regions of the graph\ignore{(not randomly scattered throughout)}.

\begin{figure}[!hbt]
  \centering
  \subfigure{
    \label{fig:strong-scaling--8020}
    \includegraphics[width=0.98\linewidth]{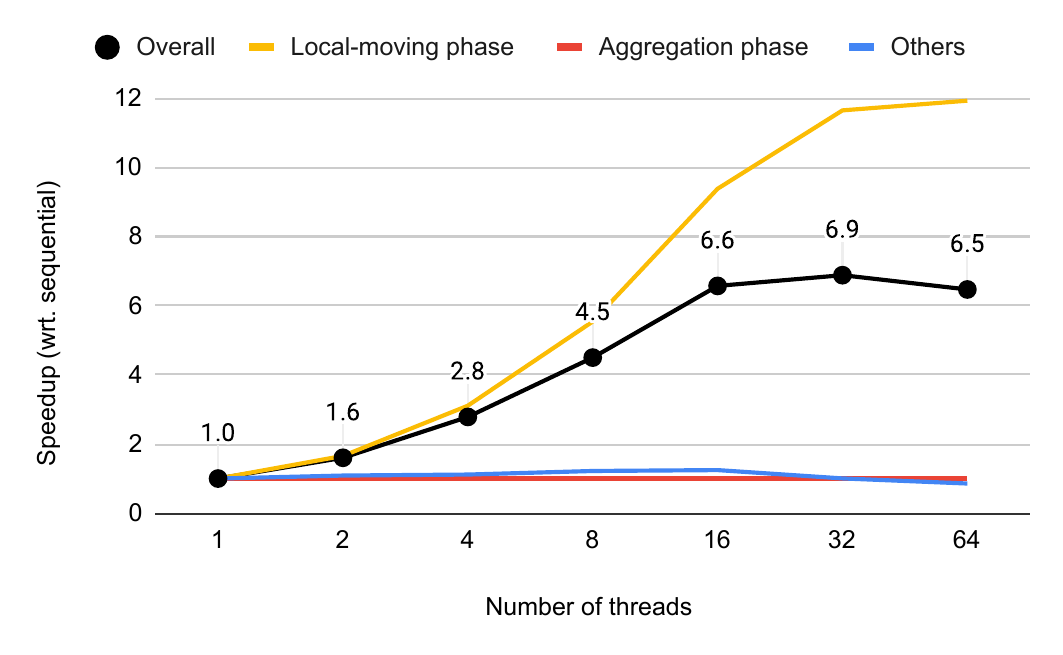}
  } \\[-2ex]
  \caption{Mean speedup of\ignore{our} \textit{Dynamic Frontier (DF)} Louvain with increasing number of threads (in multiples of $2$), on large\ignore{(static)} graphs with random batch updates of size $10^{-7}|E|$ to $0.1|E|$, composed $80\%$ edge insertions and $20\%$ insertions.}
  \label{fig:strong-scaling}
\end{figure}

\subsubsection{Strong scaling}
\label{sec:strong-scaling}

Finally, we study the strong-scaling behavior of DF Louvain on large (static) graphs, with generated random batch updates of size $10^{-7}|E|$ to $0.1|E|$. The speedup of DF Louvain is measured as the number of threads increases from $1$ to $64$ in multiples of $2$, relative to single-threaded execution. This process is repeated for each graph in the dataset (refer to Table \ref{tab:dataset}), and the results are averaged using geometric mean.

The results, depicted in Figure \ref{fig:strong-scaling}, indicate that with $16$ threads, DF Louvain achieves an average speedup of $6.6\times$ compared to single-threaded execution, showing a performance increase of $1.6\times$ for every doubling of threads. The speedup of DF Louvain is lower, likely due to the reduced work performed by the algorithm. At $32$ and $64$ threads, DF Louvain is affected by NUMA effects (the $64$-core processor used has $4$ NUMA domains), resulting in a speedup of only $6.9\times$ and $6.5\times$, respectively. The results are similar on real-world dynamic graphs\ignore{, as shown in Figure X}, but the speedup is even lower for $64$ threads to due lack of sufficient work per thread.

\section{Conclusion}
\label{sec:conclusion}
In conclusion, in this report we presented our Parallel Dynamic Frontier (DF) Louvain algorithm, which given a batch update of edge deletions or insertions, incrementally identifies and processes an approximate set of affected vertices in the graph with minimal overhead. In addition, we use a novel approach of incrementally updating weighted-degrees of vertices and total edge weights of communities, by using them as auxiliary information to DF Louvain. We also present our parallel implementations of Naive-dynamic (ND) \cite{com-aynaud10} and Delta-screening (DS) Louvain \cite{com-zarayeneh21}. On a server with a 64-core AMD EPYC-7742 processor, our experiments show that DF Louvain obtains speedups of $179\times$, $7.2\times$, and $5.3\times$ on real-world dynamic graphs, compared to Static, ND, and DS Louvain, respectively, and is $183\times$, $13.8\times$, and $8.7\times$ faster, respectively, on large graphs with random batch updates. Note that our static implementation of the Louvain algorithm, GVE-Louvain, here referred to Static Louvain, is itself $50\times$, $22\times$, $20\times$, and $5.8\times$ faster than Vite, Grappolo, NetworKit Louvain, and cuGraph Louvain (running on NVIDIA A100 GPU) \cite{sahu2023gvelouvain}. The performance of DF Louvain can be attributed to incremental marking, and marking of fewer vertices. Furthermore, DF Louvain improves performance at a rate of $1.6\times$ for every doubling of threads. DF Louvain is based on one of the fastest implementations of Static Louvain \cite{sahu2023gvelouvain} and thus offers state-of-the-art performance. However, we observe DF Louvain to obtain lower modularity score on the \textit{sx-superuser} real-world dynamic graph. We thus recommend the reader to use DF Louvain for updating community memberships on dynamic graphs, but switch to DS Louvain if lower modularity is observed.

\begin{acks}
I would like to thank Prof. Kishore Kothapalli and Prof. Dip Sankar Banerjee for their support.
\end{acks}

\bibliographystyle{ACM-Reference-Format}
\bibliography{main}

\clearpage
\appendix
\section{Appendix}
\subsection{Our Parallel Naive-dynamic (ND) Louvain}
\label{sec:our-naive}

Algorithm \ref{alg:naive} presents our multicore implementation of Naive dynamic (ND) Louvain, where vertices are assigned to communities from the previous snapshot of the graph, and all vertices are processed irrespective of edge deletions and insertions in the batch update. Algorithm \ref{alg:naive} requires several inputs, including the previous $G^{t-1}$ and current graph snapshots $G^t$, edge deletions $\Delta^{t-}$ and insertions $\Delta^{t+}$ in the batch update, the previous community membership of each vertex $C^{t-1}$, weighted degree of each vertex $K^{t-1}$, and total edge weight of each community $\Sigma^{t-1}$. The output consists of the updated community memberships $C^t$, weighted-degrees $K^t$, and total edge weight of communities $\Sigma^t$.

In the algorithm, we start by defining two lambda functions for the Louvain algorithm, \texttt{isAffected()} (lines \ref{alg:naive--isaff-begin}-\ref{alg:naive--isaff-end}) and \texttt{inAffectedR} \texttt{ange()} (lines \ref{alg:naive--isaffrng-begin}-\ref{alg:naive--isaffrng-end}), which indicate that all vertices in the graph $G^t$ are to be marked as affected, and that all such vertices can be incrementally marked as affected, respectively. Unlike existing works, we then utilize $K^{t-1}$ and $\Sigma^{t-1}$, along with the batch update $\Delta^{t-}$ and $\Delta^{t+}$, to quickly obtain $K^t$ and $\Sigma^t$ which is needed in the local-moving phase of the Louvain algorithm (line \ref{alg:naive--auxliliary}). The lambda functions and the total vertex/edge weights are then used to run the Louvain algorithm, and obtain the updated community assignments $C^t$ (line \ref{alg:naive--louvain}). Finally, $C^t$ is returned, along with $K^t$ and $\Sigma^t$ as the updated auxiliary information (line \ref{alg:naive--return}).

\subsection{Our Parallel Delta-screening (DS) Louvain}
\label{sec:our-delta}

The pseudocode of our multicore implementation of Delta-screening (DS) Louvain is given in Algorithm \ref{alg:delta}. It uses modularity-based scoring to determine an approximate region of the graph in which vertices are likely to change their community membership \cite{com-zarayeneh21}. The algorithm accepts as input the previous $G^{t-1}$ and current snapshot of the graph $G^t$, edge deletions $\Delta^{t-}$ and insertions $\Delta^{t+}$ in the batch update, the previous community memberships of vertices $C^{t-1}$, weighted degrees of vertices $K^{t-1}$, and total edge weights of communities $\Sigma^{t-1}$. It outputs the updated community memberships $C^t$, weighted-degrees $K^t$, and total edge weights of communities $\Sigma^t$. The batch update, comprising edge deletions $(i, j, w) \in \Delta^{t-}$ and insertions $(i, j, w) \in \Delta^{t+}$, is sorted separately by their source vertex ID $i$ beforehand, as a preprocessing step.

\begin{algorithm}[hbtp]
\caption{Our Parallel \textit{Naive-dynamic (ND)} Louvain.}
\label{alg:naive}
\begin{algorithmic}[1]
\Require{$G^{t-1}, G^t$: Previous, current input graph}
\Require{$\Delta^{t-}, \Delta^{t+}$: Edge deletions and insertions (batch update)}
\Require{$C^{t-1}, C^t$: Previous, current community of each vertex}
\Require{$K^{t-1}, K^t$: Previous, current weighted-degree of vertices}
\Require{$\Sigma^{t-1}, \Sigma^t$: Previous, current total edge weight of communities}
\Ensure{$isAffected(i)$: Is vertex $i$ is marked as affected?}
\Ensure{$inAffectedRange(i)$: Can $i$ be incrementally marked?}
\Ensure{$F$: Lambda functions passed to parallel Louvain (Alg. \ref{alg:louvain})}

\Statex

\Function{naiveDynamic}{$G^{t-1}, G^t, \Delta^{t-}, \Delta^{t+}, C^{t-1}, K^{t-1}, \Sigma^{t-1}$}
  \State $\rhd$ Mark affected vertices
  \Function{isAffected}{$i$} \label{alg:naive--isaff-begin}
    \Return{$1$}
  \EndFunction \label{alg:naive--isaff-end}
  \Function{inAffectedRange}{$i$} \label{alg:naive--isaffrng-begin}
    \Return{$1$}
  \EndFunction \label{alg:naive--isaffrng-end}
  \State $F \gets \{isAffected, inAffectedRange\}$ \label{alg:naive--lambdas}
  \State $\rhd$ Use $K^{t-1}$, $\Sigma^{t-1}$ as auxiliary information (Alg. \ref{alg:update})
  \State $\{K^t, \Sigma^t\} \gets updateWeights(G^t, \Delta^{t-}, \Delta^{t+}, C^{t-1}, K^{t-1}, \Sigma^{t-1})$\label{alg:naive--auxliliary}
  \State $\rhd$ Obtain updated communities (Alg. \ref{alg:louvain})
  \State $C^t \gets louvain(G^t, C^{t-1}, K^t, \Sigma^t, F)$ \label{alg:naive--louvain}
  \Return{$\{C^t, K^t, \Sigma^t\}$} \label{alg:naive--return}
\EndFunction
\end{algorithmic}
\end{algorithm}

\begin{algorithm}[hbtp]
\caption{Our Parallel \textit{Delta-screening (DS)} Louvain.}
\label{alg:delta}
\begin{algorithmic}[1]
\Require{$G^{t-1}, G^t$: Previous, current input graph}
\Require{$\Delta^{t-}, \Delta^{t+}$: Edge deletions and insertions (batch update)}
\Require{$C^{t-1}, C^t$: Previous, current community of each vertex}
\Require{$K^{t-1}, K^t$: Previous, current weighted-degree of vertices}
\Require{$\Sigma^{t-1}, \Sigma^t$: Previous, current total edge weight of communities}
\Ensure{$\delta V, \delta E, \delta C$: Is vertex, neighbors, or community affected?}
\Ensure{$H$: Hashtable mapping a community to associated weight}
\Ensure{$isAffected(i)$: Is vertex $i$ is marked as affected?}
\Ensure{$inAffectedRange(i)$: Can $i$ be incrementally marked?}
\Ensure{$F$: Lambda functions passed to parallel Louvain (Alg. \ref{alg:louvain})}

\Statex

\Function{deltaScreening}{$G^{t-1}, G^t, \Delta^{t-}, \Delta^{t+}, C^{t-1}, K^{t-1}, \Sigma^{t-1}$}
  \State $H, \delta V, \delta E, \delta C \gets \{\}$ \label{alg:delta--init}
  \State $\rhd$ Mark affected vertices
  \ForAll{$(i, j, w) \in \Delta^{t-}$ \textbf{in parallel}} \label{alg:delta--loopdel-begin}
    \If{$C^{t-1}[i] = C^{t-1}[j]$}
      \State $\delta V[i], \delta E[i], \delta C[C^{t-1}[j]] \gets 1$ \label{alg:delta--loopdelmark}
    \EndIf
  \EndFor \label{alg:delta--loopdel-end}
  \ForAll{unique source vertex $i \in \Delta^{t+}$ \textbf{in parallel}} \label{alg:delta--loopins-begin}
    \State $H \gets \{\}$
    \ForAll{$(i', j, w) \in \Delta^{t+}\ |\ i' = i$} \label{alg:delta--loopinssrc-begin}
      \If{$C^{t-1}[i] \neq C^{t-1}[j]$}
        \State $H[C^{t-1}[j]] \gets H[C^{t-1}[j]] + w$
      \EndIf
    \EndFor \label{alg:delta--loopinssrc-end}
    \State $c^* \gets$ Best community linked to $i$ in $H$ \label{alg:delta--loopinschoose}
    \State $\delta V[i], \delta E[i], \delta C[c^*] \gets 1$ \label{alg:delta--loopinsmark}
  \EndFor \label{alg:delta--loopins-end}
  \ForAll{$i \in V^t$ \textbf{in parallel}} \label{alg:delta--loopaff-begin}
    \If{$\delta E[i]$} \label{alg:delta--loopaffnei-begin}
      \ForAll{$j \in G^t.neighbors(i)$}
        \State $\delta V[j] \gets 1$
      \EndFor
    \EndIf \label{alg:delta--loopaffnei-end}
    \If{$\delta C[C^{t-1}[i]]$} \label{alg:delta--loopaffcom-begin}
      \State $\delta V[i] \gets 1$
    \EndIf \label{alg:delta--loopaffcom-end}
  \EndFor \label{alg:delta--loopaff-end}
  \Function{isAffected}{$i$} \label{alg:delta--isaff-begin}
    \Return{$\delta V[i]$}
  \EndFunction \label{alg:delta--isaff-end}
  \Function{inAffectedRange}{$i$} \label{alg:delta--isaffrng-begin}
    \Return{$\delta V[i]$}
  \EndFunction \label{alg:delta--isaffrng-end}
  \State $F \gets \{isAffected, inAffectedRange\}$ \label{alg:delta--lambdas}
  \State $\rhd$ Use $K^{t-1}$, $\Sigma^{t-1}$ as auxiliary information (Alg. \ref{alg:update})
  \State $\{K^t, \Sigma^t\} \gets updateWeights(G^t, \Delta^{t-}, \Delta^{t+}, C^{t-1}, K^{t-1}, \Sigma^{t-1})$\label{alg:delta--auxiliary}
  \State $\rhd$ Obtain updated communities (Alg. \ref{alg:louvain})
  \State $C^t \gets louvain(G^t, C^{t-1}, K^t, \Sigma^t, F)$ \label{alg:delta--louvain}
  \Return{$\{C^t, K^t, \Sigma^t\}$} \label{alg:delta--return}
\EndFunction
\end{algorithmic}
\end{algorithm}

In the algorithm, we first initialize a hashtable $H$, mapping communities to their associated weights, and affected flags $\delta V$, $\delta E$, and $\delta C$, indicating if a vertex, its neighbors, or its community is affected by the batch update (lines \ref{alg:delta--init}). Then, we parallelly iterate over edge deletions $\Delta^{t-}$ and insertions $\Delta^{t+}$. For each deletion $(i, j, w) \in \Delta^{t-}$ where $i$ and $j$ are in the same community, we mark the source vertex $i$, its neighbors, and its community as affected (lines \ref{alg:delta--loopdel-begin}-\ref{alg:delta--loopdel-end}). For each unique source vertex $i\ |\ (i, j, w) \in \Delta^{t+}$ in insertions belonging to different communities, we identify the community $c^*$ with the highest delta-modularity if $i$ moves to one of its neighboring communities and mark $i$, its neighbors, and the community $c^*$ as affected (lines \ref{alg:delta--loopins-begin}-\ref{alg:delta--loopins-end}). Deletions between different communities and insertions within the same community are disregarded. Using affected neighbor $\delta E$ and community flags $\delta C$, affected vertices in $\delta V$ are marked (lines \ref{alg:delta--loopaff-begin}-\ref{alg:delta--loopaff-end}). Next, similar to ND Louvain, we use $K^{t-1}$ and $\Sigma^{t-1}$, along with $\Delta^{t-}$ and $\Delta^{t+}$, to quickly yield $K^t$ and $\Sigma^t$ (line \ref{alg:naive--auxliliary}), define the needed lambda functions \texttt{isAffected()} (lines \ref{alg:delta--isaff-begin}-\ref{alg:delta--isaff-end}) and \texttt{inAffectedRange()} (lines \ref{alg:delta--isaffrng-begin}-\ref{alg:delta--isaffrng-end}), and run the Louvain algorithm, yielding updated community assignments $C^t$ (line \ref{alg:delta--louvain}). Finally, updated community memberships $C^t$ are returned, alongside $K^t$ and $\Sigma^t$ as updated auxiliary information (line \ref{alg:delta--return}).

\subsection{Our Dynamic-supporting Parallel Louvain}
\label{sec:our-louvain}

The main step of our Dynamic-supporting Parallel Louvain algorithm is given in Algorithm \ref{alg:louvain}. Unlike our implementation of Static Louvain \cite{sahu2023gvelouvain}, in addition to the current graph snapshot $G^t$, it accepts as input the previous community membership of each vertex $C^{t-1}$, the updated weighted-degree of each vertex $K^t$, the updated total edge weight of each community $\Sigma^t$, and a set of lambda functions $F$ which determine if a given vertex is affected, or if it can be incrementally marked as affected (it is in the affected range). It outputs the updated community memberships of vertices $C^t$.

\begin{algorithm}[hbtp]
\caption{Our Dynamic-supporting Parallel Louvain \cite{sahu2023gvelouvain}.}
\label{alg:louvain}
\begin{algorithmic}[1]
\Require{$G^t$: Current input graph}
\Require{$C^{t-1}$: Previous community of each vertex}
\Require{$K^t$: Current weighted-degree of each vertex}
\Require{$\Sigma^t$: Current total edge weight of each community}
\Require{$F$: Lambda functions passed to parallel Louvain}
\Ensure{$G'$: Current/super-vertex graph.}
\Ensure{$C, C'$: Current community of each vertex in $G^t$, $G'$}
\Ensure{$K, K'$: Current weighted-degree of each vertex in $G^t$, $G'$}
\Ensure{$\Sigma, \Sigma'$: Current total edge weight of each community in $G^t$, $G'$}
\Ensure{$\tau, \tau_{agg}$: Iteration, aggregation tolerance}

\Statex

\Function{louvain}{$G^t, C^{t-1}, K^t, \Sigma^t, F$} \label{alg:louvain--begin}
  \State $\rhd$ Mark affected vertices as unprocessed
  \ForAll{$i \in V^t$} \label{alg:louvain--mark-begin}
    \If{$F.isAffected(i)$} Mark $i$ as unprocessed
    \EndIf
  \EndFor \label{alg:louvain--mark-end}
  \State $\rhd$ Initialization phase
  \State Vertex membership: $C \gets [0 .. |V^t|)$ \label{alg:louvain--init-begin}
  \State $G' \gets G^t$ \textbf{;} $C' \gets C^{t-1}$ \textbf{;} $K' \gets K^t$ \textbf{;} $\Sigma' \gets \Sigma^t$ \label{alg:louvain--init-end}
  \State $\rhd$ Local-moving and aggregation phases
  \ForAll{$l_p \in [0 .. \text{\small{MAX\_PASSES}})$} \label{alg:louvain--passes-begin}
    \State $l_i \gets louvainMove(G', C', K', \Sigma', F)$ \Comment{Alg. \ref{alg:louvainlm}} \label{alg:louvain--local-move}
    \If{$l_i \le 1$} \textbf{break} \Comment{Globally converged?} \label{alg:louvain--globally-converged}
    \EndIf
    \State $|\Gamma|, |\Gamma_{old}| \gets$ Number of communities in $C$, $C'$
    \If{$|\Gamma|/|\Gamma_{old}| > \tau_{agg}$} \textbf{break} \Comment{Low shrink?} \label{alg:louvain--aggregation-tolerance}
    \EndIf
    \State $C' \gets$ Renumber communities in $C'$ \label{alg:louvain--renumber}
    \State $C \gets$ Lookup dendrogram using $C$ to $C'$ \label{alg:louvain--lookup}
    \State $G' \gets$ Aggregate communities in $G'$ using $C'$ \Comment{Alg. \ref{alg:louvainag}} \label{alg:louvain--aggregate}
    \State $\Sigma' \gets K' \gets$ Weight of each vertex in $G'$ \label{alg:louvain--reset-weights}
    \State Mark all vertices in $G'$ as unprocessed \label{alg:louvain--reset-affected}
    \State $C' \gets [0 .. |V'|)$
    \State $\tau \gets \tau / \text{\small{TOLERANCE\_DROP}}$ \Comment{Threshold scaling} \label{alg:louvain--threshold-scaling}
  \EndFor \label{alg:louvain--passes-end}
  \State $C \gets$ Lookup dendrogram using $C$ to $C'$ \label{alg:louvain--lookup-last}
  \Return{$C$} \label{alg:louvain--return}
\EndFunction \label{alg:louvain--end}
\end{algorithmic}
\end{algorithm}


In the algorithm, we commence by marking affected vertices as unprocessed (lines \ref{alg:louvain--mark-begin}-\ref{alg:louvain--mark-end}). Subsequently, the initialization phase follows, wherein we initialize the community membership of each vertex $C$ in $G^t$. Additionally, we initialize the total edge weight of each vertex $K'$, the total edge weight of each community $\Sigma'$, and the community membership $C'$ of each vertex in the current/super-vertex graph $G'$ (lines \ref{alg:louvain--init-begin}-\ref{alg:louvain--init-end}). After initialization, we conduct a series of passes (lines \ref{alg:louvain--passes-begin}-\ref{alg:louvain--passes-end}) of the local-moving and aggregation phases (limited to $MAX\_PASSES$). Within each pass, in line \ref{alg:louvain--local-move}, we execute the local-moving phase of the Louvain algorithm (Algorithm \ref{alg:louvainlm}), which optimizes community assignments. If the local-moving phase converges within a single iteration, it implies global convergence, prompting the termination of passes (line \ref{alg:louvain--globally-converged}). Conversely, if the drop in community count $|\Gamma|$ is deemed insignificant, indicating diminishing returns, we halt at the current pass (line \ref{alg:louvain--aggregation-tolerance}). Should the convergence conditions not be met, we progress to the aggregation phase. Here, we renumber communities (line \ref{alg:louvain--renumber}), update top-level community memberships $C$ using dendrogram lookup (line \ref{alg:louvain--lookup}), execute the aggregation phase (Algorithm \ref{alg:louvainag}), and adjust the convergence threshold for subsequent passes, i.e., perform threshold scaling (line \ref{alg:louvain--threshold-scaling}). The subsequent pass commences in line \ref{alg:louvain--passes-begin}. Following all passes, we do a final update of the top-level community membership $C$ of each vertex in $G^t$ via dendrogram lookup (line \ref{alg:louvain--lookup-last}), before ultimately returning it (line \ref{alg:louvain--return}).

\subsubsection{Local-moving phase of our Parallel Louvain}

The pseudocode detailing the local-moving phase of our Parallel Louvain algorithm is presented in Algorithm \ref{alg:louvainlm}. This phase iteratively moves vertices among communities in order to maximize modularity. Here, the \texttt{louvainMove()} function operates on the current graph $G'$, community membership $C'$, total edge weight of each vertex $K'$, total edge weight of each community $\Sigma'$, and a set of lambda functions as inputs, yielding the number of iterations performed $l_i$.

\begin{algorithm}[hbtp]
\caption{Local-moving phase of our Parallel Louvain \cite{sahu2023gvelouvain}.}
\label{alg:louvainlm}
\begin{algorithmic}[1]
\Require{$G'$: Input/super-vertex graph}
\Require{$C'$: Community membership of each vertex}
\Require{$K'$: Total edge weight of each vertex}
\Require{$\Sigma'$: Total edge weight of each community}
\Require{$F$: Lambda functions passed to parallel Louvain}
\Ensure{$H_t$: Collision-free per-thread hashtable}
\Ensure{$l_i$: Number of iterations performed}
\Ensure{$\tau$: Per iteration tolerance}

\Statex

\Function{louvainMove}{$G', C', K', \Sigma', F$} \label{alg:louvainlm--move-begin}
  \ForAll{$l_i \in [0 .. \text{\small{MAX\_ITERATIONS}})$} \label{alg:louvainlm--iterations-begin}
    \State Total delta-modularity per iteration: $\Delta Q \gets 0$ \label{alg:louvainlm--init-deltaq}
    \ForAll{unprocessed $i \in V'$ \textbf{in parallel}} \label{alg:louvainlm--loop-vertices-begin}
      \State Mark $i$ as processed (prune) \label{alg:louvainlm--prune}
      \If{\textbf{not} $F.inAffectedRange(i)$} \textbf{continue} \label{alg:louvainlm--affrng}
      \EndIf
      \State $H_t \gets scanCommunities(\{\}, G', C', i, false)$ \label{alg:louvainlm--scan}
      \State $\rhd$ Use $H_t, K', \Sigma'$ to choose best community
      \State $c^* \gets$ Best community linked to $i$ in $G'$ \label{alg:louvainlm--best-community-begin}
      \State $\delta Q^* \gets$ Delta-modularity of moving $i$ to $c^*$ \label{alg:louvainlm--best-community-end}
      \If{$c^* = C'[i]$} \textbf{continue} \label{alg:louvainlm--best-community-same}
      \EndIf
      \State $\Sigma'[C'[i]] -= K'[i]$ \textbf{;} $\Sigma'[c^*] += K'[i]$ \textbf{atomic} \label{alg:louvainlm--perform-move-begin}
      \State $C'[i] \gets c^*$ \textbf{;} $\Delta Q \gets \Delta Q + \delta Q^*$ \label{alg:louvainlm--perform-move-end}
      \State Mark neighbors of $i$ as unprocessed \label{alg:louvainlm--remark}
    \EndFor \label{alg:louvainlm--loop-vertices-end}
    \If{$\Delta Q \le \tau$} \textbf{break} \Comment{Locally converged?} \label{alg:louvainlm--locally-converged}
    \EndIf
  \EndFor \label{alg:louvainlm--iterations-end}
  \Return{$l_i$} \label{alg:louvainlm--return}
\EndFunction \label{alg:louvainlm--move-end}

\Statex

\Function{scanCommunities}{$H_t, G', C', i, self$}
  \ForAll{$(j, w) \in G'.edges(i)$}
    \If{$self$ \textbf{or} $i \neq j$} $H_t \gets H_t[C'[j]] + w$
    \EndIf
  \EndFor
  \Return{$H_t$}
\EndFunction
\end{algorithmic}
\end{algorithm}


Lines \ref{alg:louvainlm--iterations-begin}-\ref{alg:louvainlm--iterations-end} encapsulate the primary loop of the local-moving phase. In line \ref{alg:louvainlm--init-deltaq}, we initialize the total delta-modularity per iteration $\Delta Q$. Subsequently, in lines \ref{alg:louvainlm--loop-vertices-begin}-\ref{alg:louvainlm--loop-vertices-end}, we concurrently iterate over unprocessed vertices. For each vertex $i$, we perform vertex pruning by marking $i$ as processed (line \ref{alg:louvainlm--prune}). Next, we verify if $i$ falls within the affected range (i.e., it is permitted to be incrementally marked as affected), and if not, we proceed to the next vertex (line \ref{alg:louvainlm--affrng}). For each unskipped vertex $i$, we scan communities connected to $i$ (line \ref{alg:louvainlm--scan}), excluding itself, ascertain the optimal community $c*$ to move $i$ to (line \ref{alg:louvainlm--best-community-begin}), compute the delta-modularity of moving $i$ to $c*$ (line \ref{alg:louvainlm--best-community-end}), update the community membership of $i$ (lines \ref{alg:louvainlm--perform-move-begin}-\ref{alg:louvainlm--perform-move-end}), and mark its neighbors as unprocessed (line \ref{alg:louvainlm--remark}) if a superior community is identified. It's worth noting that this practice of marking neighbors of $i$ as unprocessed, which is part of the vertex pruning optimization, also aligns with algorithm of DF Louvain --- which marks its neighbors as affected, when a vertex changes its community. Thus, vertex pruning facilitates incremental expansion of the set of affected vertices without requiring any extra code. In line \ref{alg:louvainlm--locally-converged}, we examine whether the local-moving phase has achieved convergence (locally); if so, the loop is terminated (or if $MAX\_ITERATIONS$ is reached). Finally, in line \ref{alg:louvainlm--return}, we return the number of iterations performed by the local-moving phase $l_i$.

\subsubsection{Aggregation phase of our Parallel Louvain}

Finally, the pseudocode for the aggregation phase is depicted in Algorithm \ref{alg:louvainag}, where communities are amalgamated into super-vertices. Here, the \texttt{louvai} \texttt{nAggregate()} function receives the current graph $G'$ and the community membership $C'$\ignore{as inputs}, and yields the super-vertex graph $G''$.

\begin{algorithm}[hbtp]
\caption{Aggregation phase of our Parallel Louvain \cite{sahu2023gvelouvain}.}
\label{alg:louvainag}
\begin{algorithmic}[1]
\Require{$G'$: Input/super-vertex graph}
\Require{$C'$: Community membership of each vertex}
\Ensure{$G'_{C'}$: Community vertices (CSR)}
\Ensure{$G''$: Super-vertex graph (weighted CSR)}
\Ensure{$*.offsets$: Offsets array of a CSR graph}
\Ensure{$H_t$: Collision-free per-thread hashtable}

\Statex

\Function{louvainAggregate}{$G', C'$}
  \State $\rhd$ Obtain vertices belonging to each community
  \State $G'_{C'}.offsets \gets countCommunityVertices(G', C')$ \label{alg:louvainag--coff-begin}
  \State $G'_{C'}.offsets \gets exclusiveScan(G'_{C'}.offsets)$ \label{alg:louvainag--coff-end}
  \ForAll{$i \in V'$ \textbf{in parallel}} \label{alg:louvainag--comv-begin}
    \State Add edge $(C'[i], i)$ to CSR $G'_{C'}$ atomically
  \EndFor \label{alg:louvainag--comv-end}
  \State $\rhd$ Obtain super-vertex graph
  \State $G''.offsets \gets communityTotalDegree(G', C')$ \label{alg:louvainag--yoff-begin}
  \State $G''.offsets \gets exclusiveScan(G''.offsets)$ \label{alg:louvainag--yoff-end}
  \State $|\Gamma| \gets$ Number of communities in $C'$
  \ForAll{$c \in [0, |\Gamma|)$ \textbf{in parallel}} \label{alg:louvainag--y-begin}
    \If{degree of $c$ in $G'_{C'} = 0$} \textbf{continue}
    \EndIf
    \State $H_t \gets \{\}$
    \ForAll{$i \in G'_{C'}.edges(c)$}
      \State $H_t \gets scanCommunities(H, G', C', i, true)$
    \EndFor
    \ForAll{$(d, w) \in H_t$}
      \State Add edge $(c, d, w)$ to CSR $G''$ atomically
    \EndFor
  \EndFor \label{alg:louvainag--y-end}
  \Return $G''$ \label{alg:louvainag--return}
\EndFunction
\end{algorithmic}
\end{algorithm}

In the algorithm, we begin by obtaining the offsets array for the community vertices CSR $G'_{C'}.offsets$ in lines \ref{alg:louvainag--coff-begin}-\ref{alg:louvainag--coff-end}. This involves first counting the number of vertices belonging to each community using \texttt{countCommunityVertices()}, followed by performing an exclusive scan operation on the array. Subsequently, in lines \ref{alg:louvainag--comv-begin}-\ref{alg:louvainag--comv-end}, we concurrently iterate over all vertices, atomically populating vertices associated with each community into the community graph CSR $G'_{C'}$. Next, we derive the offsets array for the super-vertex graph CSR by overestimating the degree of each super-vertex in lines \ref{alg:louvainag--yoff-begin}-\ref{alg:louvainag--yoff-end}, by computing the total degree of each community using \texttt{communityTotalDegree()} and then performing an exclusive scan on the array. As a result, the super-vertex graph CSR exhibits holes/intervals between the edges and weights array of each super-vertex. Following this, in lines \ref{alg:louvainag--y-begin}-\ref{alg:louvainag--y-end}, we iterate over all communities $c \in [0, |\Gamma|)$. Here, we add all communities $d$ (along with their associated edge weight $w$) linked to each vertex $i$ belonging to community $c$ (via \texttt{scanCommunities()} defined in Algorithm \ref{alg:louvainlm}) to the per-thread hashtable $H_t$. Once $H_t$ encompasses all communities (alongside their weights) linked to community $c$, we atomically append them as edges to super-vertex $c$ within the super-vertex graph $G''$. Finally, in line \ref{alg:louvainag--return}, we return the super-vertex graph $G''$.

\subsection{Updating vertex/community weights}
\label{sec:our-update}

We now discuss the parallel algorithm for obtaining the updated the weighted degree of each vertex $K^t$ and total edge weight of each community $\Sigma^t$, given the previous community memberships of vertices $C^{t-1}$, weighted-degrees of vertices $K^{t-1}$, total edge weights of communities, and the batch update (consisting of edge deletions $\Delta^{t-}$ and insertions $\Delta^{t+}$). Its psuedocode is in Algorithm \ref{alg:update}.

\begin{algorithm}[hbtp]
\caption{Updating vertex/community weights in parallel.}
\label{alg:update}
\begin{algorithmic}[1]
\Require{$G^t$: Current input graph}
\Require{$\Delta^{t-}, \Delta^{t+}$: Edge deletions and insertions (batch update)}
\Require{$C^{t-1}$: Previous community of each vertex}
\Require{$K^{t-1}$: Previous weighted-degree of each vertex}
\Require{$\Sigma^{t-1}$: Previous total edge weight of each community}
\Ensure{$K$: Updated weighted-degree of each vertex}
\Ensure{$\Sigma$: Updated total edge weight of each community}
\Ensure{$work_{th}$: Work-list of current thread}

\Statex

\Function{updateWeights}{$G^t, \Delta^{t-}, \Delta^{t+}, C^{t-1}, K^{t-1}, \Sigma^{t-1}$}
  \State $K \gets K^{t-1}$ \textbf{;} $\Sigma \gets \Sigma^{t-1}$ \label{alg:update--init}
  \ForAll{\textbf{threads in parallel}} \label{alg:update--loopdel-begin}
    \ForAll{$(i, j, w) \in \Delta^{t-}$}
      \State $c \gets C^{t-1}[i]$ \label{alg:update--delc}
      \If{$i \in work_{th}$} $K[i] \gets K[i] - w$ \label{alg:update--delk}
      \EndIf
      \If{$c \in work_{th}$} $\Sigma[c] \gets \Sigma[c] - w$ \label{alg:update--delsigma}
      \EndIf
    \EndFor \label{alg:update--loopdel-end}
    \ForAll{$(i, j, w) \in \Delta^{t+}$} \label{alg:update--loopins-begin}
      \State $c \gets C^{t-1}[i]$
      \If{$i \in work_{th}$} $K[i] \gets K[i] + w$
      \EndIf
      \If{$c \in work_{th}$} $\Sigma[c] \gets \Sigma[c] + w$
      \EndIf
    \EndFor
  \EndFor \label{alg:update--loopins-end}
  \Return $\{K, \Sigma\}$ \label{alg:update--return}
\EndFunction
\end{algorithmic}
\end{algorithm}

In the algorithm, we begin by initializing $K$ and $\Sigma$, the weighted-degree of each vertex, and the total edge weight of each community (line \ref{alg:update--init}). Then, across multiple threads, we iterate over both the sets of edge deletions $\Delta^{t-}$ (lines \ref{alg:update--loopdel-begin}-\ref{alg:update--loopdel-end}) and edge insertions $\Delta^{t+}$ (lines \ref{alg:update--loopins-begin}-\ref{alg:update--loopins-end}). For each edge deletion $(i, j, w)$ in $\Delta^{t-}$, we determine the community $c$ of vertex $i$ based on the previous community assignment $C^{t-1}$ (line \ref{alg:update--delc}). If $i$ belongs to the current thread's work-list, we decrement its weighted-degree by $w$ (line \ref{alg:update--delk}), and if community $c$ belongs to the work-list, we decrement its total edge weight by $w$ (line \ref{alg:update--delsigma}). Similarly, for each edge insertion $(i, j, w)$ in $\Delta^{t+}$, we adjust the vertex $i$'s weighted-degree and its community's total edge weight accordingly. Finally, we return the updated values of $K$ and $\Sigma$ for each vertex and community for further processing (line \ref{alg:update--return}).

\ignore{\subsubsection{Comparison with respect to Riedy and Bader \cite{com-riedy13}}}

\ignore{Riedy and Bader propose a batch parallel dynamic algorithm for community detection. They compare the run time of their dynamic algorithm to that of a static recomputation. On the graphs \verb|caidaRouterLevel|, \verb|coPapersDBLP|, and \verb|eu-2005|, from \cite{com-riedy13} and at the batch size of $0.1|E|, 0.03|E|,$ and $0.1|E|$ respectively, they report a speedup of $40\times$, $1.08\times$, and $327\times$, respectively, over their corresponding static algorithm performing a full recomputation. On these three graphs and batch sizes, \FroLou{} achieves a speedup of $6.1\times$, $10.9\times$, and $4.2\times$, respectively, compared to a full static recomputation. This might compare unfavorably with the speedups claimed by Riedy and Bader \cite{com-riedy13}. However, note from Section \ref{sec:introduction} that the algorithm of Riedy and Bader \cite{com-riedy13} does not identify cascading changes to communities. In addition, as their source code is not available, we could not do a more direct comparison.}

\subsection{Correctness of DF Louvain}
\label{sec:correctness}

\begin{figure}[hbtp]
  \centering
  \subfigure{
    \label{fig:dynamic-frontier-detailed}
    \includegraphics[width=0.78\linewidth]{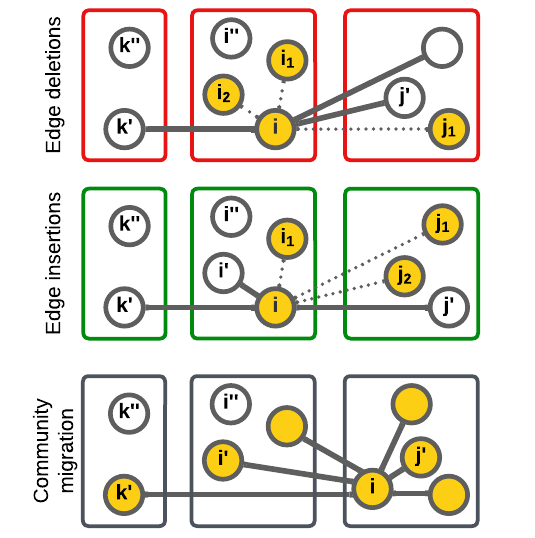}
  } \\[-2ex]
  \caption{A detailed illustration for presenting arguments on the correctness of \textit{Dynamic Frontier (DF)} Louvain.}
  \label{fig:frontier-approach}
\end{figure}

We now provide arguments for the correctness of Dynamic Frontier (DF) Louvain. To help with this, we refer the reader to Figure \ref{fig:frontier-approach}. Here, pre-existing edges are represented by solid lines, and $i$ represents a source vertex of edge deletions/insertions in the batch update. Edge deletions in the batch update with $i$ as the source vertex are shown in the top row (denoted by dashed lines), edge insertions are shown in the middle row (also denoted by dashed lines), and community migration of vertex $i$ is is shown in the bottom row. Vertices $i_n$ and $j_n$ represent the destination vertices (of edge deletions or insertions). Vertices $i'$, $j'$, and $k'$ signify neighboring vertices of vertex $i$. Finally, vertices $i''$, $j''$, and $k''$ represent non-neighbor vertices (to vertex $i$). Yellow highlighting is used to indicate vertices marked as affected, initially or in the current iteration of the Louvain algorithm.\ignore{We understand this figure is dense, but we tried to capture several details for the correctness arguments.}

Given a batch update consisting of edge deletions $\Delta^{t-}$ and insertions $\Delta^{t+}$, we now show that DF Louvain marks the essential vertices, which have an incentive to change their community membership, as affected. For any given vertex $i$ in the original graph (before the batch update), the delta-modularity of moving it from its current community $d$ to a new community $c$ is given by Equation \ref{eq:delta-modularity-clone}. We now consider the direct effect of each individual edge deletion $(i, j)$ or insertion $(i, j, w)$ in the batch update, on the delta-modularity of the a vertex, as well as the indirect cascading effect of migration of a vertex (to another community) on other vertices.

\begin{equation}
\label{eq:delta-modularity-clone}
  \Delta Q_{i: d \rightarrow c}
  = \frac{1}{m} (K_{i \rightarrow c} - K_{i \rightarrow d}) - \frac{K_i}{2m^2} (K_i + \Sigma_c - \Sigma_d)
\end{equation}

\subsubsection{On edge deletion}

\begin{lemma}
\label{thm:louvain--mark-deletion}
Given an edge deletion $(i, j)$ between vertices $i$ and $j$ belonging to the same community $d$, vertex $i$ (and $j$) should be marked as affected.
\end{lemma}

Consider the case of edge deletion $(i, j)$ of weight $w$ between vertices $i$ and $j$ belonging to the same community $C_i = C_j = d$ (see Figure \ref{fig:frontier-approach}, where $j = i_1$). Let $i''$ be a vertex belonging $i$'s community $C_{i''} = d$, and let $k''$ be a vertex belonging to another community $C_{k''} = b$. As shown below in Case \textbf{(1)}, the delta-modularity of vertex $i$ moving from its original community $d$ to another community $b$ has a significant positive factor $w/m$. There is thus a chance that vertex $i$ would change its community membership, and we should mark it as affected. The same argument applies for vertex $j$, as the edge is undirected. On the other hand, for the Cases \textbf{(2)}-\textbf{(3)}, there is only a small positive change in delta-modularity for vertex $k''$. Thus, there is little incentive for vertex $k''$ to change its community membership, and no incentive for a change in community membership of vertex $i''$.

Note that it is possible that the community $d$ would split due to the edge deletion. However, this is unlikely, given that one would need a large number of edge deletions between vertices belonging to the same community for the community to split. One can take care of such rare events by running Static Louvain, say every $1000$ batch updates, which also helps us ensure high-quality communities. The same applies to Delta-screening (DS) Louvain.

\begin{enumerate}
  \item $\Delta Q_{i:d \rightarrow b}^{new} = \Delta Q_{i:d \rightarrow b} + [\frac{w}{m}] + \frac{w}{2m^2} (\Sigma_c - \Sigma_d + w)$
  \item $\Delta Q_{i'':d \rightarrow b}^{new} = \Delta Q_{i'':d \rightarrow b} - \frac{wK_{i''}}{m^2}$
  \item $\Delta Q_{k'':b \rightarrow d}^{new} = \Delta Q_{k'':b \rightarrow d} + \frac{wK_{k''}}{m^2}$
\end{enumerate}

Now, consider the case of edge deletion $(i, j)$ between vertices $i$ and $j$ belonging to different communities, i.e., $C_i = d$, $C_j = c$ (see Figure \ref{fig:frontier-approach}, where $j = j_2$ or $j_3$). Let $i''$ be a vertex belonging to $i$'s community $C_{i''} = d$, $j''$ be a vertex belonging to $j$'s community $C_{j''} = c$, and $k''$ be a vertex belonging another community $C_{k''} = b$. As shown in Cases \textbf{(4)}-\textbf{(8)}, due to the absence of any significant positive change in delta-modularity, there is little to no incentive for vertices $i$, $j$, $k''$, $i''$, and $j''$ to change their community membership.

\begin{enumerate}[start=4]
  \item $\Delta Q_{i:d \rightarrow c}^{new} = \Delta Q_{i:d \rightarrow c} - \frac{w}{m} + \frac{w}{2m^2} (2K_i + \Sigma_c - \Sigma_d - w)$
  \item $\Delta Q_{i:d \rightarrow b}^{new} = \Delta Q_{i:d \rightarrow b} + \frac{w}{2m^2} (K_i + \Sigma_b - \Sigma_d)$
  \item $\Delta Q_{i'':d \rightarrow c}^{new} = \Delta Q_{i'':d \rightarrow c}$
  \item $\Delta Q_{i'':d \rightarrow b}^{new} = \Delta Q_{i'':d \rightarrow b} - \frac{wK_{i''}}{2m^2}$
  \item $\Delta Q_{k'':b \rightarrow d/c}^{new} = \Delta Q_{k'':b \rightarrow d/c} + \frac{wK_{k''}}{m^2}$ \hfill $\diamond$
\end{enumerate}

\subsubsection{On edge insertion}

\begin{lemma}
\label{thm:louvain--mark-insertion}
Given an edge insertion $(i, j, w)$ between vertices $i$ and $j$ belonging to different communities $d$ and $c$, vertex $i$ (and $j$) should be marked as affected.
\end{lemma}

Let us consider the case of edge insertion $(i, j, w)$ between vertices $i$ and $j$ belonging to different communities $C_i = d$ and $C_j = c$ respectively (see Figure \ref{fig:frontier-approach}, where $j = j_3$). Let $i''$ be a vertex belonging $i$'s community $C_{i''} = d$, $j''$ be a vertex belonging to $j$'s community $C_{j''} = c$, and $k''$ be a vertex belonging to another community $C_{k''} = b$. As shown below in Case \textbf{(9)}, we have a significant positive factor $w/m$ (and a small negative factor) which increases the delta-modularity of vertex $i$ moving to $j$'s community after the insertion of the edge $(i, j)$. There is, therefore, incentive for vertex $i$ to change its community membership. Accordingly, we mark $i$ as affected. Again, the same argument applies for vertex $j$, as the edge is undirected. Further, we observe from other Cases \textbf{(10)}-\textbf{(13)} there is only a small change in delta-modularity. Thus, there is hardly any to no incentive for a change in community membership of vertices $i''$, $j''$, and $k''$.

\begin{enumerate}[start=9]
  \item $\Delta Q_{i:d \rightarrow c}^{new} = \Delta Q_{i:d \rightarrow c} + [\frac{w}{m}] - \frac{w}{2m^2} (2K_i + \Sigma_c - \Sigma_d + w)$
  \item $\Delta Q_{i:d \rightarrow b}^{new} = \Delta Q_{i:d \rightarrow b} - \frac{w}{2m^2} (K_i + \Sigma_b - \Sigma_d)$
  \item $\Delta Q_{i'':d \rightarrow c}^{new} = \Delta Q_{i'':d \rightarrow c}$
  \item $\Delta Q_{i'':d \rightarrow b}^{new} = \Delta Q_{i'':d \rightarrow b} + \frac{wK_{i''}}{2m^2}$
  \item $\Delta Q_{k'':b \rightarrow d/c}^{new} = \Delta Q_{k'':b \rightarrow d/c} - \frac{wK_{k''}}{2m^2}$
\end{enumerate}

Now, consider the case of edge insertion $(i, j, w)$ between vertices $i$ and $j$ belonging to the same community $C_i = C_j = d$ (see Figure \ref{fig:frontier-approach}, where $j = i_1$ or $i_2$). From Cases \textbf{(14)}-\textbf{(16)}, we note that it is little to no incentive for vertices $i''$, $k''$, $i$, and $j$ to change their community membership. Note that it is possible for the insertion of edges within the same community to cause it to split into two more strongly connected communities, but it is very unlikely.

\begin{enumerate}[start=14]
  \item $\Delta Q_{i:d \rightarrow b}^{new} = \Delta Q_{i:d \rightarrow b} - \frac{w}{m} - \frac{w}{2m^2} (\Sigma_c - \Sigma_d - w)$
  \item $\Delta Q_{i'':d \rightarrow b}^{new} = \Delta Q_{i'':d \rightarrow b} + \frac{wK_{i''}}{m^2}$
  \item $\Delta Q_{k'':b \rightarrow d}^{new} = \Delta Q_{k'':b \rightarrow d} - \frac{wK_{k''}}{m^2}$ \hfill $\diamond$
\end{enumerate}

\subsubsection{On vertex migration to another community}

\begin{lemma}
\label{thm:louvain--remark}
When a vertex $i$ changes its community membership, and vertex $j$ is its neighbor, $j$ should be marked as affected.
\end{lemma}

We considered the direct effects of deletion and insertion of edges above. Now we consider its indirect effects by studying the impact of change in community membership of one vertex on the other vertices. Consider the case where a vertex $i$ changes its community membership from its previous community $d$ to a new community $c$ (see Figure \ref{fig:frontier-approach}). Let $i'$ be a neighbor of $i$ and $i''$ be a non-neighbor of $i$ belonging to $i$'s previous community $C_{i'} = C_{i''} = d$, $j'$ be a neighbor of $i$ and $j''$ be a non-neighbor of $i$ belonging to $i$'s new community $C_{j'} = C_{j''} = c$, $k'$ be a neighbor of $i$ and $k''$ be a non-neighbor of $i$ belonging to another community $C_{k'} = C_{k''} = b$. From Cases \textbf{(17)}-\textbf{(22)}, we note that neighbors $i'$ and $k'$ have an incentive to change their community membership (as thus necessitate marking), but not $j'$. However, to keep the algorithm simple, we simply mark all the neighbors of vertex $i$ as affected.

\begin{enumerate}[start=17]
  \item $\Delta Q_{i':d \rightarrow c}^{new} = \Delta Q_{i':d \rightarrow c} + [\frac{2w_{ii'}}{m}] - \frac{K_iK_{i'}}{m^2}$
  \item $\Delta Q_{i':d \rightarrow b}^{new} = \Delta Q_{i':d \rightarrow b} + [\frac{w_{ii'}}{m}] - \frac{K_iK_{i'}}{2m^2}$
  \item $\Delta Q_{j':c \rightarrow d}^{new} = \Delta Q_{j':c \rightarrow d} - \frac{2w_{ij'}}{m} + \frac{K_iK_{j'}}{m^2}$
  \item $\Delta Q_{j':c \rightarrow b}^{new} = \Delta Q_{j':c \rightarrow b} - \frac{w_{ij'}}{m} + \frac{K_iK_{j'}}{2m^2}$
  \item $\Delta Q_{k':b \rightarrow d}^{new} = \Delta Q_{k':b \rightarrow d} - \frac{w_{ik'}}{m} + \frac{K_iK_{k'}}{2m^2}$
  \item $\Delta Q_{k':b \rightarrow c}^{new} = \Delta Q_{k':b \rightarrow c} + [\frac{w_{ik'}}{m}] - \frac{K_iK_{k'}}{2m^2}$
\end{enumerate}

Further, from Cases \textbf{(23)}-\textbf{(28)}, we note that there is hardly any incentive for a change in community membership of vertices $i''$, $j''$, and $k''$. This is due to the change in delta-modularity being insignificant. There could still be an indirect cascading impact, where a common neighbor between vertices $i$ and $j$ would change its community, which could eventually cause vertex $j$ to change its community as well \cite{com-zarayeneh21}. However, this case is automatically taken care of as we perform marking of affected vertices during the community detection process.

\begin{enumerate}[start=23]
  \item $\Delta Q_{i'':d \rightarrow c}^{new} = \Delta Q_{i'':d \rightarrow c} + \frac{K_iK_{i''}}{m^2}$
  \item $\Delta Q_{i'':d \rightarrow b}^{new} = \Delta Q_{i'':d \rightarrow b} - \frac{K_iK_{i''}}{2m^2}$
  \item $\Delta Q_{j'':c \rightarrow d}^{new} = \Delta Q_{j'':c \rightarrow d} + \frac{K_iK_{j''}}{m^2}$
  \item $\Delta Q_{j'':c \rightarrow b}^{new} = \Delta Q_{j'':c \rightarrow b} + \frac{K_iK_{j''}}{2m^2}$
  \item $\Delta Q_{k'':b \rightarrow d}^{new} = \Delta Q_{k'':b \rightarrow d} + \frac{K_iK_{k''}}{2m^2}$
  \item $\Delta Q_{k'':b \rightarrow c}^{new} = \Delta Q_{k'':b \rightarrow c} - \frac{K_iK_{k''}}{2m^2}$ \hfill $\diamond$
\end{enumerate}

\subsubsection{Overall}

Finally, based on Observations \ref{thm:louvain--mark-deletion}, \ref{thm:louvain--mark-insertion}, and \ref{thm:louvain--remark}, we can state the following for DF Louvain.

\begin{theorem}
\label{thm:louvain}
Given a batch update, DF Louvain marks vertices having an incentive to change their community\ignore{membership} as affected. \qed
\end{theorem}

We note that with DF Louvain, without any direct link to vertices in the frontier, outlier vertices may not be marked as affected --- even if they have the potential to change community. Such outliers may be weakly connected to multiple communities, and if the current community becomes weakly (or less strongly) connected, they may leave and join some other community. It may also be noted that DS Louvain is also an approximate method and can miss certain outliers. In practice, however, we see little to no impact of this approximation of the affected subset of the graph on the final quality (modularity) of the communities obtained, as shown in Section \ref{sec:evaluation}.

\begin{figure*}[!hbt]
  \centering
  \subfigure[Runtime on consecutive batch updates of size $10^{-5}|E_T|$]{
    \label{fig:temporal-sx-mathoverflow--runtime5}
    \includegraphics[width=0.48\linewidth]{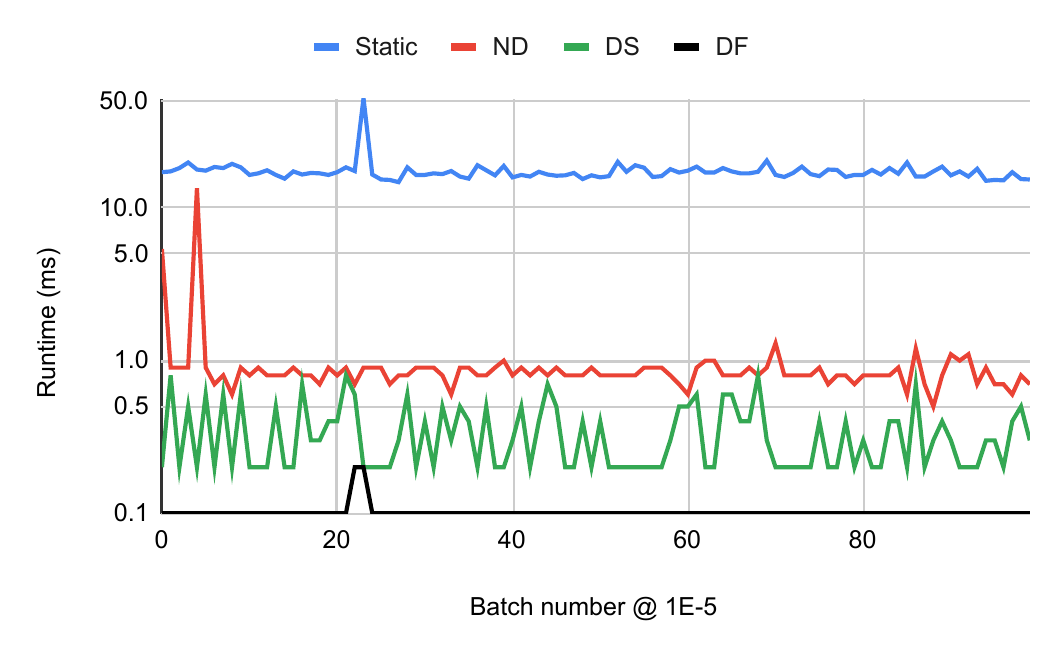}
  }
  \subfigure[Modularity of communities obtained on consecutive batch updates of size $10^{-5}|E_T|$]{
    \label{fig:temporal-sx-mathoverflow--modularity5}
    \includegraphics[width=0.48\linewidth]{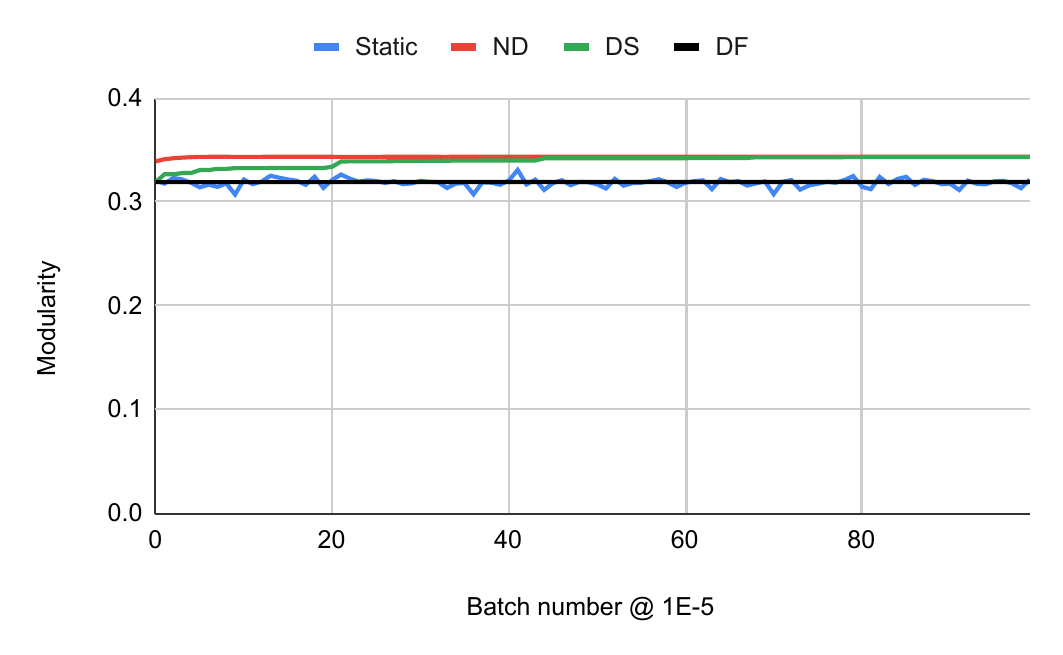}
  } \\[2ex]
  \subfigure[Runtime on consecutive batch updates of size $10^{-4}|E_T|$]{
    \label{fig:temporal-sx-mathoverflow--runtime4}
    \includegraphics[width=0.48\linewidth]{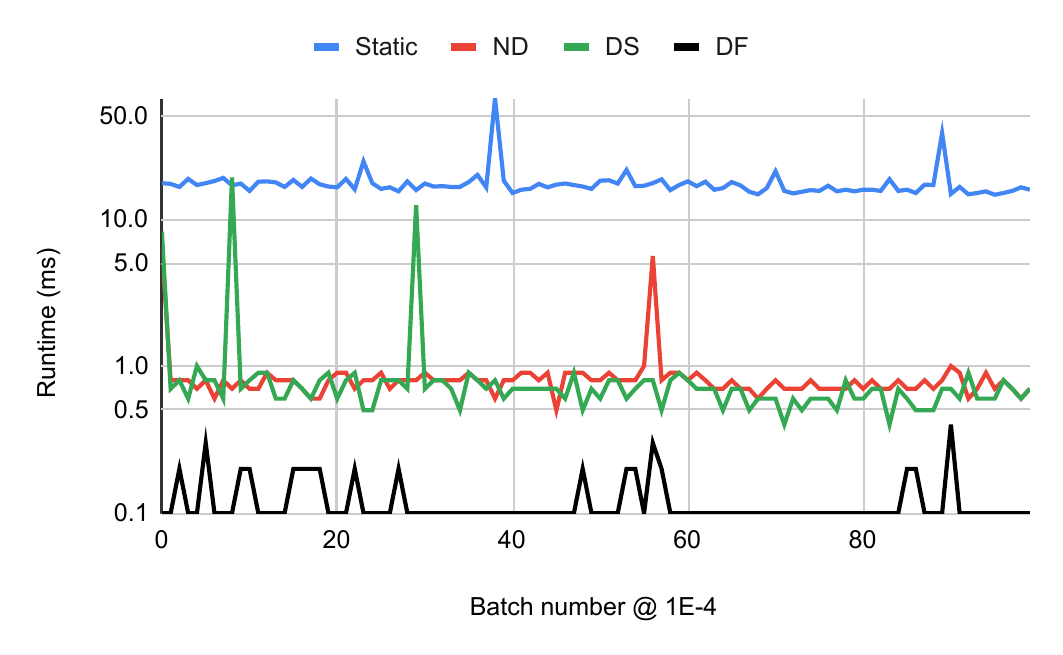}
  }
  \subfigure[Modularity of communities obtained on consecutive batch updates of size $10^{-4}|E_T|$]{
    \label{fig:temporal-sx-mathoverflow--modularity4}
    \includegraphics[width=0.48\linewidth]{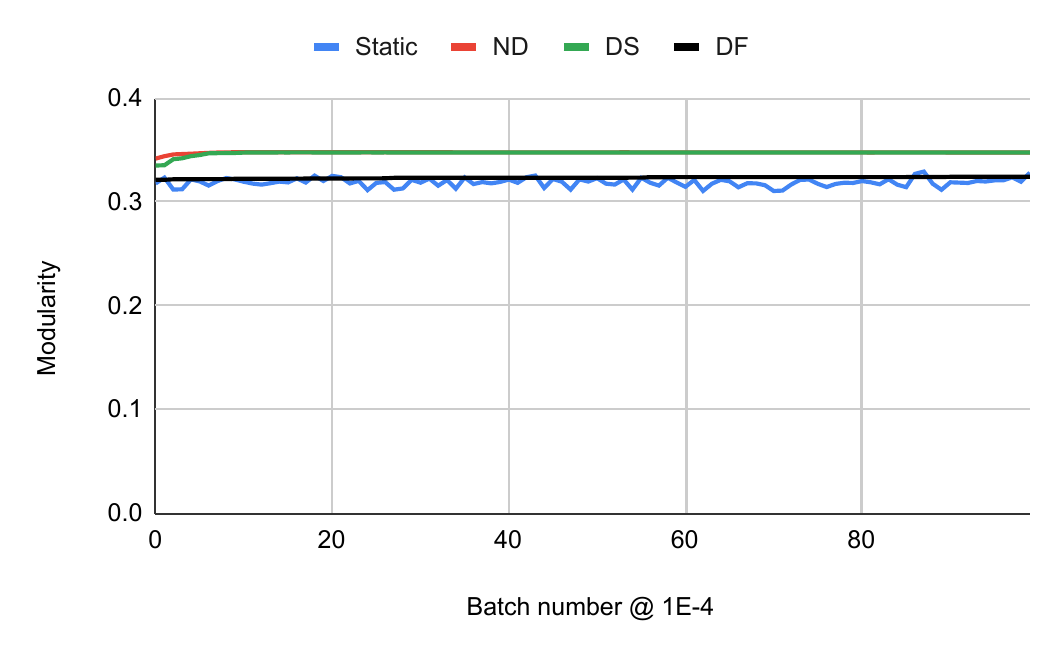}
  } \\[2ex]
  \subfigure[Runtime on consecutive batch updates of size $10^{-3}|E_T|$]{
    \label{fig:temporal-sx-mathoverflow--runtime3}
    \includegraphics[width=0.48\linewidth]{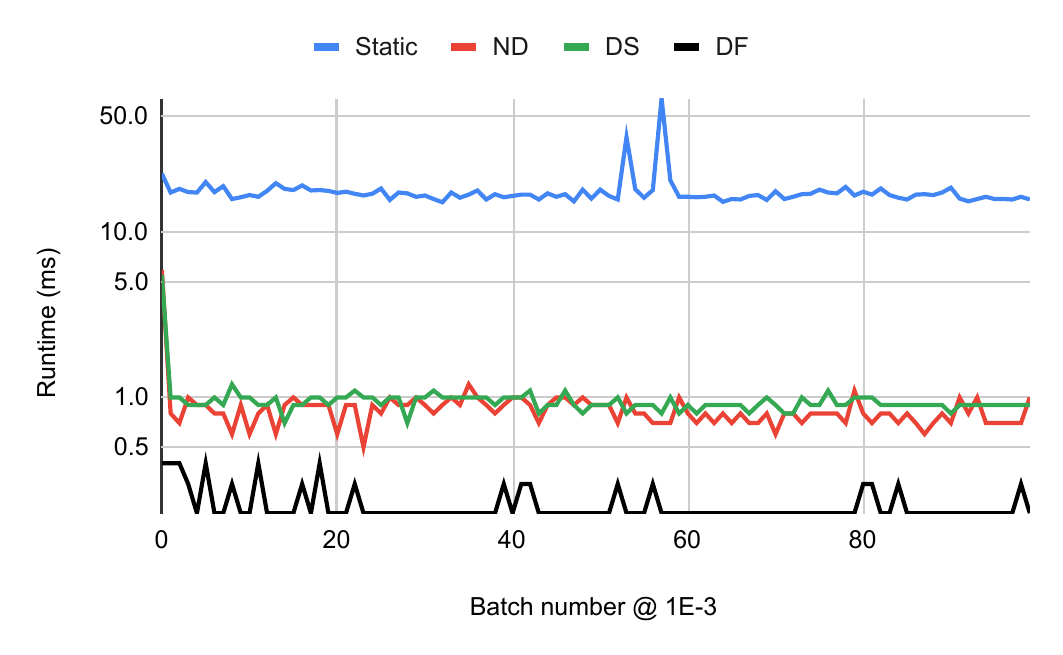}
  }
  \subfigure[Modularity of communities obtained on consecutive batch updates of size $10^{-3}|E_T|$]{
    \label{fig:temporal-sx-mathoverflow--modularity3}
    \includegraphics[width=0.48\linewidth]{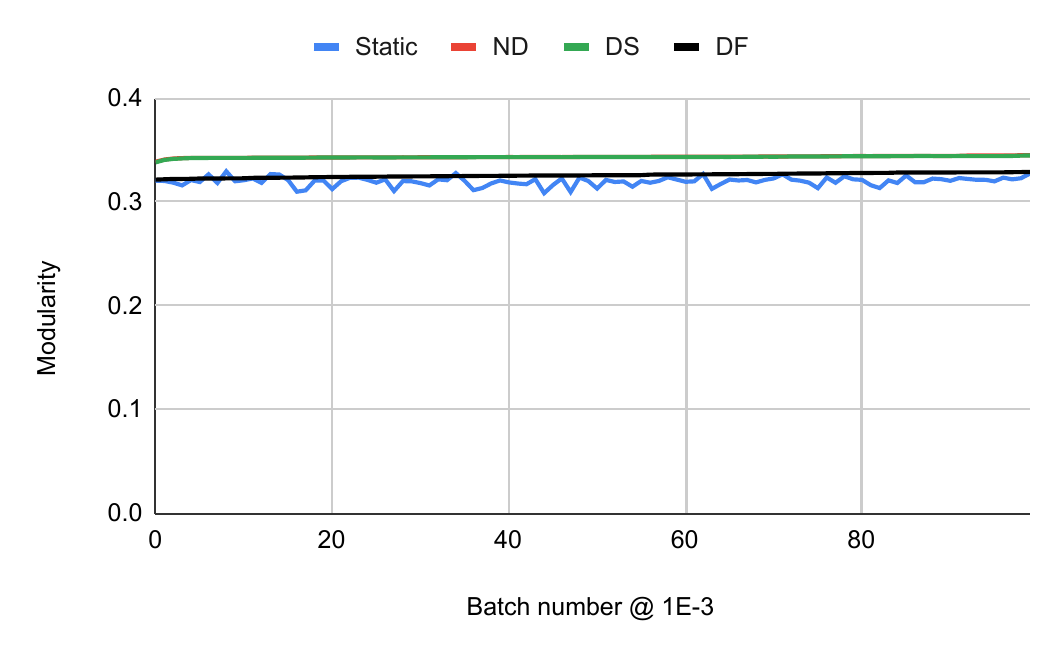}
  } \\[-2ex]
  \caption{Runtime and Modularity of communities obtained with \textit{Static}, \textit{Naive-dynamic (ND)}, \textit{Delta-screening (DS)}, and \textit{Dynamic Frontier (DF)} Louvain on the \textit{sx-mathoverflow} dynamic graph. The size of batch updates range from $10^{-5}|E_T|$ to $10^{-3}|E_T|$.}
  \label{fig:temporal-sx-mathoverflow}
\end{figure*}

\begin{figure*}[!hbt]
  \centering
  \subfigure[Runtime on consecutive batch updates of size $10^{-5}|E_T|$]{
    \label{fig:temporal-sx-askubuntu--runtime5}
    \includegraphics[width=0.48\linewidth]{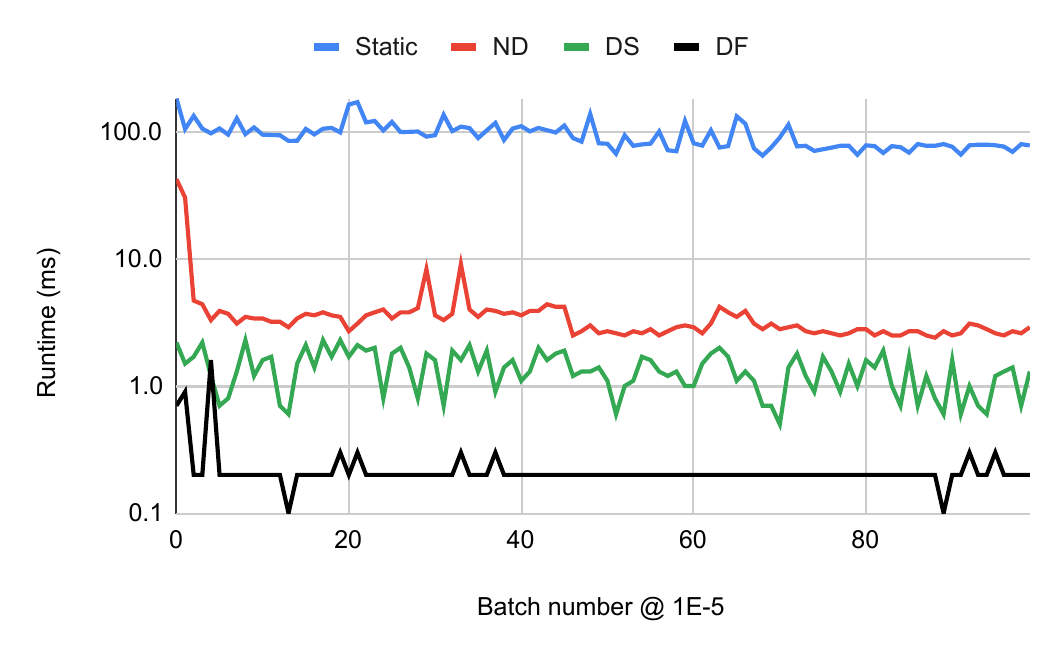}
  }
  \subfigure[Modularity of communities obtained on consecutive batch updates of size $10^{-5}|E_T|$]{
    \label{fig:temporal-sx-askubuntu--modularity5}
    \includegraphics[width=0.48\linewidth]{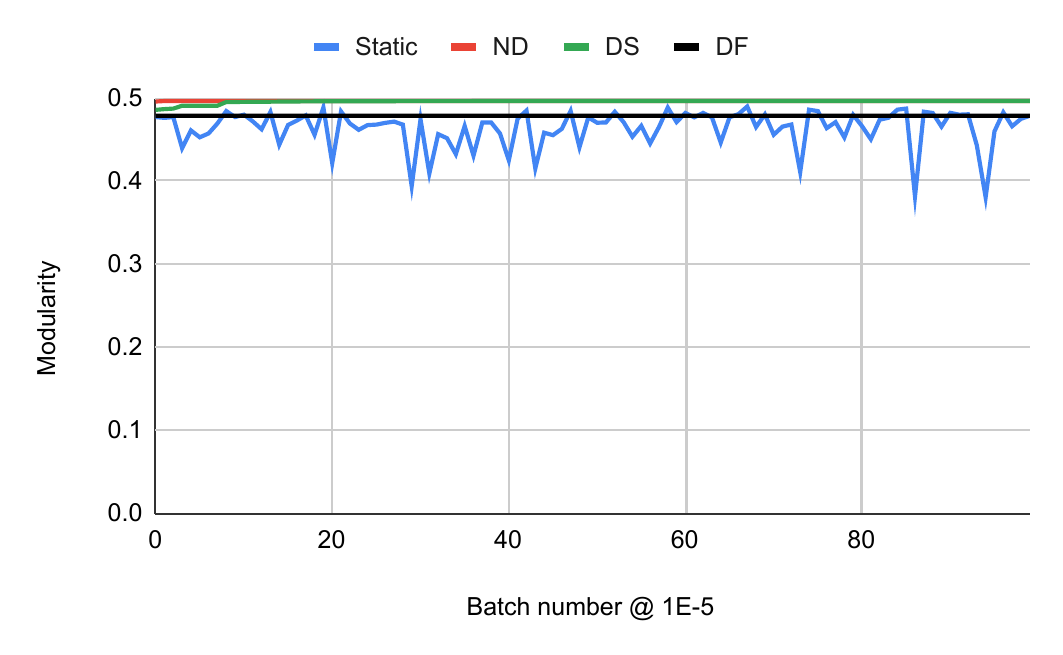}
  } \\[2ex]
  \subfigure[Runtime on consecutive batch updates of size $10^{-4}|E_T|$]{
    \label{fig:temporal-sx-askubuntu--runtime4}
    \includegraphics[width=0.48\linewidth]{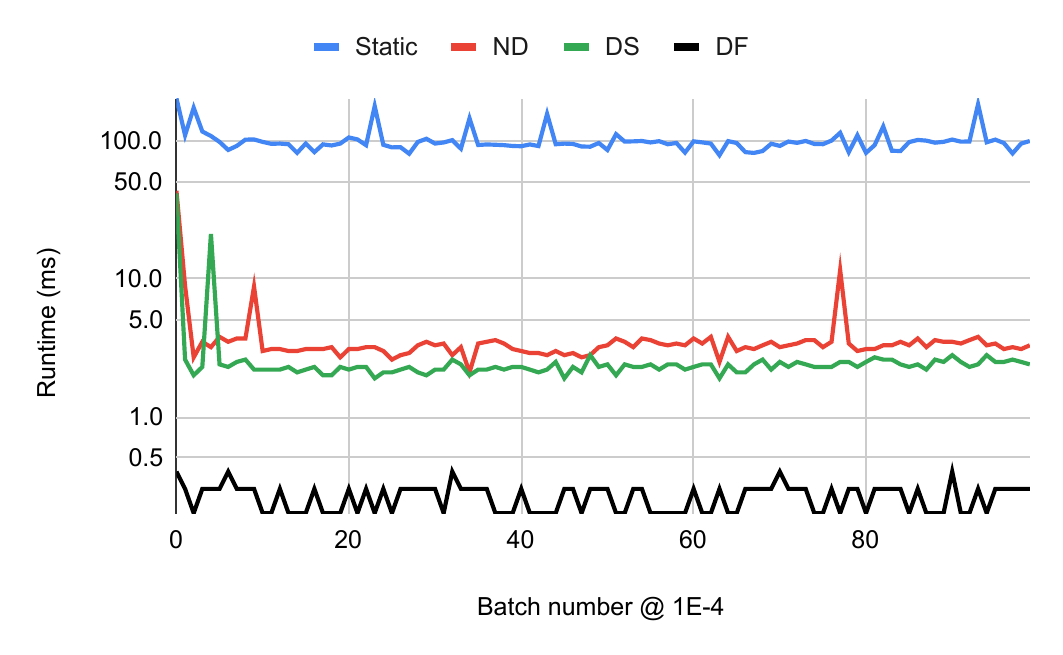}
  }
  \subfigure[Modularity of communities obtained on consecutive batch updates of size $10^{-4}|E_T|$]{
    \label{fig:temporal-sx-askubuntu--modularity4}
    \includegraphics[width=0.48\linewidth]{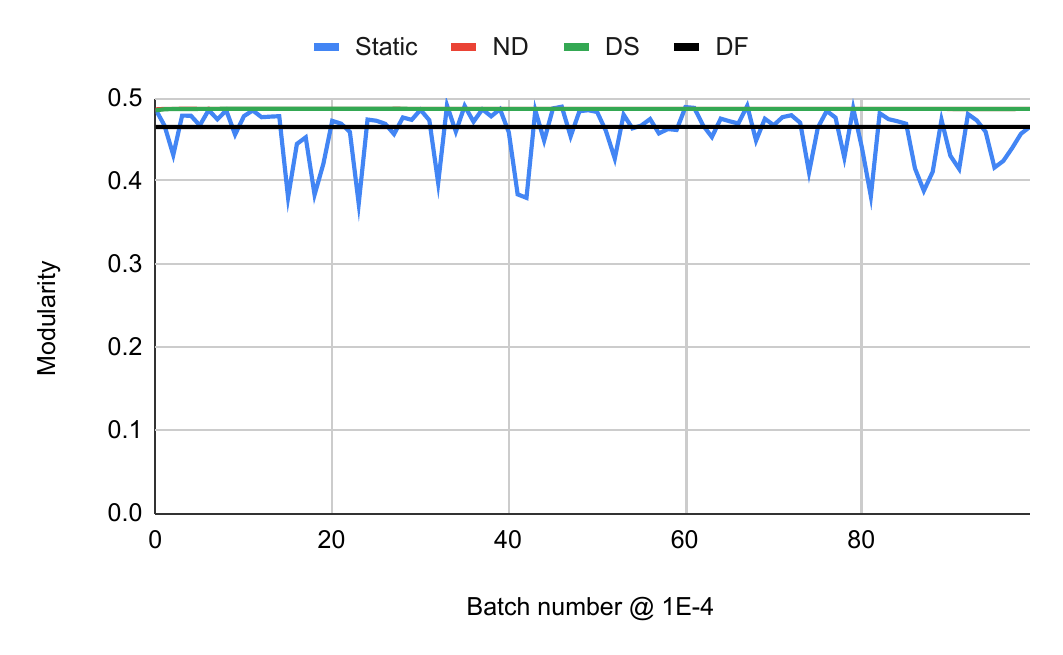}
  } \\[2ex]
  \subfigure[Runtime on consecutive batch updates of size $10^{-3}|E_T|$]{
    \label{fig:temporal-sx-askubuntu--runtime3}
    \includegraphics[width=0.48\linewidth]{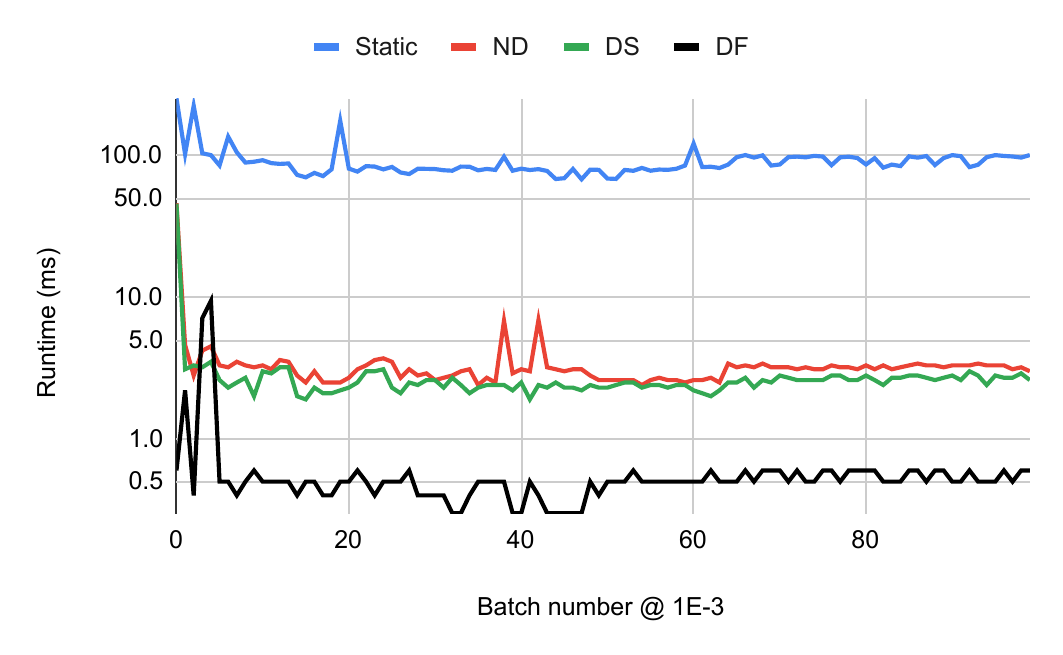}
  }
  \subfigure[Modularity of communities obtained on consecutive batch updates of size $10^{-3}|E_T|$]{
    \label{fig:temporal-sx-askubuntu--modularity3}
    \includegraphics[width=0.48\linewidth]{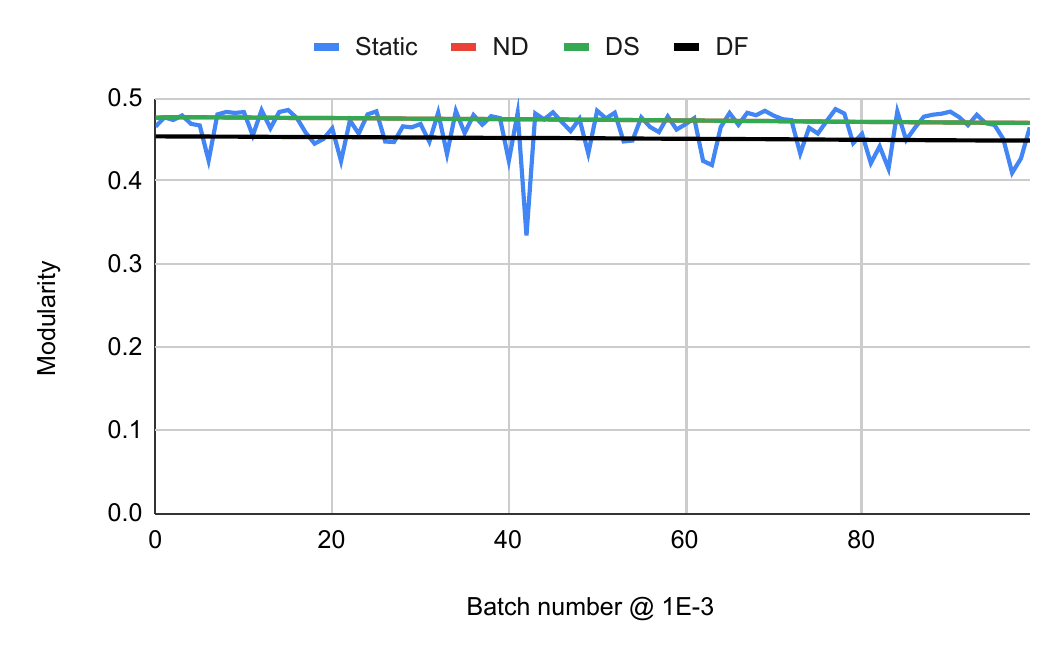}
  } \\[-2ex]
  \caption{Runtime and Modularity of communities obtained with \textit{Static}, \textit{Naive-dynamic (ND)}, \textit{Delta-screening (DS)}, and \textit{Dynamic Frontier (DF)} Louvain on the \textit{sx-askubuntu} dynamic graph. The size of batch updates range from $10^{-5}|E_T|$ to $10^{-3}|E_T|$.}
  \label{fig:temporal-sx-askubuntu}
\end{figure*}

\begin{figure*}[!hbt]
  \centering
  \subfigure[Runtime on consecutive batch updates of size $10^{-5}|E_T|$]{
    \label{fig:temporal-sx-superuser--runtime5}
    \includegraphics[width=0.48\linewidth]{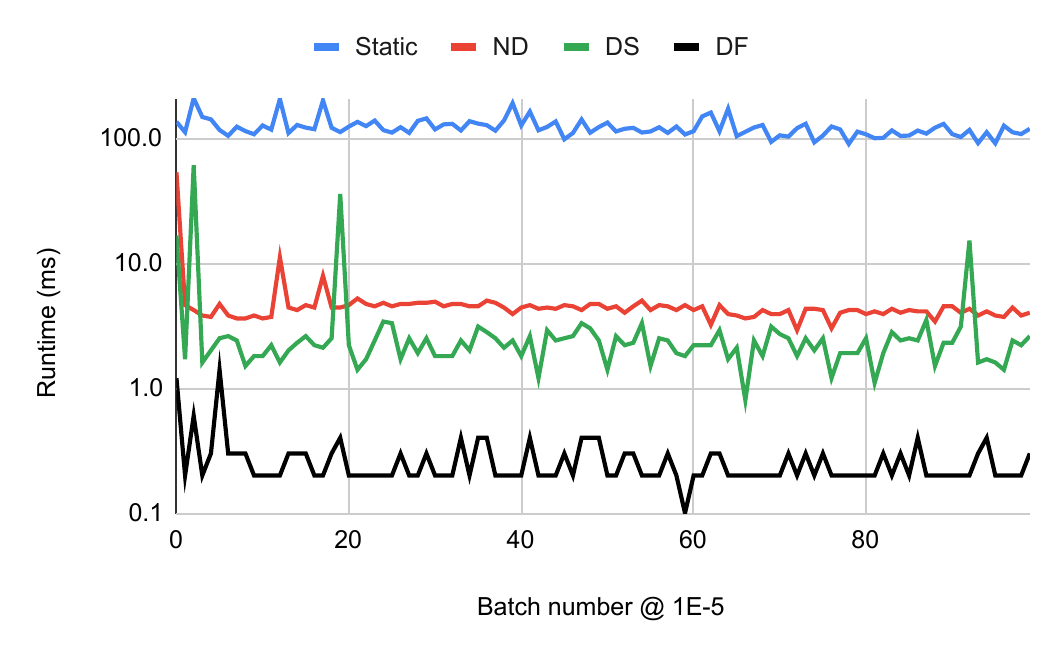}
  }
  \subfigure[Modularity of communities obtained on consecutive batch updates of size $10^{-5}|E_T|$]{
    \label{fig:temporal-sx-superuser--modularity5}
    \includegraphics[width=0.48\linewidth]{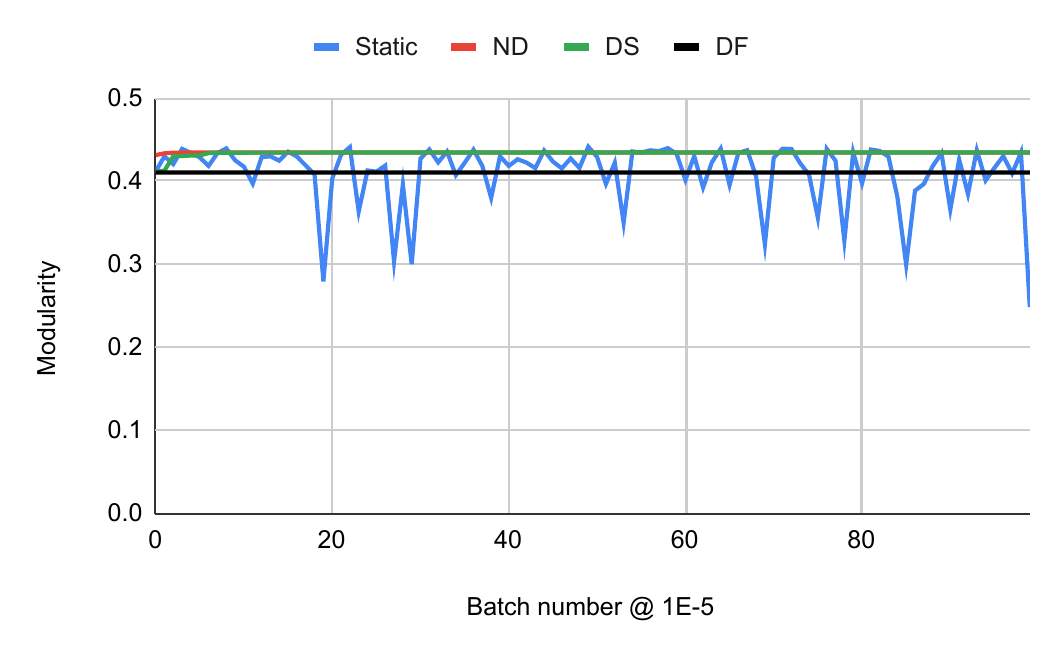}
  } \\[2ex]
  \subfigure[Runtime on consecutive batch updates of size $10^{-4}|E_T|$]{
    \label{fig:temporal-sx-superuser--runtime4}
    \includegraphics[width=0.48\linewidth]{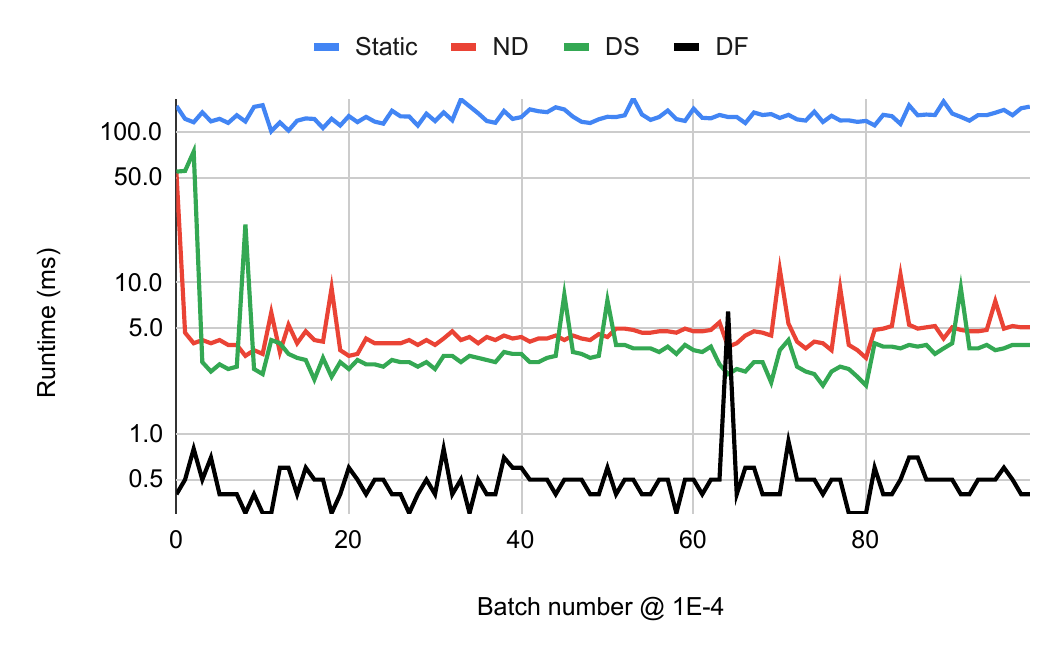}
  }
  \subfigure[Modularity of communities obtained on consecutive batch updates of size $10^{-4}|E_T|$]{
    \label{fig:temporal-sx-superuser--modularity4}
    \includegraphics[width=0.48\linewidth]{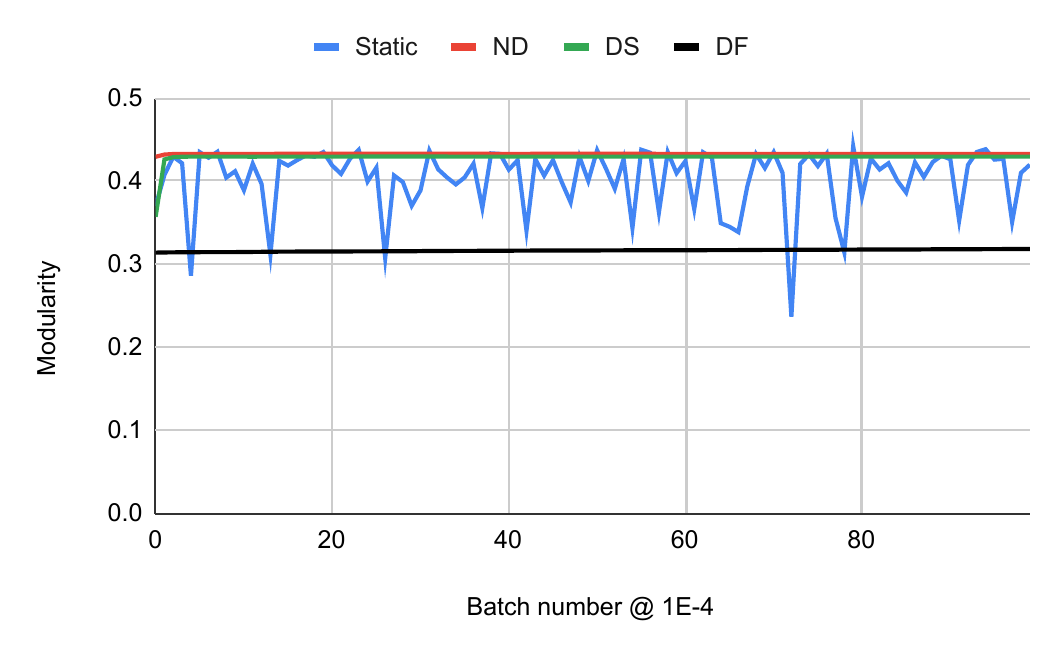}
  } \\[2ex]
  \subfigure[Runtime on consecutive batch updates of size $10^{-3}|E_T|$]{
    \label{fig:temporal-sx-superuser--runtime3}
    \includegraphics[width=0.48\linewidth]{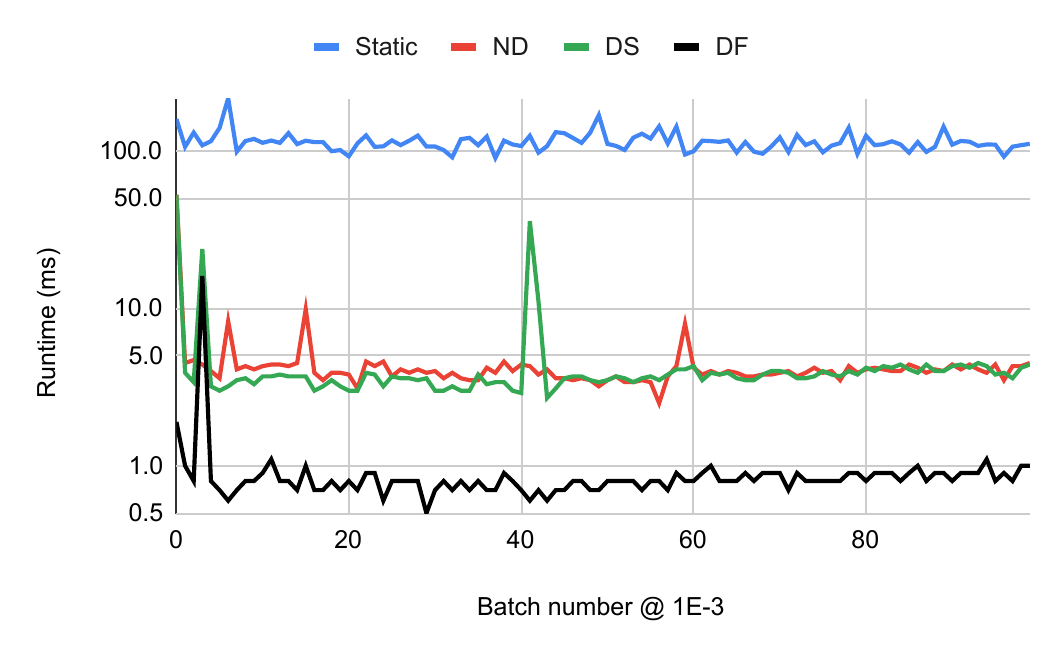}
  }
  \subfigure[Modularity of communities obtained on consecutive batch updates of size $10^{-3}|E_T|$]{
    \label{fig:temporal-sx-superuser--modularity3}
    \includegraphics[width=0.48\linewidth]{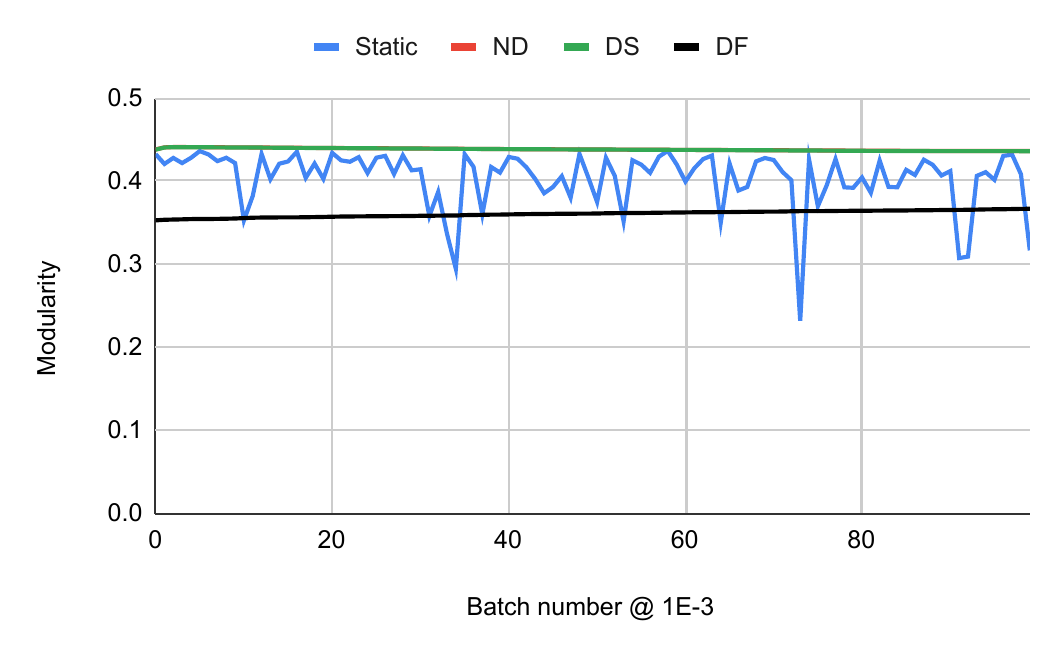}
  } \\[-2ex]
  \caption{Runtime and Modularity of communities obtained with \textit{Static}, \textit{Naive-dynamic (ND)}, \textit{Delta-screening (DS)}, and \textit{Dynamic Frontier (DF)} Louvain on the \textit{sx-superuser} dynamic graph. The size of batch updates range from $10^{-5}|E_T|$ to $10^{-3}|E_T|$.}
  \label{fig:temporal-sx-superuser}
\end{figure*}

\begin{figure*}[!hbt]
  \centering
  \subfigure[Runtime on consecutive batch updates of size $10^{-5}|E_T|$]{
    \label{fig:temporal-wiki-talk-temporal--runtime5}
    \includegraphics[width=0.48\linewidth]{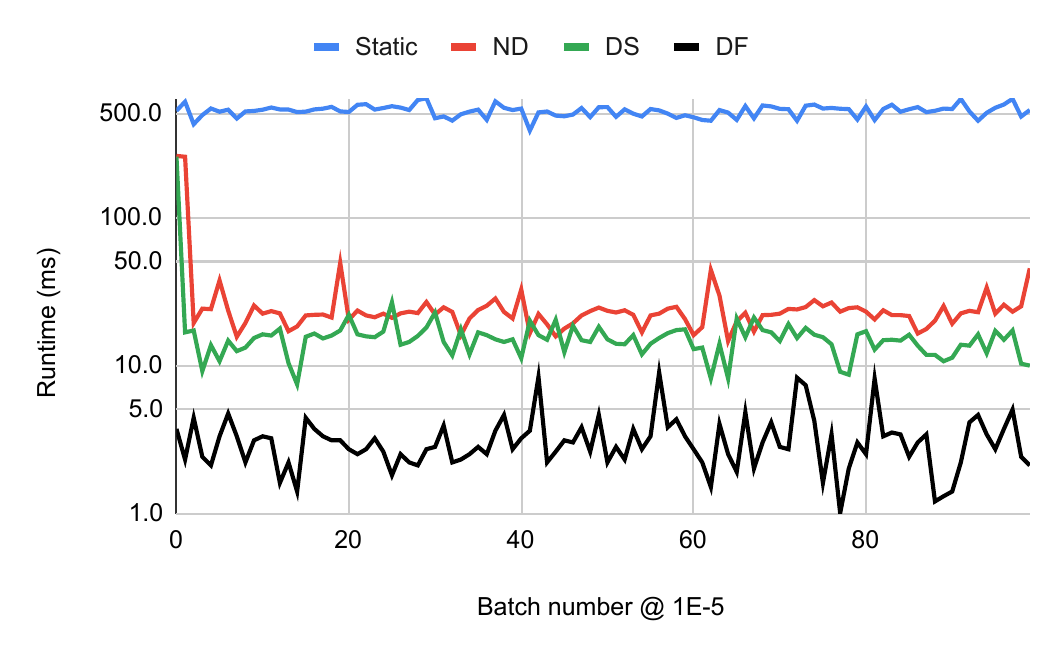}
  }
  \subfigure[Modularity of communities obtained on consecutive batch updates of size $10^{-5}|E_T|$]{
    \label{fig:temporal-wiki-talk-temporal--modularity5}
    \includegraphics[width=0.48\linewidth]{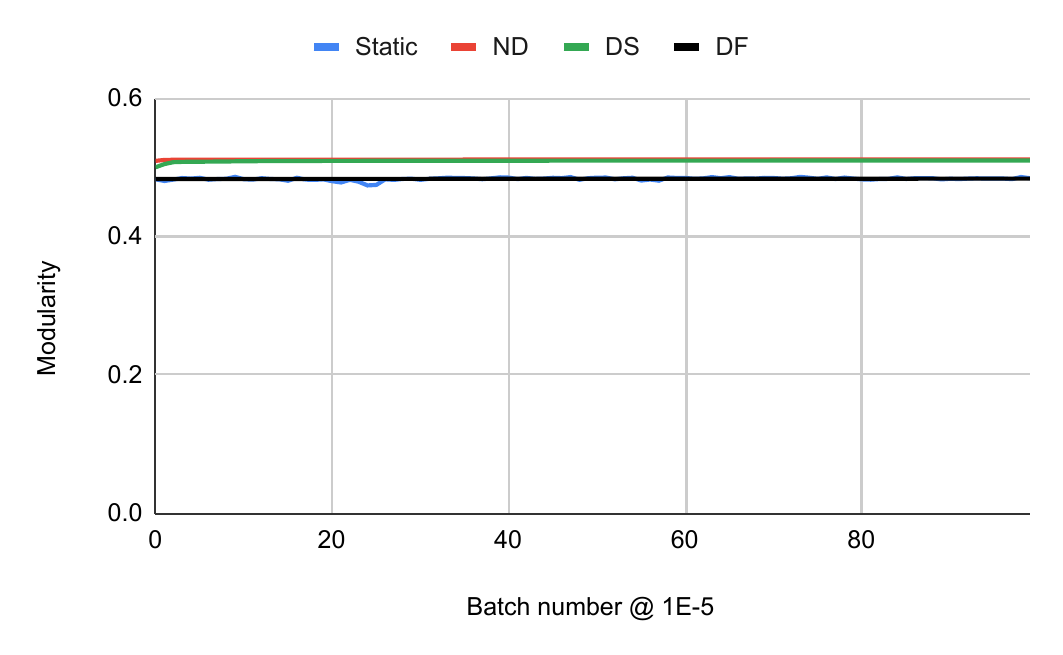}
  } \\[2ex]
  \subfigure[Runtime on consecutive batch updates of size $10^{-4}|E_T|$]{
    \label{fig:temporal-wiki-talk-temporal--runtime4}
    \includegraphics[width=0.48\linewidth]{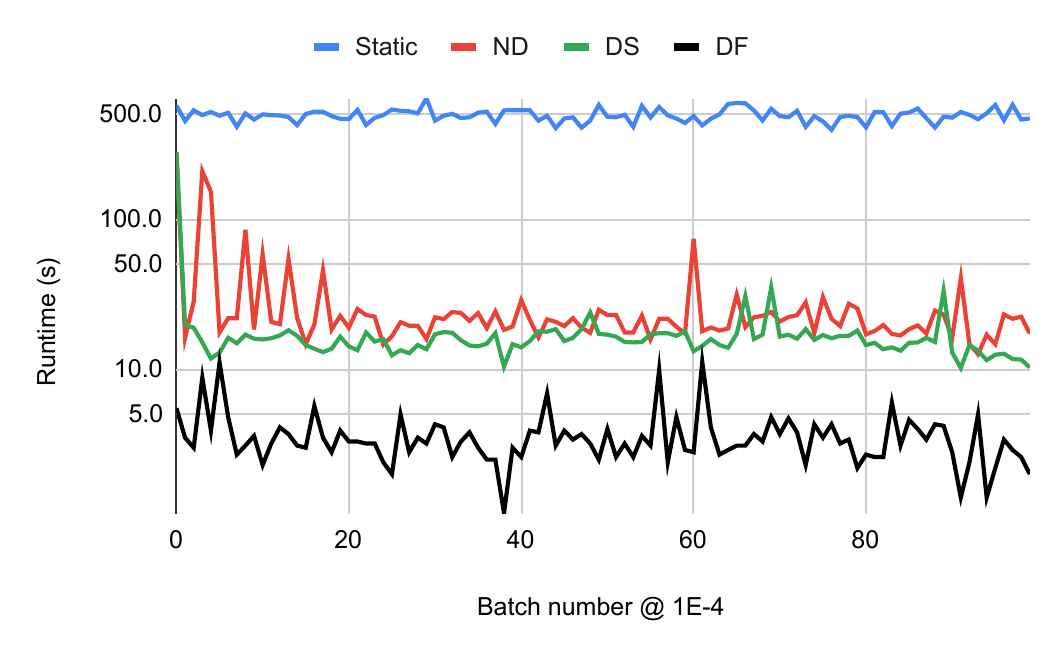}
  }
  \subfigure[Modularity of communities obtained on consecutive batch updates of size $10^{-4}|E_T|$]{
    \label{fig:temporal-wiki-talk-temporal--modularity4}
    \includegraphics[width=0.48\linewidth]{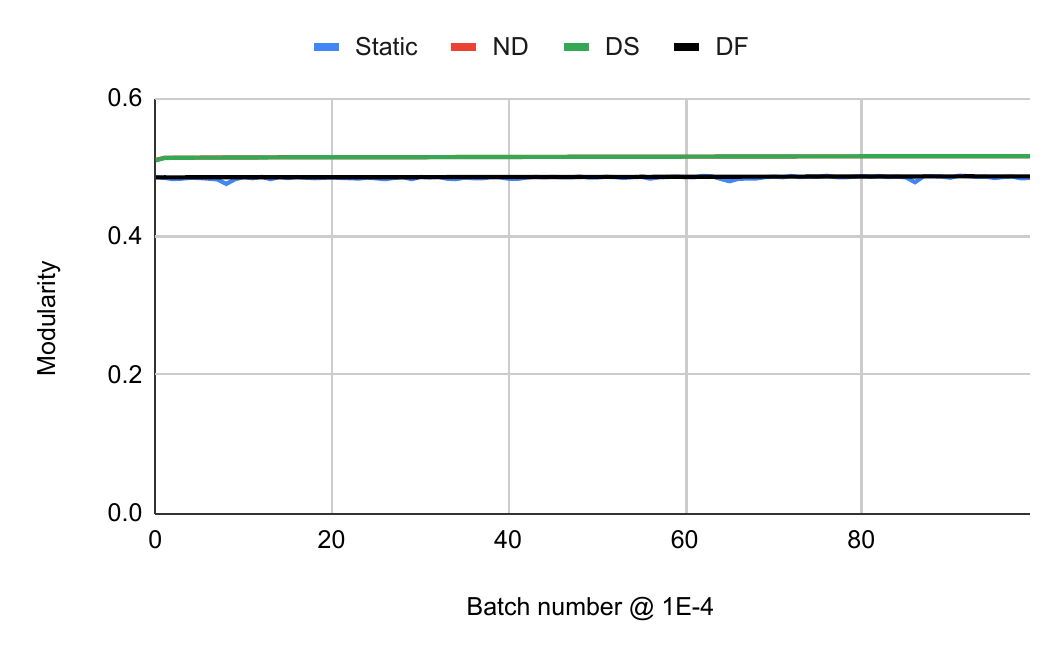}
  } \\[2ex]
  \subfigure[Runtime on consecutive batch updates of size $10^{-3}|E_T|$]{
    \label{fig:temporal-wiki-talk-temporal--runtime3}
    \includegraphics[width=0.48\linewidth]{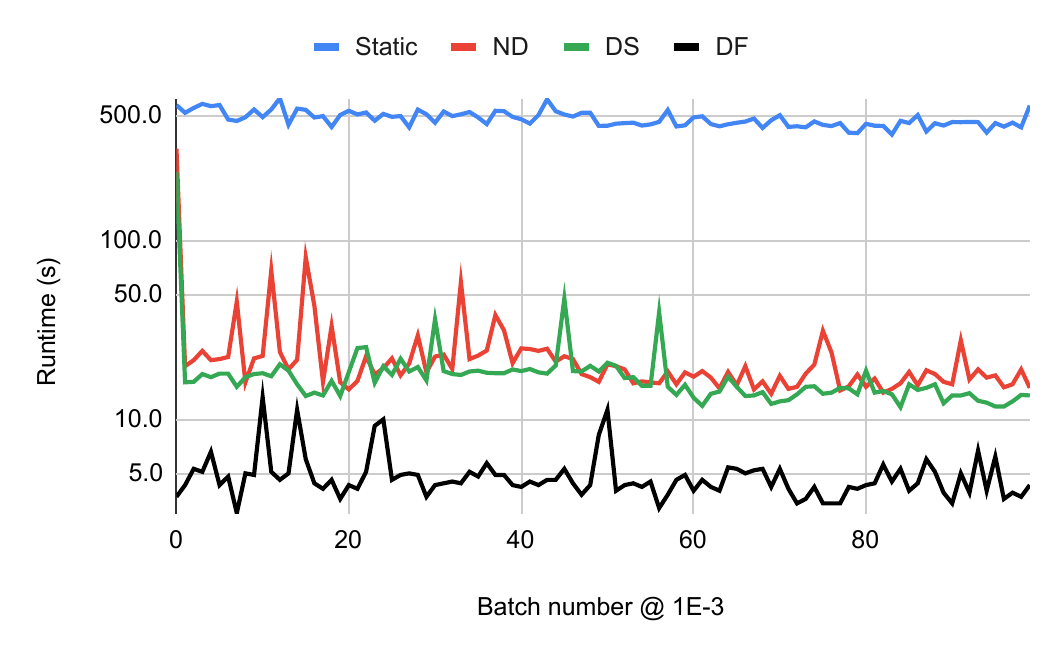}
  }
  \subfigure[Modularity of communities obtained on consecutive batch updates of size $10^{-3}|E_T|$]{
    \label{fig:temporal-wiki-talk-temporal--modularity3}
    \includegraphics[width=0.48\linewidth]{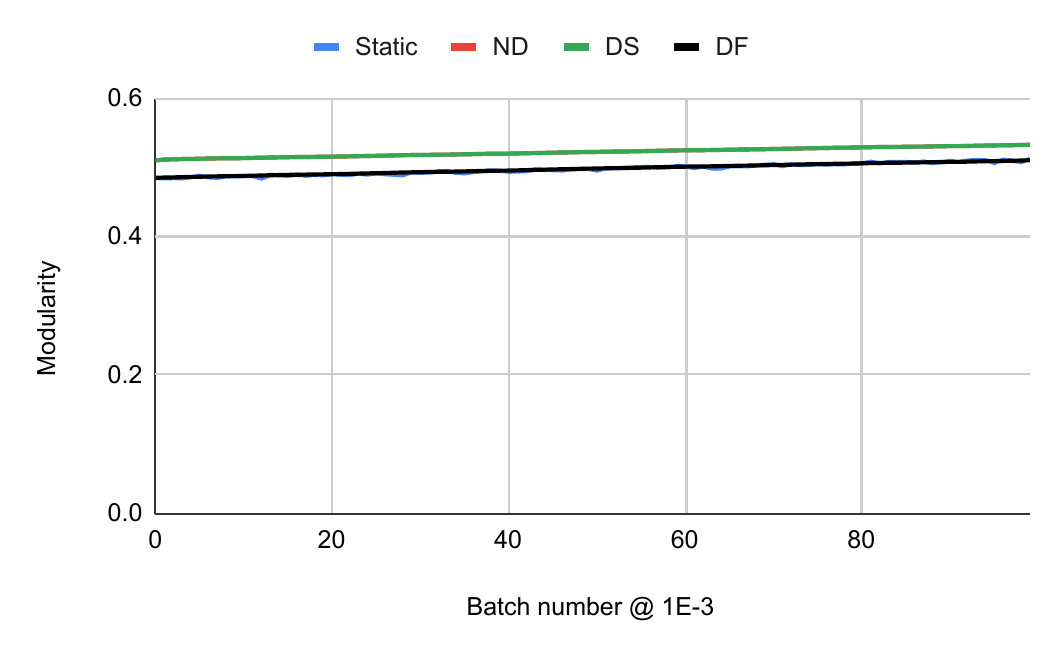}
  } \\[-2ex]
  \caption{Runtime and Modularity of communities obtained with \textit{Static}, \textit{Naive-dynamic (ND)}, \textit{Delta-screening (DS)}, and \textit{Dynamic Frontier (DF)} Louvain on the \textit{wiki-talk-temporal} dynamic graph. The size of batch updates range from $10^{-5}|E_T|$ to $10^{-3}|E_T|$.}
  \label{fig:temporal-wiki-talk-temporal}
\end{figure*}

\begin{figure*}[!hbt]
  \centering
  \subfigure[Runtime on consecutive batch updates of size $10^{-5}|E_T|$]{
    \label{fig:temporal-sx-stackoverflow--runtime5}
    \includegraphics[width=0.48\linewidth]{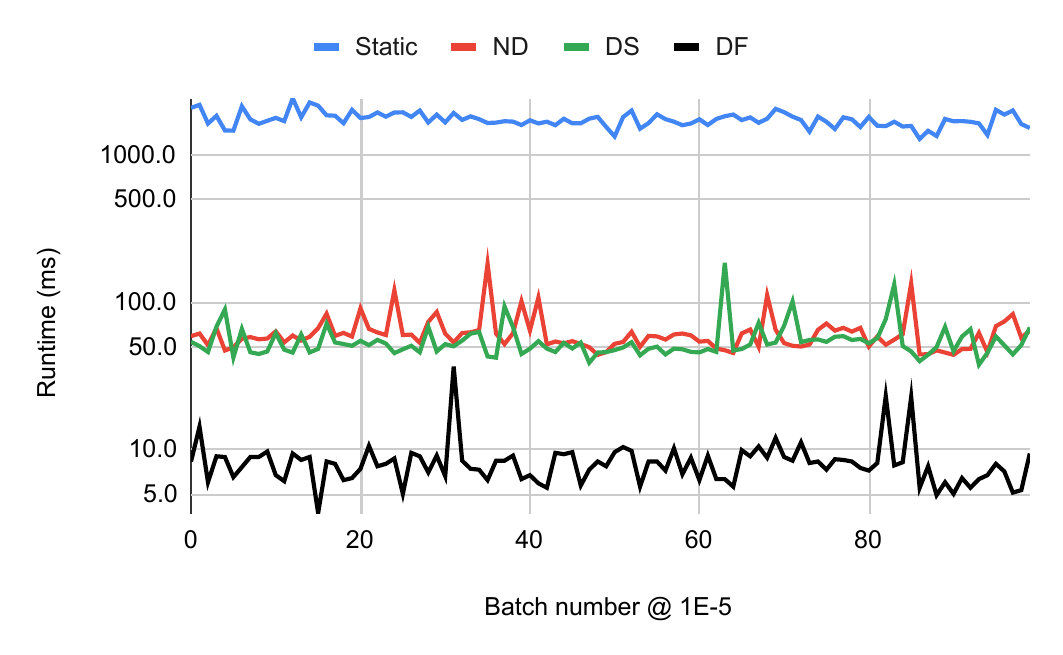}
  }
  \subfigure[Modularity of communities obtained on consecutive batch updates of size $10^{-5}|E_T|$]{
    \label{fig:temporal-sx-stackoverflow--modularity5}
    \includegraphics[width=0.48\linewidth]{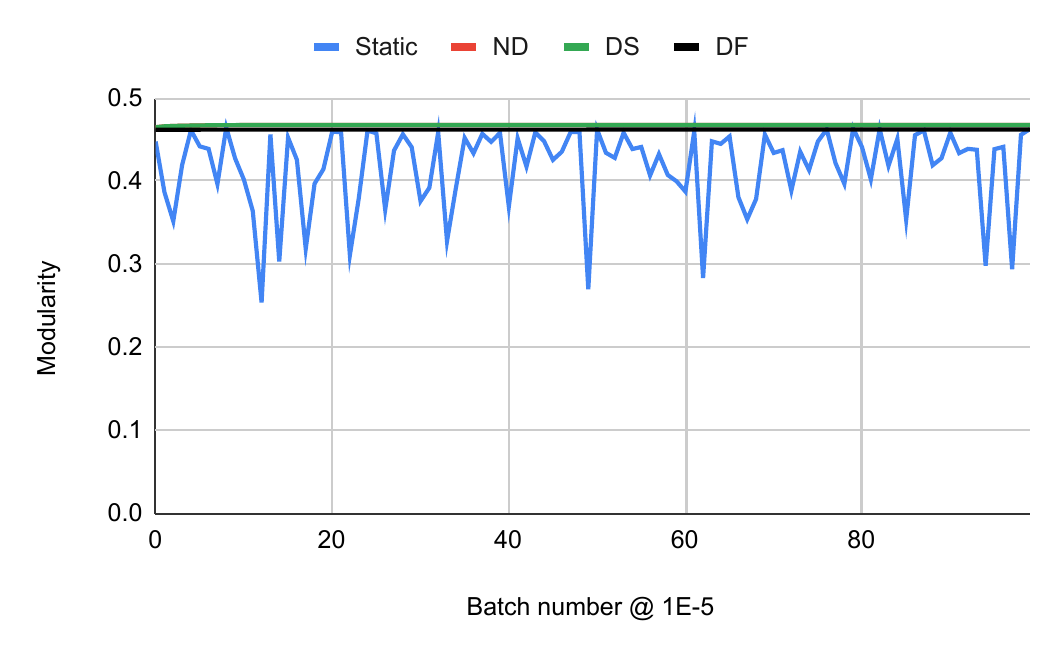}
  } \\[2ex]
  \subfigure[Runtime on consecutive batch updates of size $10^{-4}|E_T|$]{
    \label{fig:temporal-sx-stackoverflow--runtime4}
    \includegraphics[width=0.48\linewidth]{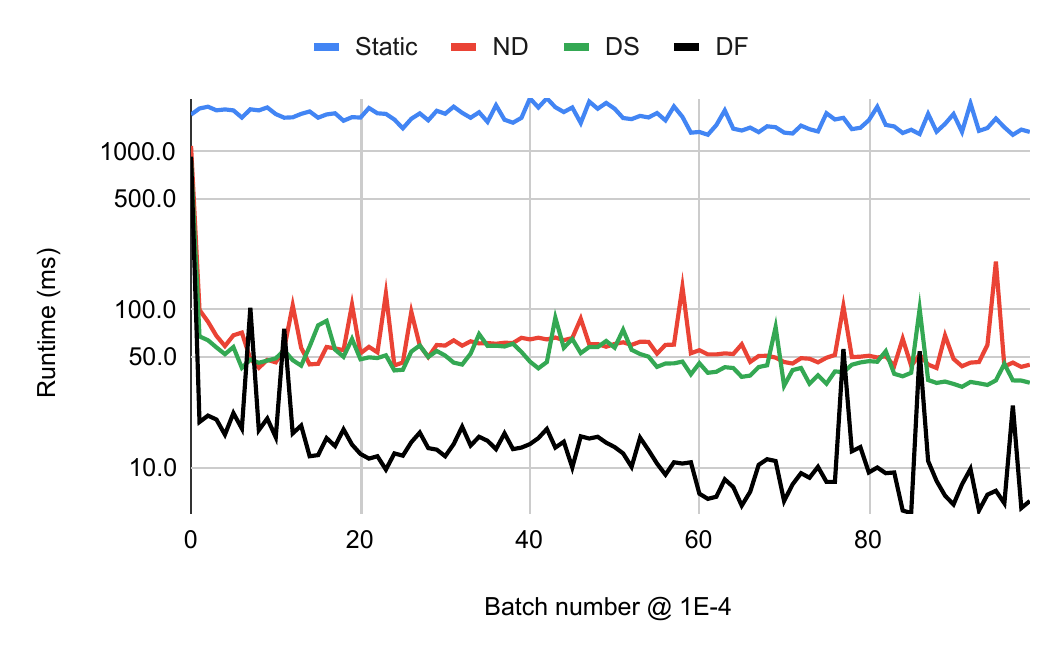}
  }
  \subfigure[Modularity of communities obtained on consecutive batch updates of size $10^{-4}|E_T|$]{
    \label{fig:temporal-sx-stackoverflow--modularity4}
    \includegraphics[width=0.48\linewidth]{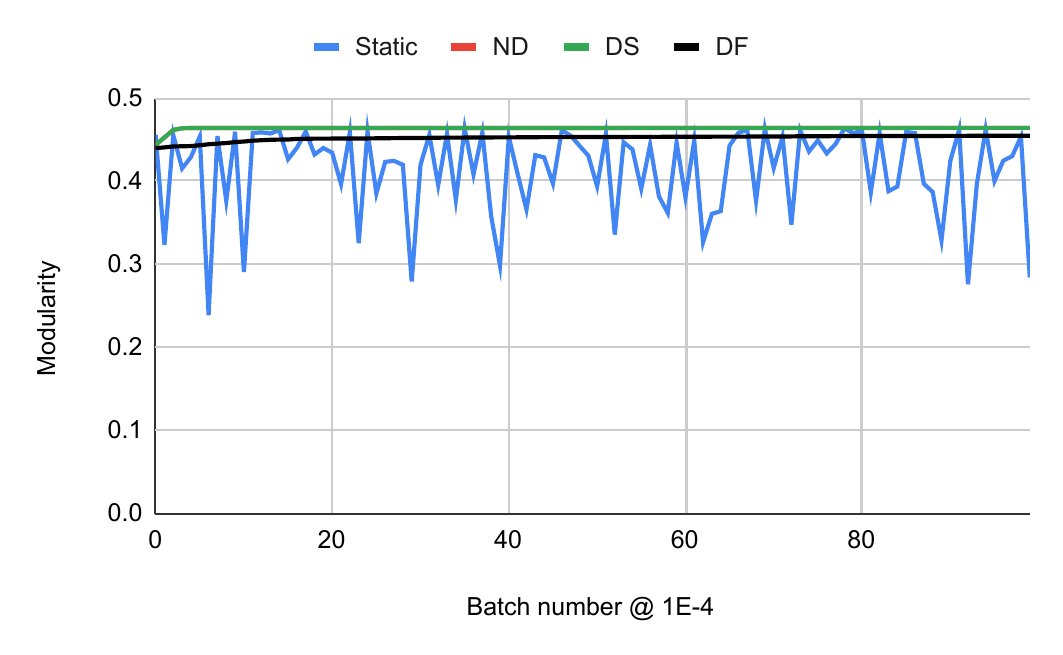}
  } \\[2ex]
  \subfigure[Runtime on consecutive batch updates of size $10^{-3}|E_T|$]{
    \label{fig:temporal-sx-stackoverflow--runtime3}
    \includegraphics[width=0.48\linewidth]{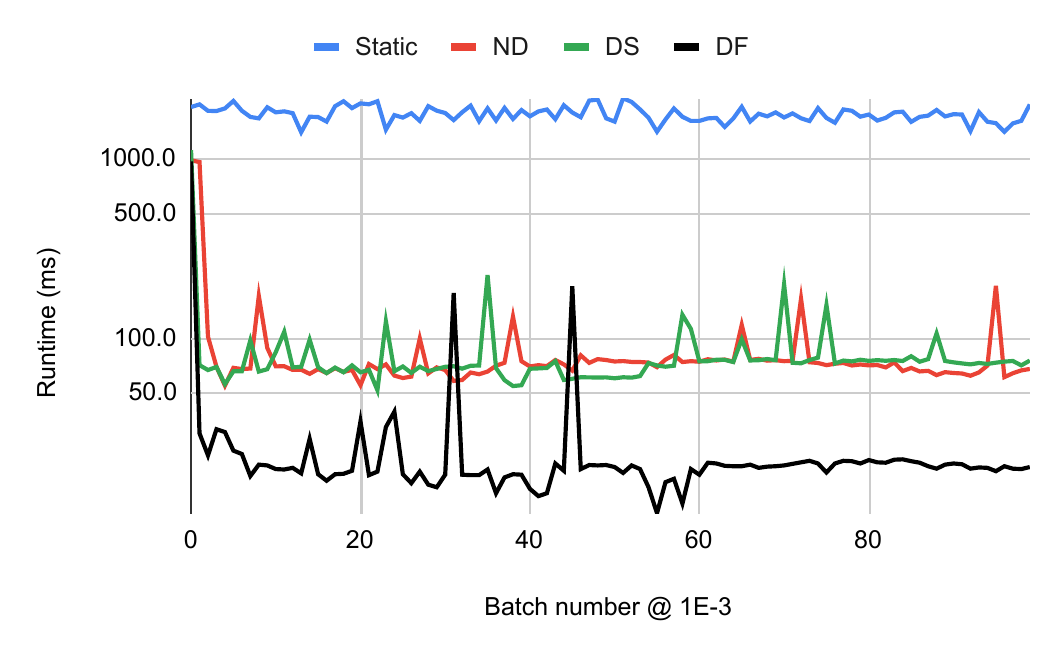}
  }
  \subfigure[Modularity of communities obtained on consecutive batch updates of size $10^{-3}|E_T|$]{
    \label{fig:temporal-sx-stackoverflow--modularity3}
    \includegraphics[width=0.48\linewidth]{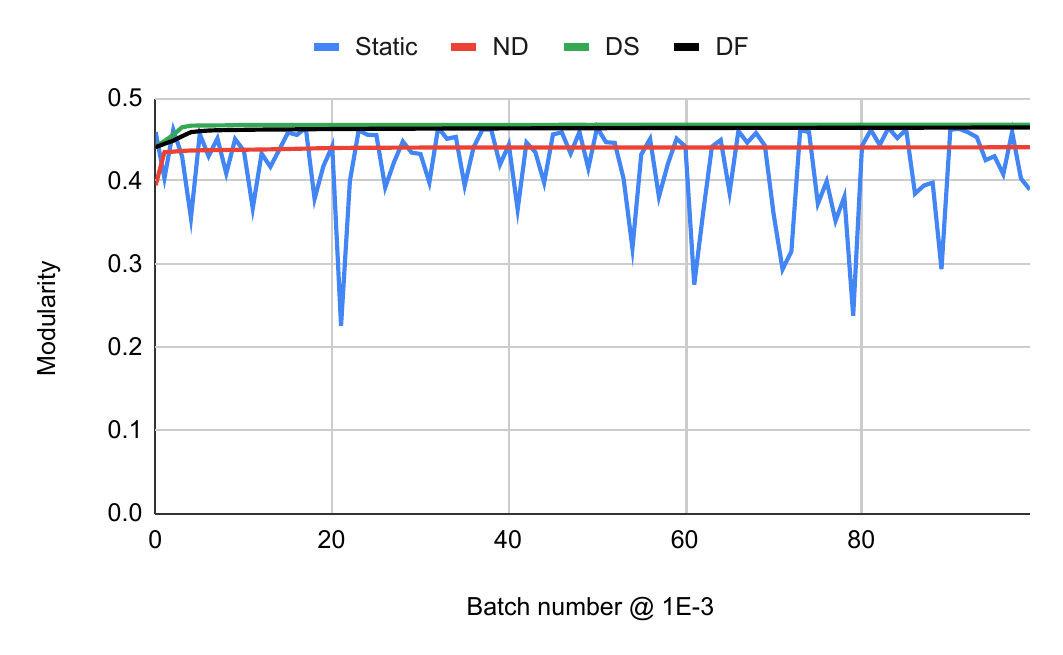}
  } \\[-2ex]
  \caption{Runtime and Modularity of communities obtained with \textit{Static}, \textit{Naive-dynamic (ND)}, \textit{Delta-screening (DS)}, and \textit{Dynamic Frontier (DF)} Louvain on the \textit{sx-stackoverflow} dynamic graph. The size of batch updates range from $10^{-5}|E_T|$ to $10^{-3}|E_T|$.}
  \label{fig:temporal-sx-stackoverflow}
\end{figure*}

\end{document}